\documentclass[preprint,12pt]{elsarticle}

\usepackage[T1]{fontenc}
\usepackage[utf8]{inputenc}

\DeclareUnicodeCharacter{00A0}{ }   
\DeclareUnicodeCharacter{2010}{-}   
\DeclareUnicodeCharacter{2011}{-}   
\DeclareUnicodeCharacter{2012}{-}   
\DeclareUnicodeCharacter{2013}{--}  
\DeclareUnicodeCharacter{2014}{---} 
\DeclareUnicodeCharacter{2018}{`}   
\DeclareUnicodeCharacter{2019}{'}   
\DeclareUnicodeCharacter{201C}{``}  
\DeclareUnicodeCharacter{201D}{''}  
\DeclareUnicodeCharacter{2212}{-}   

\usepackage{lmodern}

\usepackage{amsmath,amssymb,amsfonts,amsthm}
\usepackage{mathtools}

\usepackage{graphicx}
\usepackage{subcaption}
\usepackage{booktabs,multirow,array}
\usepackage[table]{xcolor}
\usepackage{float}
\usepackage{stfloats}
\usepackage{wrapfig}
\usepackage{tikz}

\usepackage[ruled,vlined]{algorithm2e}

\usepackage{microtype}
\usepackage{balance}
\usepackage{url}
\usepackage{optidef}
\usepackage{verbatim}
\usepackage{multicol}

\newtheorem{remark}{Remark}
\newtheorem{assumption}{Assumption}

\newtheorem{proposition}{Proposition}

\newcommand{\AlltoAll}{All\nobreakdash-to\nobreakdash-All}

\usepackage{hyperref}
\hypersetup{
  colorlinks=true,
  linkcolor=blue,
  citecolor=blue,
  urlcolor=blue,
  breaklinks=true
}

\usepackage[nameinlink,noabbrev]{cleveref}
\crefname{equation}{Eq.}{Eqs.}
\Crefname{equation}{Equation}{Equations}

\begin{document}
\begin{frontmatter}

\title{Federated Sinkhorn: Communication, Performance, and Privacy for Distributed Entropic Optimal Transport}

\author[1]{Jeremy Kulcsar\corref{cor1}}
\ead{jeremy.kulcsar@hsbc.com.hk}

\author[4]{Vyacheslav Kungurtsev}
\ead{kunguvya@fel.cvut.cz}

\author[2,4,5]{Georgios Korpas}
\ead{georgios.korpas@hsbc.com.sg}

\author[3]{Giulio Giaconi}
\ead{giulio.giaconi@hsbc.com}

\author[3]{William Shoosmith}
\ead{william.shoosmith@hsbc.com}

\cortext[cor1]{Corresponding author.}

\address[1]{Emerging Technologies, Innovation and Ventures, HSBC, 1 Queen's Road Central, Central, Hong Kong SAR, China}
\address[2]{Emerging Technologies, Innovation and Ventures, HSBC, 20 Pasir Panjang Road, 117440 Singapore}
\address[3]{Emerging Technologies, Innovation and Ventures, HSBC, 8 Canada Square, Canary Wharf, London E14 5HQ, United Kingdom}
\address[4]{Department of Computer Science, Faculty of Electrical Engineering, Czech Technical University in Prague, Karlovo n\'am\v{e}st\'{\i} 13, 120 00 Prague 2, Czech Republic}
\address[5]{Archimedes Research Unit on AI, Data Science and Algorithms, Athena Research Center, 1 Artemidos Street, 15125 Marousi, Greece}

\begin{abstract}
We study distributed Sinkhorn iterations for entropy-regularized optimal transport when the Gibbs kernel operator is \emph{row-partitioned} across $c$ workers and cannot be centralized. We present \emph{Federated Sinkhorn}, two exact synchronous protocols that exchange only scaling-vector slices: (i) an All-to-All scheme implemented by \textsc{Allgather}, and (ii) a Star (parameter-server) scheme implemented by client$\rightarrow$server sends and server$\rightarrow$client broadcasts. For both, we derive closed-form per-iteration compute, communication, and memory costs under an $\alpha$--$\beta$ latency--bandwidth model, and show that the distributed iterates match centralized Sinkhorn under standard positivity assumptions. Multi-node CPU/GPU experiments validate the model and show that repeated global scaling exchange quickly becomes the dominant bottleneck as $c$ increases. We also report an optional bounded-delay asynchronous schedule and an optional privacy measurement layer for communicated log-scalings.
\end{abstract}

\begin{keyword}
optimal transport \sep Sinkhorn iterations \sep federated computing \sep distributed optimization \sep asynchronous communication \sep differential privacy \sep MPI
\end{keyword}

\end{frontmatter}


\section{Introduction}

Optimal Transport (OT) compares probability distributions and is widely used in machine learning, economics, and the physical sciences~\cite{villani2009optimal,peyre2019computational}. In discrete form, OT is a large linear program; entropic regularization yields a strictly convex objective solvable by fast scaling iterations~\cite{cuturi2013sinkhorn}:
\begin{equation}
\min_{P\in\mathbb{R}^{n\times n}_+}\;
\langle P,C\rangle
+\varepsilon_{\scriptscriptstyle\mathrm{OT}}
\sum_{i,j=1}^{n} P_{ij}(\log P_{ij}-1)
\quad\text{s.t.}\quad
P\mathbf{1}=a,\;\;P^\top\mathbf{1}=b.
\label{eq:otprob}
\end{equation}
Here, $C\in\mathbb{R}^{n\times n}$ is the cost matrix and $a,b\in\mathbb{R}^n_+$ are prescribed marginals with equal mass. Writing $K=\exp(-C/\varepsilon_{\scriptscriptstyle\mathrm{OT}})$, the optimum admits the diagonal scaling form
\[
P^\star=\operatorname{diag}(u)\,K\,\operatorname{diag}(v),
\qquad
u\leftarrow a\oslash(Kv),\;\;
v\leftarrow b\oslash(K^\top u),
\]
i.e., the Sinkhorn--Knopp iteration~\cite{cuturi2013sinkhorn,sinkhorn1967concerning}. In practice, stabilized log-domain updates are commonly used to avoid numerical underflow for small $\varepsilon_{\scriptscriptstyle\mathrm{OT}}$.

\subsection{Motivation and deployment topologies}
Sinkhorn iterations maintain only low-dimensional state (the scalings $u,v$) while their dominant cost is repeated application of the kernel operators $Kv$ and $K^\top u$. When the cost/kernel operator cannot be pooled on one machine (due to size, administrative boundaries, or policy), a natural alternative is to keep the operator sharded and exchange only scaling-vector state to enforce global marginals.

Communication topology is then a first-order design choice. In HPC settings with routable peer-to-peer connectivity, scaling slices can be exchanged efficiently via collectives (All-to-All). In cross-silo deployments where direct client-to-client connections are impractical (firewalls/NAT/operational constraints), Star-Network (parameter-server) orchestration is often the only feasible option. In both cases, raw cost data remain local; only scaling state is communicated.

\subsection{System setting and scope}
We consider $c$ workers connected by an MPI-style runtime. Worker $j$ owns an index set $I_j$ (typically $|I_j|=m$ and $n=cm$), the corresponding marginal slices $(a_{I_j}, b_{I_j})$, and access to the kernel rows $K_{I_j,:}$ (materialized or implicit). Each worker can apply both $x\mapsto K_{I_j,:}x$ and the restricted transpose action $x\mapsto (K_{:,I_j})^\top x$ for the same index set (e.g., when $K$ is symmetric/shared-support or stored in a layout exposing the needed tiles). Communication exchanges only scaling slices (and a few scalars); raw samples/cost data remain local.

We focus on row-partitioned entropic OT. Fully cross-silo cost evaluation where each $C_{ij}$ depends jointly on private features held by different parties without a trusted evaluation service or secure computation is out of scope.

\subsection{Contributions}
\begin{itemize}
\item \textbf{Federated Sinkhorn topologies.} Exact synchronous Sinkhorn protocols for row-partitioned kernels under All-to-All and Star orchestration, plus an optional bounded-delay asynchronous schedule.
\item \textbf{Topology-aware cost model.} Closed-form per-iteration \\ compute/communication/memory costs under an $\alpha$--$\beta$ latency--bandwidth model and calibration.
\item \textbf{Implementation and evaluation.} Multi-node CPU/GPU experiments validating the model and identifying communication bottlenecks; optional privacy measurement for communicated log-scalings is reported in the appendix.
\end{itemize}

\section{Background and Related Work}

\subsection{Entropic OT and Sinkhorn.}
Entropy-regularized OT yields the scaling form $P^\star=\mathrm{diag}(u)K\mathrm{diag}(v)$ and can be solved by alternating updates of the positive scaling vectors $u$ and $v$ (Sinkhorn--Knopp) \cite{cuturi2013sinkhorn,sinkhorn1967concerning,peyre2019computational}.
For small $\varepsilon_{\mathrm{OT}}$, stabilized log-domain implementations (and related continuation heuristics) are standard to prevent numerical underflow \cite{peyre2019computational,schmitzer2019stabilized}.

\subsection{Distributed Sinkhorn and scalable OT.}
Each Sinkhorn iteration is dominated by kernel applications $Kv$ and $K^\top u$ \cite{cuturi2013sinkhorn,peyre2019computational}.
In distributed-memory settings the operator can be partitioned across workers, but the scalings are global state; repeated exchange and synchronization of scaling information (or matvec outputs) is therefore a primary systems bottleneck \cite{bertsekas1989parallel}.
From an algorithms viewpoint, Sinkhorn is a special case of matrix scaling/balancing, connecting to classical work on scaling of nonnegative matrices \cite{sinkhorn1967concerning,franklin1989scaling,berman1994nonnegative}.
Complementary scalability approaches develop stochastic and approximate OT/Sinkhorn methods for large-scale problems \cite{genevay2016stochastic,altschuler2017nearlinear,peyre2019computational}.

\subsection{Topology, asynchrony, and privacy.}
Centralized (parameter-server/star) and decentralized schemes trade simpler reasoning under synchrony for improved utilization under asynchrony, at the cost of stale information and more delicate convergence behavior \cite{bertsekas1989parallel,recht2011hogwild,lian2018asynchronous}.
Bounded-delay (stale-synchronous) models provide a tractable middle ground when staleness can be uniformly bounded \cite{bertsekas1989parallel}.
Federated learning similarly emphasizes data locality and communication efficiency under practical connectivity constraints \cite{mcmahan2017communication,konecny2016federated}.
Privacy of communicated updates is commonly addressed with secure aggregation \cite{bonawitz2017practical} and/or differential privacy mechanisms with explicit composition accounting (e.g., R\'enyi-DP) \cite{dwork2014algorithmic,mironov2017renyi}; such tools have also been applied in settings using Sinkhorn-based objectives, e.g., Sinkhorn divergences in private generative modeling \cite{cao2021don}.

\subsection{Our focus.}
We quantify how All-to-All vs.\ Star orchestration changes per-iteration costs, provide a simple $\alpha$--$\beta$ model to predict when communication dominates, and empirically validate the resulting scaling behavior.
Convergence details for a bounded-delay asynchronous variant and optional privacy measurement are provided in the appendix.

\section{Federated Sinkhorn Algorithms}
\label{sec:alg}

\subsection{Problem structure and data layout}
\label{subsec:pb_struct}

We consider entropy‑regularized OT in a setting where data are partitioned across $c$ clients that jointly solve \cref{eq:otprob} without sharing raw samples. We focus on the common case $n = c m$, with client $j$ owning a local index set $\mathcal{I}_j\subset\{1,\dots,n\}$ of size $m$, local marginals $a_j,b_j\in\mathbb{R}_+^m$, and the corresponding row block $C_{\mathcal{I}_j,:}$ of the cost matrix $C\in\mathbb{R}^{n\times n}$. Stacking these blocks gives the global histograms $a=[a_1;\ldots;a_c]$ and $b=[b_1;\ldots;b_c]$.\footnote{We write $x_{\mathcal{I}_j}$ for the subvector of $x\in\mathbb{R}^n$ indexed by $\mathcal{I}_j$ and $X_{\mathcal{I}_j,:}$ (resp.\ $X_{:,\mathcal{I}_j}$) for the row (resp.\ column) block of a matrix $X$.}

\begin{wrapfigure}{r}{0.50\textwidth}
  \centering
  \includegraphics[
    width=0.3\textwidth
  ]{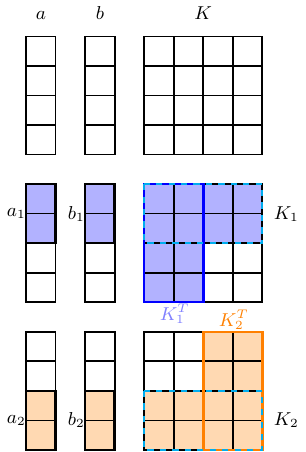}
  \caption{Partition with $n=cm$: client $j$ owns indices $\mathcal{I}_j$, local marginals $a_j,b_j$, and rows $C_{\mathcal{I}_j,:}$ (hence $K_{\mathcal{I}_j,:}$). Clients compute $(Kv)_{\mathcal{I}_j}$ and $(K^\top u)_{\mathcal{I}_j}$ locally, exchanging only slices of the scaling vectors.}
  \vspace{2.5em}
  \label{fig:sliced_inputs}
\end{wrapfigure}

We work with the Gibbs kernel $K=\exp(-C/\varepsilon_{\mathrm{OT}})$ but do not assume it is explicitly materialized. Client $j$ must be able to apply the local operator $x\mapsto K_{I_j,:}x$ and the matching restricted transpose action $x\mapsto (K_{:,I_j})^\top x$ (see Assumption~\ref{assump:transpose}). Each client maintains only local scaling slices $(u_{I_j}, v_{I_j})$ and the corresponding matvec buffers
\[
q_{I_j} := (Kv)_{I_j},\qquad r_{I_j} := (K^\top u)_{I_j}.
\]
A federated Sinkhorn step is then purely local elementwise rescaling,
\[
u_{I_j}\leftarrow a_{I_j}\oslash q_{I_j},\qquad v_{I_j}\leftarrow b_{I_j}\oslash r_{I_j},
\]
with communication used only to keep the matvecs consistent across shards (Figure~\ref{fig:sliced_inputs}).

\begin{remark}[Local steps before communication]
    Unless stated otherwise we synchronize every iteration ($w=1$); the effect of local steps is reported in ~\ref{app:local_steps}.
\end{remark}

\begin{remark}[Theoretical foundations]
    A section on convergence properties discussing propositions along with their proofs can be found in~\ref{sec:con}.
\end{remark}

\begin{assumption}[Local transpose access]\label{assump:transpose}
For its index set $I_j$, client $j$ can apply both $x\mapsto K_{I_j,:}x$ and the restricted transpose action $x\mapsto (K_{:,I_j})^\top x$ (equivalently $K^\top_{I_j,:}x$). This holds, for example, when $K$ is symmetric/shared-support or when the kernel/cost is stored in a layout exposing the needed tiles. Fully cross-silo cost evaluation requiring secure computation is out of scope.
\end{assumption}

\subsection{Synchronous variants}
\label{subsec:sync}

We now instantiate the federated updates under two synchronous orchestration patterns: a decentralized \AlltoAll\ topology where all clients play symmetric roles, and a centralized star topology with a distinguished coordinator. In the \AlltoAll\ case, clients keep their local kernel shards and apply matvecs locally. In the star case, the coordinator has access to the kernel operator (or can evaluate $Kv$ and $K^\top u$) and orchestrates broadcast + point-to-point send/recv while clients perform only local elementwise rescalings using their marginals. In both cases, clients exchange only scaling-vector slices (or matvec slices); raw samples/cost data remain local to their owners.

\begin{figure}[!htbp]
  \centering
  \begin{subfigure}[t]{0.49\linewidth}
    \centering
    \begin{tikzpicture}
      \node[anchor=south west, inner sep=0] (image) at (0,0)
        {\includegraphics[width=\linewidth]{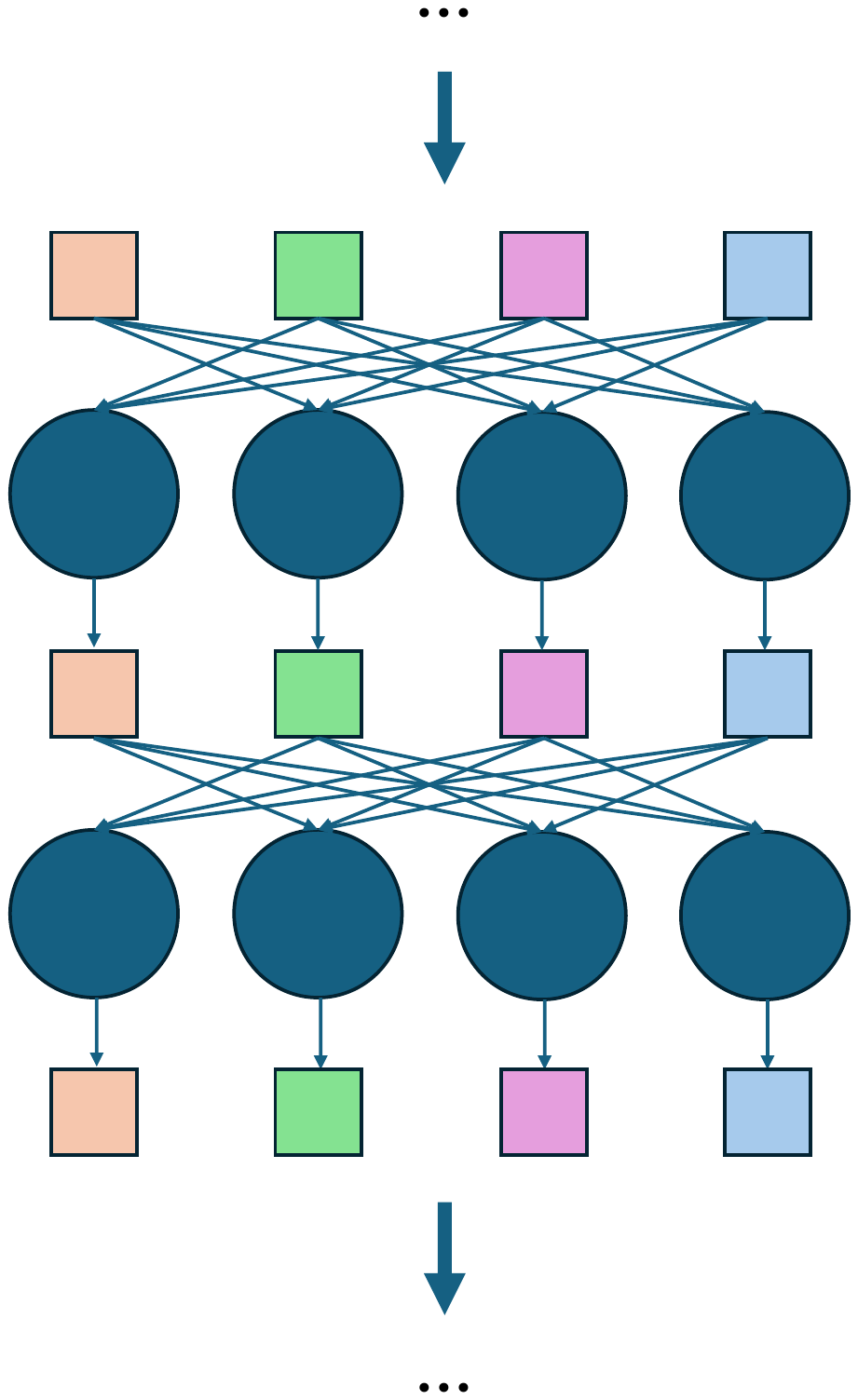}};
      \begin{scope}[x={(image.south east)}, y={(image.north west)}]
        \node[font=\bfseries\scriptsize, text=black] at (0.2155, 0.743) {$u_1$};
        \node[font=\bfseries\scriptsize, text=black] at (0.412, 0.743) {$u_2$};
        \node[font=\bfseries\scriptsize, text=black] at (0.605, 0.743) {$u_3$};
        \node[font=\bfseries\scriptsize, text=black] at (0.8, 0.743) {$u_4$};

        \node[font=\bfseries\scriptsize, text=white] at (0.2155, 0.612) {$b_1, K^T_1$};
        \node[font=\bfseries\scriptsize, text=white] at (0.412, 0.612) {$b_2, K^T_2$};
        \node[font=\bfseries\scriptsize, text=white] at (0.605, 0.612) {$b_3, K^T_3$};
        \node[font=\bfseries\scriptsize, text=white] at (0.8, 0.612) {$b_4, K^T_4$};

        \node[font=\bfseries\scriptsize, text=black] at (0.2155, 0.488) {$v_1$};
        \node[font=\bfseries\scriptsize, text=black] at (0.412, 0.488) {$v_2$};
        \node[font=\bfseries\scriptsize, text=black] at (0.605, 0.488) {$v_3$};
        \node[font=\bfseries\scriptsize, text=black] at (0.8, 0.488) {$v_4$};

        \node[font=\bfseries\scriptsize, text=white] at (0.2155, 0.356) {$a_1, K_1$};
        \node[font=\bfseries\scriptsize, text=white] at (0.412, 0.356) {$a_2, K_2$};
        \node[font=\bfseries\scriptsize, text=white] at (0.605, 0.356) {$a_3, K_3$};
        \node[font=\bfseries\scriptsize, text=white] at (0.8, 0.356) {$a_4, K_4$};

        \node[font=\bfseries\scriptsize, text=black] at (0.2155, 0.233) {$u_1$};
        \node[font=\bfseries\scriptsize, text=black] at (0.412, 0.233) {$u_2$};
        \node[font=\bfseries\scriptsize, text=black] at (0.605, 0.233) {$u_3$};
        \node[font=\bfseries\scriptsize, text=black] at (0.8, 0.233) {$u_4$};
      \end{scope}
    \end{tikzpicture}
    \caption{All-to-All process. Each node $i$ receives slices of $u$ from other nodes, concatenates them into a global $u$, uses its local $K^T_i$ and $b_i$ to compute $r_i=K^T_i u$ and $v_i=b_i / r_i$, and sends slice $v_i$ to other nodes. Then, each node does the same respective operations with $v$, $K_i$ and $a_i$.}
    \label{fig:All-to-All}
  \end{subfigure}\hfill
  \begin{subfigure}[t]{0.45\linewidth}
    \centering
    \begin{tikzpicture}
      \node[anchor=south west, inner sep=0] (image) at (0,0)
        {\includegraphics[width=1.2\linewidth]{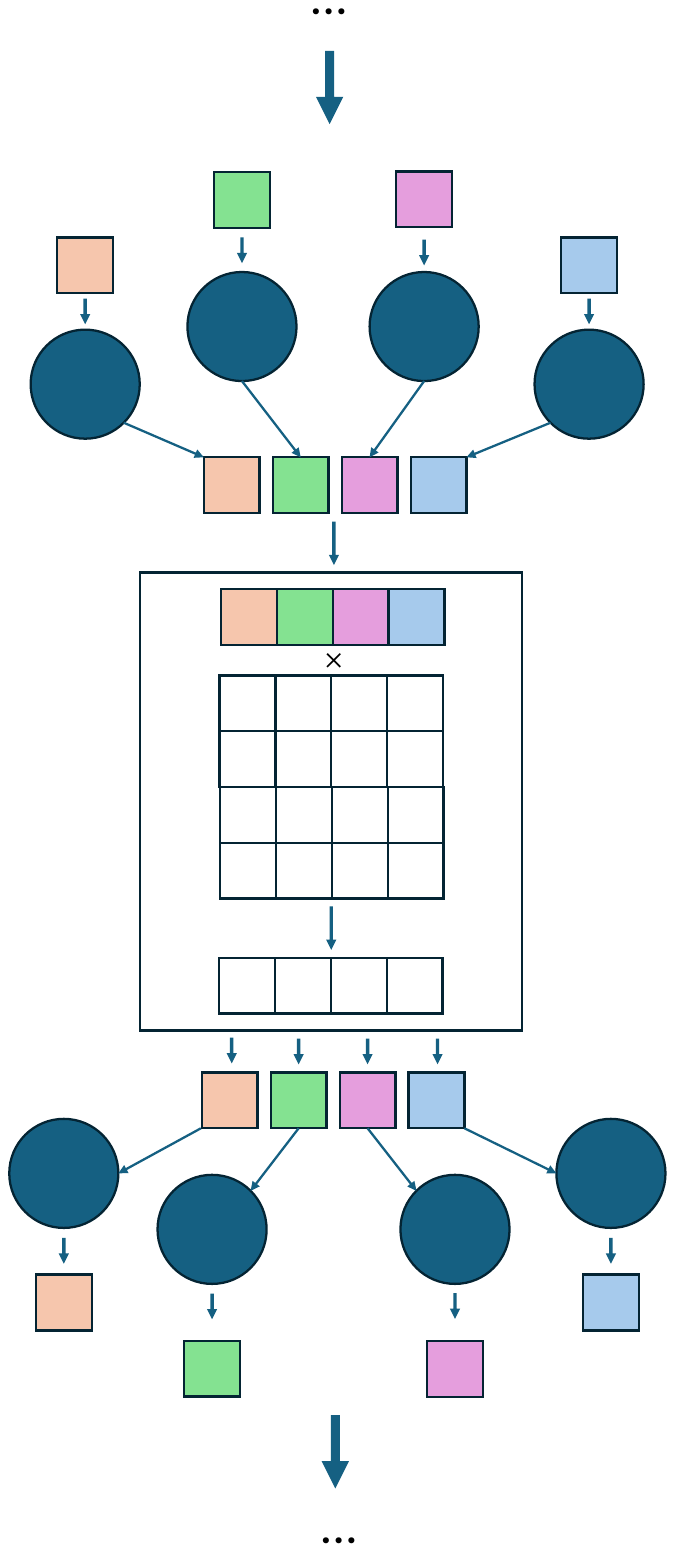}};
      \begin{scope}[x={(image.south east)}, y={(image.north west)}]
        \node[font=\bfseries, text=black] at (0.307, 0.787) {\fontsize{7}{10}\selectfont $r_1$};
        \node[font=\bfseries, text=black] at (0.432, 0.8235) {\fontsize{7}{10}\selectfont $r_2$};
        \node[font=\bfseries, text=black] at (0.5785, 0.8235) {\fontsize{7}{10}\selectfont $r_3$};
        \node[font=\bfseries, text=black] at (0.712, 0.787) {\fontsize{7}{10}\selectfont $r_4$};

        \node[font=\bfseries, text=black] at (0.425, 0.662) {\fontsize{7}{10}\selectfont $u_1$};
        \node[font=\bfseries, text=black] at (0.48, 0.662) {\fontsize{7}{10}\selectfont $u_2$};
        \node[font=\bfseries, text=black] at (0.535, 0.662) {\fontsize{7}{10}\selectfont $u_3$};
        \node[font=\bfseries, text=black] at (0.59, 0.662) {\fontsize{7}{10}\selectfont $u_4$};

        \node[font=\bfseries, text=black] at (0.62, 0.585) {\fontsize{7}{10}\selectfont $u$};
        \node[font=\bfseries, text=black] at (0.62, 0.488) {\fontsize{7}{10}\selectfont $K$};
        \node[font=\bfseries, text=black] at (0.62, 0.375) {\fontsize{7}{10}\selectfont $q$};

        \node[font=\bfseries, text=black] at (0.425, 0.311) {\fontsize{7}{10}\selectfont $q_1$};
        \node[font=\bfseries, text=black] at (0.48, 0.311) {\fontsize{7}{10}\selectfont $q_2$};
        \node[font=\bfseries, text=black] at (0.535, 0.311) {\fontsize{7}{10}\selectfont $q_3$};
        \node[font=\bfseries, text=black] at (0.59, 0.311) {\fontsize{7}{10}\selectfont $q_4$};

        \node[font=\bfseries, text=black] at (0.289, 0.196) {\fontsize{7}{10}\selectfont $v_1$};
        \node[font=\bfseries, text=black] at (0.407, 0.158) {\fontsize{7}{10}\selectfont $v_2$};
        \node[font=\bfseries, text=black] at (0.603, 0.158) {\fontsize{7}{10}\selectfont $v_3$};
        \node[font=\bfseries, text=black] at (0.728, 0.196) {\fontsize{7}{10}\selectfont $v_4$};
      \end{scope}
    \end{tikzpicture}
    \caption{Star-Network process. Clients send their local scaling slices to the server (send/recv). The server uses the global kernel $K$ to compute $q = Kv$ and broadcasts $q$ to all clients; each client then uses only its local slice $q_{I_j}$ to update $u_{I_j}$ and sends that slice back to the server. The same pattern is repeated with $r = K^\top u$ to update $v_{I_j}$.}
    \label{fig:star-network}
  \end{subfigure}

  \caption{Two distributed update schemes shown side by side.}
  \label{fig:schemes-side-by-side}
\end{figure}

\subsubsection{Synchronous All-to-All}
\label{subsubsec:topo1}

In the All‑to‑All variant, all clients play identical roles and coordinate via collectives. At each iteration, every client $j$ holds its local scalings $u_{\mathcal{I}_j},v_{\mathcal{I}_j}$ and a current global view of $(u,v)$ obtained from the previous collective. The synchronous update consists of:
(i) an optional collective to refresh $v$,
(ii) a local matvec $q_{\mathcal{I}_j} \gets K_{\mathcal{I}_j,:} v$ and row scaling $u_{\mathcal{I}_j} \gets a_j \oslash q_{\mathcal{I}_j}$,
(iii) an optional collective to refresh $u$, and
(iv) a local matvec $r_{\mathcal{I}_j} \gets (K_{:,\mathcal{I}_j})^\top u$ and column scaling $v_{\mathcal{I}_j} \gets b_j \oslash r_{\mathcal{I}_j}$.

Algorithm~\ref{alg:fedsinksyncall} summarizes the procedure. For a strictly positive $K$, the resulting global iterates $(u,v)$ coincide with those of centralized Sinkhorn, and thus inherit its linear convergence (Proposition~\ref{prop:sync}).

\begin{algorithm}[!htbp]
\DontPrintSemicolon
\caption{Synchronous Federated Sinkhorn (All-to-All)}
\label{alg:fedsinksyncall}
\small
\KwIn{Client $j$ holds $a_{j},b_{j},C_{\mathcal{I}_j,:}$}
\KwOut{$u,v\in\mathbb{R}_{++}^n$ and $P=\mathrm{diag}(u)K\mathrm{diag}(v)$}
Initialize $u\gets \mathbf{1}$, $v\gets \mathbf{1}$ (identical at all clients)\;
\For{$k=1,2,\dots,\texttt{IterMax}$}{
  Compute $q_{\mathcal{I}_j}\gets K_{\mathcal{I}_j,:}\,v$\;
  Update $u_{\mathcal{I}_j}\gets a_{j} \oslash q_{\mathcal{I}_j}$\;
  \textbf{Allgather} $u_{\mathcal{I}_j}$ to form global $u$ at each client\;
  Compute $r_{\mathcal{I}_j}\gets (K_{:,\mathcal{I}_j})^\top u$\;
  Update $v_{\mathcal{I}_j}\gets b_{j} \oslash r_{\mathcal{I}_j}$\;
  \textbf{Allgather} $v_{\mathcal{I}_j}$ to form global $v$ at each client\;
}
\end{algorithm}

\subsubsection{Synchronous Star-Network}
\label{subsubsec:topo2}

In the star topology that we name Star-Network, a coordinator stores (or can evaluate) the kernel operator and maintains global $(u,v)$ (cf. Fig.~\ref{fig:star-network}). Each iteration has two rounds: (i) clients send their current $v_{I_j}$, the server computes $q=Kv$ and broadcasts $q$, and each client updates $u_{I_j}=a_{I_j}\oslash q_{I_j}$; (ii) clients send $u_{I_j}$, the server computes $r=K^\top u$ and broadcasts $r$, and clients update $v_{I_j}=b_{I_j}\oslash r_{I_j}$. This reproduces centralized Sinkhorn iterates exactly under standard positivity assumptions. Full pseudocode can be found in~\ref{app:sync_start_netowkr_pseudo}.

\subsection{Asynchronous All-to-All}
\label{subsec:async}

We also consider a barrier-free variant where each client updates using its most recent cached slices of $(u,v)$ and periodically exchanges updated slices via non-blocking point-to-point messages with a relaxed (damped) scaling update.

Concretely, in the centralized view the damped updates read
\[
u^{(t+1)} = (1-\eta) u^{(t)} + \eta\, a \oslash (K v^{(t)}),
\qquad
v^{(t+1)} = (1-\eta) v^{(t)} + \eta\, b \oslash (K^{\top} u^{(t+1)}),
\]

Under bounded staleness and strictly positive $K$, the undamped updates ($\eta=1$) converge to the Sinkhorn scaling solution (~\ref{sec:con}). In practice we optionally under-relax ($\eta<1$) to reduce oscillations under stale reads (cf. Section~\ref{subsec:async_exp}). Detailed pseudocode, delay statistics, and robustness sweeps are deferred to~\ref{app:async_details}.

\begin{remark}[Geometric damping]
    Besides arithmetic under relaxation in the value domain, one can also consider geometric damping of the multiplicative scalings, e.g.
    \[
    u^{t+1} \leftarrow (u^{t})^{1-\eta}\odot\bigl(a\oslash(Kv^{t})\bigr)^{\eta},\qquad
    v^{t+1} \leftarrow (v^{t})^{1-\eta}\odot\bigl(b\oslash(K^{\top}u^{t+1})\bigr)^{\eta},
    \]
    which corresponds to linear interpolation in log scalings. This preserves positivity and is more naturally aligned with the projective geometry underlying Hilbert metric analyses. We do not evaluate this variant in our experiments and leave it for future work.
\end{remark}

\section{Communication, compute, and memory model}
\label{sec:comm-comp-model}

We develop a performance model that decomposes per-iteration cost into (i) local operator time, (ii) communication, and (iii) memory footprint. The model is intentionally coarse (not cycle-accurate): it is calibrated from a small set of microbenchmarks and used to interpret the scaling trends in Section~\ref{sec:experiments}.

\subsection{Local computation cost}
\label{subsec:comp-cost}

On client~$j$, one Sinkhorn iteration applies two kernel operators and rescales:
\[
q_{\mathcal{I}_j} \leftarrow K_{\mathcal{I}_j,:} v,
\qquad
u_{\mathcal{I}_j} \leftarrow a_j \oslash q_{\mathcal{I}_j},
\qquad
r_{\mathcal{I}_j} \leftarrow (K_{:,\mathcal{I}_j})^\top u,
\qquad
v_{\mathcal{I}_j} \leftarrow b_j \oslash r_{\mathcal{I}_j}.
\]

For a dense kernel, each local matvec touches $mn$ entries. For sparse/structured operators we use the standard sparse-linear-algebra notation $\mathrm{nnz}(K_{\mathcal{I}_j,:})$ for the number of entries that actually participate (stored nonzeros or on-the-fly evaluated terms); in particular, $\mathrm{nnz}(K_{\mathcal{I}_j,:}) = mn$ in the dense case. In practice, the two operator applications dominate; elementwise updates are typically bandwidth-light and fused.

We model local compute by a single measured matvec time:
\[
t_{\mathrm{mv}}(m,n) \;\approx\; \text{time to apply } x\mapsto K_{\mathcal{I}_j,:}x
\text{ (or } x\mapsto (K_{:,\mathcal{I}_j})^\top x\text{)},
\]
obtained from a device microbenchmark. We approximate the per-iteration local compute time on client~$j$ as
\begin{equation}
T_{\mathrm{comp}}^{(j)}
\;\approx\;
2\,t_{\mathrm{mv}}(m,n) \;+\; t_{\mathrm{ew}}(m),
\label{eq:comp-model}
\end{equation}
where $t_{\mathrm{ew}}(m)$ aggregates the elementwise updates.

\begin{remark}[Materialized vs.\ on-the-fly kernels]
If $K$ is applied implicitly (e.g.\ $K_{ij}=\exp(-C_{ij}/\varepsilon_{\mathrm{OT}})$ evaluated on demand), then $t_{\mathrm{mv}}$ includes kernel evaluation; the model is unchanged.
\end{remark}

\subsection{Communication model and collective primitives}
\label{subsec:comm-model}

We use the standard latency--bandwidth ($\alpha$--$\beta$) model in \emph{bytes}:
sending a message of size $B$ bytes costs
\begin{equation}
T_{\mathrm{p2p}}(B) \;=\; \alpha + \beta B,
\label{eq:alphabeta}
\end{equation}
with $\alpha$ the latency-like startup cost and $\beta$ the inverse bandwidth (seconds/byte).
This convention matches our microbenchmarks, which report time as a function of payload bytes.

Federated Sinkhorn exchanges only scaling-vector state.
For $k$ targets, the global scaling state has length $n k$; for dtype with $b_{\mathrm{dtype}}$ bytes per scalar (e.g., $b_{\mathrm{dtype}}=8$ for float64), the payload size is
\[
B(n,k) \;=\; b_{\mathrm{dtype}}\, n k .
\]
We model each relevant collective primitive $\mathrm{X}\in\{\mathrm{AG},\mathrm{BC},\mathrm{SR}\}$ as
\begin{equation}
T_{\mathrm{X}}(n,k,c)
\;\approx\;
\alpha_{\mathrm{X}}(c) \;+\; \beta_{\mathrm{X}}(c)\,B(n,k),
\label{eq:collective-affine-bytes}
\end{equation}
where $\mathrm{AG}$ denotes Allgather, $\mathrm{BC}$ denotes a server$\to$clients broadcast, and $\mathrm{SR}$ denotes clients$\to$server communication.
We estimate $(\alpha_{\mathrm{X}}(c),\beta_{\mathrm{X}}(c))$ from microbenchmarks on the target system and use them to interpret the scaling trends in Section~\ref{sec:experiments}. (For optional tree-based closed forms as a mental model, see~\ref{app:collectives}.)

\subsection{Per-iteration wall-time by topology}
\label{subsec:per-topology}

Combining \cref{eq:comp-model} with the collective models above yields clean per-topology per-iteration wall-time approximations.

\subsubsection{Synchronous All-to-All}
\label{subsubsec:cost-a2a-sync}

Each iteration performs two Allgathers (refresh $v$ before $Kv$, and refresh $u$ before $K^\top u$), giving
\begin{equation}
T_{\mathrm{iter}}^{\mathrm{A2A,sync}}
\;\approx\;
2\,t_{\mathrm{mv}}(m,n) \;+\; 2\,T_{\mathrm{AG}}(n,k,c).
\label{eq:titer-a2a-sync}
\end{equation}

\subsubsection{Synchronous Star-Network}
\label{subsubsec:cost-star}

\begin{equation}
T_{\mathrm{iter}}^{\mathrm{Star}}
\;\approx\;
2\,t_{\mathrm{mv}}^{\mathrm{srv}}(n,n)
\;+\;
2\,T_{\mathrm{BC}}(n,k,c)
\;+\;
2\,T_{\mathrm{SR}}(n,k,c),
\label{eq:titer-star}
\end{equation}
where $t_{\mathrm{mv}}^{\mathrm{srv}}(n,n)$ is the server matvec time.
In our reference Star implementation, the server receives $(c-1)$ client slices per uplink phase (one slice from each non-server rank), each of size approximately $B(n,k)/c$ bytes.
Accordingly, we model the uplink as
\begin{equation}
T_{\mathrm{SR}}(n,k,c)
\;\approx\;
(c-1)\Bigl(\alpha_{\mathrm{SR}}(c) + \beta_{\mathrm{SR}}(c)\,\tfrac{B(n,k)}{c}\Bigr),
\label{eq:tsr-star-slices}
\end{equation}
while the downlink broadcast uses the full state size $B(n,k)$:
\begin{equation}
T_{\mathrm{BC}}(n,k,c)
\;\approx\;
\alpha_{\mathrm{BC}}(c) + \beta_{\mathrm{BC}}(c)\,B(n,k).
\label{eq:tbc-star}
\end{equation}

\subsubsection{Asynchronous All-to-All (overlap-limited)}
\label{subsubsec:cost-a2a-async}

In steady state, non-blocking communication can overlap with compute, so the effective cycle time is limited by the slower of local compute and global dissemination:
\begin{equation}
T_{\mathrm{iter}}^{\mathrm{A2A,async}}
\;\approx\;
\max\!\bigl\{2\,t_{\mathrm{mv}}(m,n),\;2\,T_{\mathrm{AG}}(n,k,c)\bigr\}.
\label{eq:titer-a2a-async}
\end{equation}

\subsection{Memory footprint}
\label{subsec:memory-model}

In All-to-All variants, each client must hold global $u,v\in\mathbb{R}^n$ (assembled from Allgathers) to apply its local operators, plus local buffers of size $\Theta(m)$ (e.g., $q_{\mathcal{I}_j},r_{\mathcal{I}_j}$). Thus the working memory for vectors is $\Theta(n)$ per client, in addition to the local kernel representation (materialized, streamed, or implicit).

In the star variant\footnote{Implementation note: our reference code allocates full-length $u$ and $v$ on each rank for simplicity of indexing; this is not required by the Star-Network algorithm and does not change the message pattern.
}, clients store only slices and buffers of size $\Theta(m)$, while the server stores global $u,v\in\mathbb{R}^n$ and the kernel representation (materialized or implicit), which dominates memory.

\subsection{Model calibration and use in the experiments}
\label{subsec:calibration}

We measure the device matvec time $t_{\mathrm{mv}}$ and fit $(\alpha_X(c),\beta_X(c))$ for the collectives used in each topology on the target system via communication microbenchmarks. The benchmark protocol and fitted coefficients are reported in~\ref{app:com_details}. These parameters are then plugged into \crefrange{eq:titer-a2a-sync}{eq:titer-a2a-async} to predict compute/communication splits in Section~\ref{sec:experiments}.

\section{Experimental Evaluation}
\label{sec:experiments}

\subsection{Implementation and hardware}
\label{subsec:implementation}

We implement the distributed variants in Python using mpi4py for message passing and PyTorch for CPU/GPU tensor operations. Unless stated otherwise we use float64. Our reference implementation stages GPU tensors through host memory for MPI communication (no CUDA-aware MPI/GPUDirect), so reported runtimes include staging overhead. All experiments use symmetric costs ($C=C^\top$, hence $K^\top=K$), satisfying the transpose-access assumption. Additional implementation details are deferred to~\ref{app:impl-details}.

\subsection{HPC-oriented 2D baseline}

For context, we also implement an HPC-style 2D block distribution of $K$ that replaces global scaling gathers with structured row/column reductions. Because this baseline assumes a kernel layout that is typically unavailable in strict row-partitioned federated deployments, we treat it as an optimistic upper bound and report details in~\ref{app:exp_details}.

\subsection{Synchronous federation: correctness and scaling}
\label{subsec:sync-federation}

\subsubsection{Computation and communication times}

Unless stated otherwise, our node-count scaling experiments are strong-scaling: we fix the global problem size $n$ and iteration budget and vary the number of nodes $c$, reporting a compute/communication breakdown.

\begin{figure}[!htbp]
    \centering
    \includegraphics[width=0.6\linewidth]{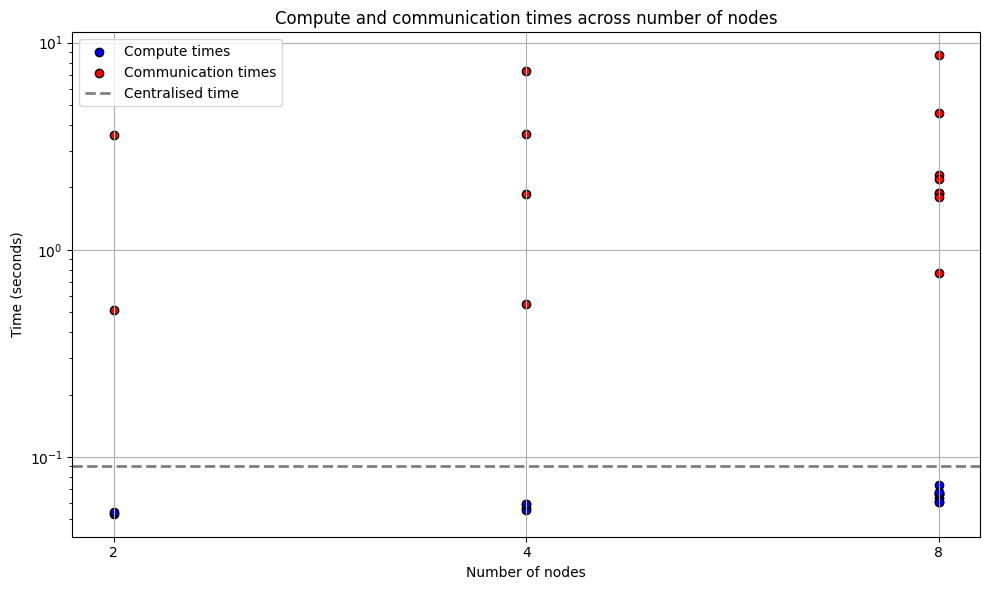}
    \caption{Per-iteration compute/communication breakdown for synchronous All-to-All federation at $n=10\,000$ and $T=250$. Communication dominates as $c$ increases, consistent with the $\alpha$--$\beta$ model (\cref{eq:titer-a2a-sync}). Each dot is one node; the centralized baseline is shown for reference.}
    \label{fig:sync_node_times}
\end{figure}

For $n=10\,000$ and $T=250$ iterations, we profiled the synchronous All-to-All variant by
separating the time spent in the two local GPU matvecs from the time spent in collective
communication. Figure~\ref{fig:sync_node_times} shows that the local matvec time is roughly
flat (and can slightly increase) as $c$ grows, while the Allgather synchronization becomes
dominant and increases rapidly with $c$. This matches the $\alpha$--$\beta$ model in
Section~\ref{sec:comm-comp-model}: the per-iteration cost in \cref{eq:titer-a2a-sync} contains a bandwidth-dominated Allgather term that scales with the global vector length. Practically, this means that for this hardware and problem size, end-to-end runtime is interconnect-limited, and \cref{eq:titer-a2a-sync} is best viewed as a provisioning tool to predict when $(n,c)$ is compute-versus communication-dominated.

\subsubsection{Comparison between serial and vectorized resolutions}

Following the standard multitarget formulation, we stack $N$ target histograms as $b\in\mathbb{R}^{n\times N}_+$ and update matrix scalings $u,v\in\mathbb{R}^{n\times N}_+$ with a shared kernel $K$. On a synthetic instance with $n=5000$ and $N=500$ (15 iterations), solving all $N$ problems jointly is essentially as fast as solving one, whereas solving them sequentially scales linearly in $N$ (Table~\ref{tab:serial_vs_vectorized}). This makes multitarget vectorization a key baseline (and a natural fit for federation) when many OT problems share the same cost matrix.

\begin{table}[h!]
  \centering
  \small
  \setlength{\tabcolsep}{6pt}
  \begin{tabular}{lr}
    \toprule
    Configuration & Time (s) \\
    \midrule
    1 problem (serial)          & 0.32 \\
    500 problems (vectorized)   & 0.31 \\
    500 problems (sequential)   & 11.56 \\
    \bottomrule
  \end{tabular}
  \caption{Isolated GPU compute time for 15 Sinkhorn iterations on the multitarget instance.}
  \label{tab:serial_vs_vectorized}
\end{table}

\subsubsection{Performance for different values of $N$}
\label{subsubsec:dif_val_N}

\begin{figure}[t]
  \centering
  \begin{subfigure}[b]{0.49\linewidth}
    \centering
    \includegraphics[width=\linewidth]{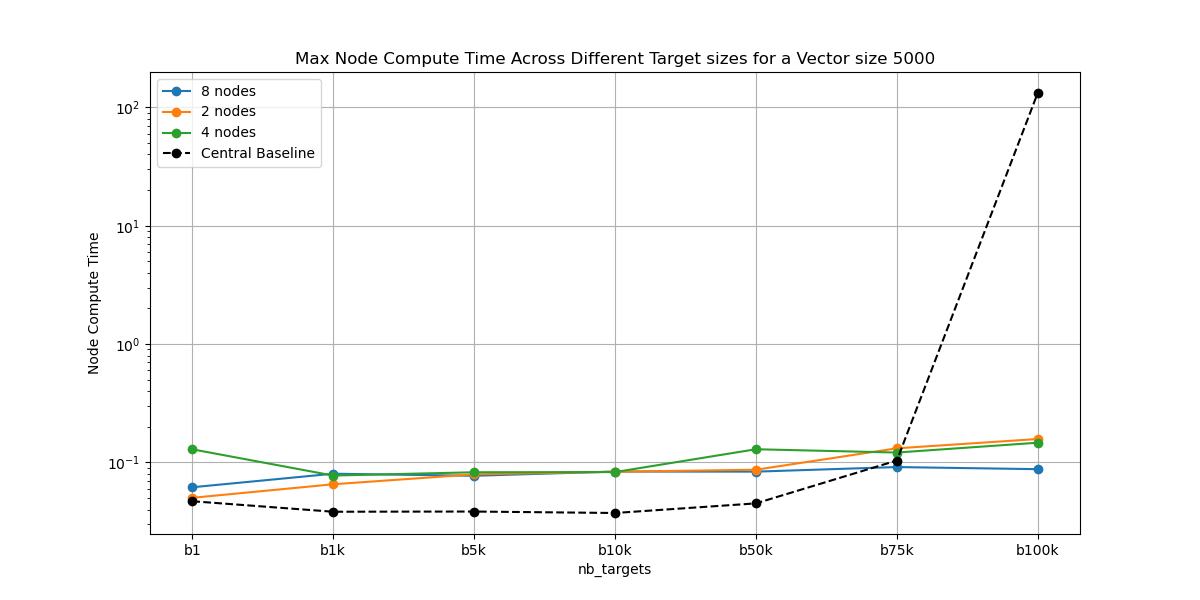}
    \caption{Compute time.}
    \label{fig:compute_time_multitarget}
  \end{subfigure}
  \hfill
  \begin{subfigure}[b]{0.49\linewidth}
    \centering
    \includegraphics[width=\linewidth]{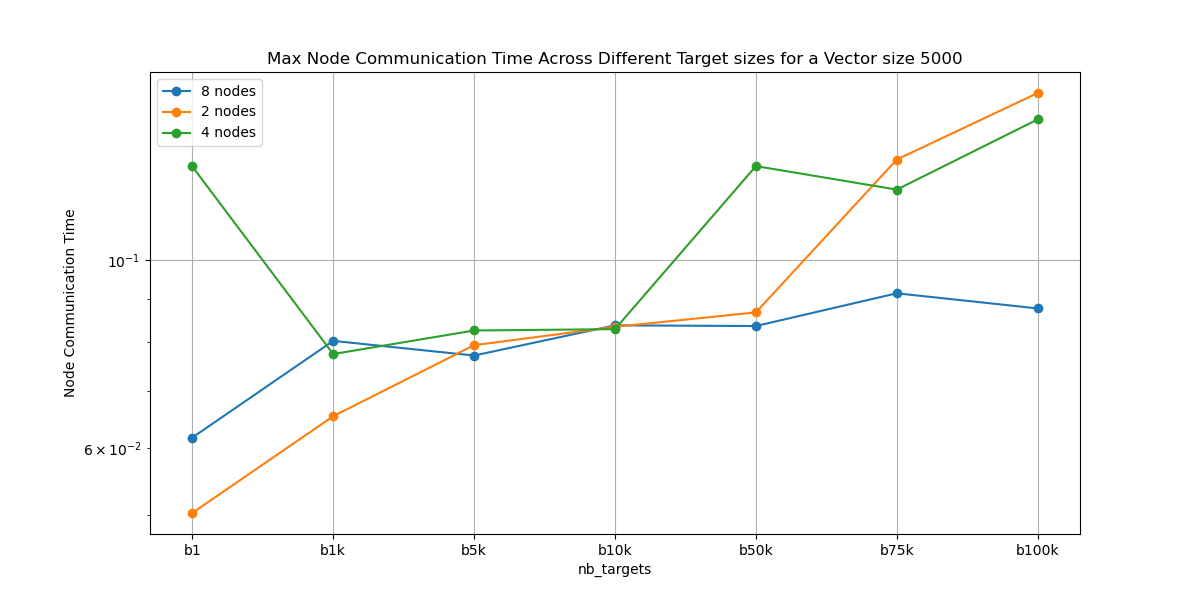}
    \caption{Communication time.}
    \label{fig:communication_time_multitarget}
  \end{subfigure}
  \caption{Isolated compute and communication time for $N \in [1, 1000, 5000, 10\,000, 50\,000, 75\,000, 100\,000]$, comparing centralized and synchronous federated settings.}
  \label{fig:multitarget_times}
\end{figure}

To probe larger multitarget regimes, we switch to \texttt{float32} and sweep $N \in [1, 1000, 5000, 10\,000, 50\,000, 75\,000, 100\,000]$. Figure~\ref{fig:multitarget_times} shows a regime change: beyond roughly $N\approx 75\,000$ the single-device centralized run slows sharply, while the federated variants keep nearly flat compute time because each node handles only a slice of the $n\times N$ state. At the same time, the communication cost also grows with $N$ and can dominate wall-clock time on our current cluster, indicating that multitarget federation becomes attractive primarily when single-device memory/throughput limits are reached and/or communication can be improved.

\subsection{Asynchronous federation: behavior and damping}
\label{subsec:async_exp}

\subsubsection{Non-determinism}

The asynchronous All-to-All variant follows the same local update rules as the synchronous scheme but removes global barriers: each node continuously reads the most recent slices it has received, performs local matvecs and damped updates, and publishes fresh slices without waiting for all peers. As a result, different runs on the same hardware can see different message arrival orders and hence different optimization trajectories.

\begin{figure}[!htbp]
  \centering
  \includegraphics[width=0.65\linewidth]{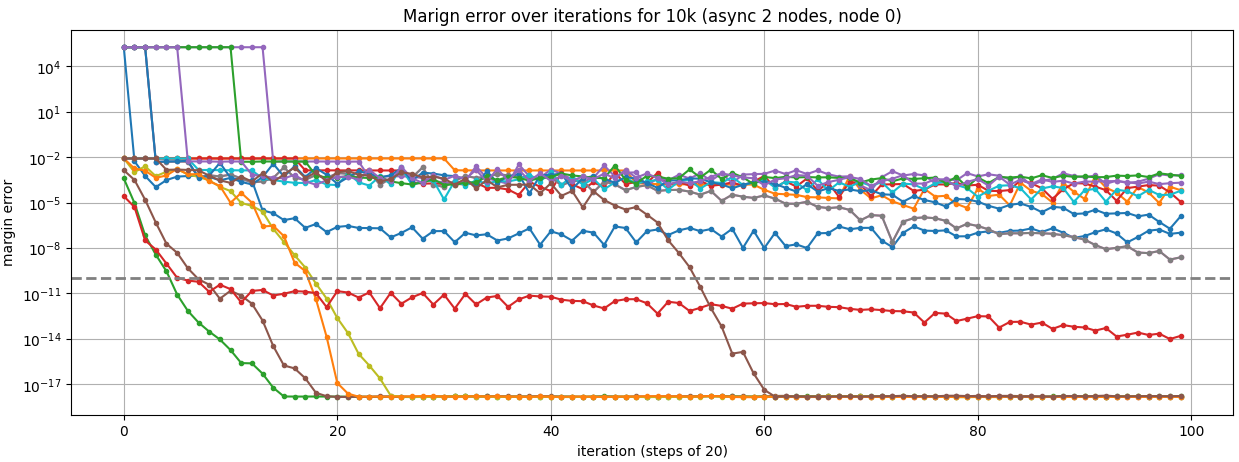}
  \caption{Run-to-run variability of asynchronous All-to-All federation on CPU (15 runs, $n=10\,000$,
  identical initialization).}
  \label{fig:async_multirun}
\end{figure}

We illustrate this by running the asynchronous algorithm $15$ times on a synthetic problem with $n=10\,000$ and identical initialization, and plotting the marginal error on $a$ at one node, $\text{err}_a=\|P\mathbf{1}-a\|_2$, versus iterations. Figure~\ref{fig:async_multirun} shows substantial run-to-run variability, motivating damping and the bounded-delay analysis; additional representative single-run trajectories are deferred to~\ref{app:async_details}.

\subsubsection{Influence of step size $\eta$}
\label{subsubsec:stepsizesec}

We treat the damping parameter $\eta\in(0,1]$ as a robustness--speed knob: smaller $\eta$ reduces the magnitude of each update under stale information but can slow progress. To reduce GPU scheduling and communication jitter in timing measurements, we perform this study on CPUs at $n=10\,000$.

For each node count $c\in\{2,4,8\}$ and each $\eta\in\{0.1,0.25,0.5\}$, we ran $15$ trials with identical initialization and recorded the time to reach $\text{err}_a \le 10^{-10}$ (up to a maximum of $8000$ iterations). Table~\ref{tab:convg_stepsizes} shows that $\eta=0.5$ is typically the fastest choice in this regime, although results are not strictly monotone across $c$ due to stochastic message timing and run-to-run variability. A broader robustness sweep with randomized inputs is reported in~\ref{app:async_details} (see Table~\ref{tab:summary_stability_all} and Figure~\ref{fig:robust}). These results are consistent with (i) the bounded‑delay convergence guarantee for the undamped asynchronous iteration Proposition~\ref{prop:async} and (ii) the common empirical behavior of asynchronous fixed‑point methods, where under‑relaxation can reduce oscillations under stale information at the cost of slower per‑update progress.

\begin{table}[!htbp]
    \centering
    \begin{tabular}{cccc}
        \toprule
         & $\eta=0.1$ & $\eta=0.25$ & $\eta=0.5$ \\
        \midrule
        2 nodes & 105.02 & 20.41 & 13.19 \\
        4 nodes & 24.11 & 10.04 & 6.62 \\
        8 nodes & 33.45 & 40.84 & 15.12 \\
        \bottomrule
    \end{tabular}
    \caption{Average time to reach $\text{err}_a \le 10^{-10}$ on CPU for $n = 10\,000$, over 15 runs per configuration.}
    \label{tab:convg_stepsizes}
\end{table}

\subsubsection{Computation and communication times}
\label{subsubsec:comp_comm_times}

We repeated the compute/communication breakdown for the asynchronous All-to-All scheme at $n=10\,000$ and $T=250$ iterations, varying $c\in\{2,4,8\}$. Communication dominates wall-clock time and becomes more variable as $c$ grows due to non-deterministic message delays; full per-node breakdowns are deferred to the appendix (Figure~\ref{fig:async_node_times}).

\subsubsection{Delays in iterations}
\label{subsubsec:delays_in_async}

Delay (staleness) distributions for the asynchronous protocol are reported in~\ref{app:delay_distributions}.

\subsection{Sensitivity to problem size and hardware architecture}
\label{subsec:pb_size}

To assess sensitivity to problem characteristics and hardware, we sweep input parameters over $n \in \{10^3, 5\cdot 10^3, 10^4, 2.5\cdot 10^4\}$, the number of target histograms $N$, sparsity in off-diagonal blocks of $C$, and conditioning of marginals and costs. For each configuration we record the iterations and wall-clock time to reach marginal error below $10^{-15}$ when convergence is achieved; detailed tables are deferred to~\ref{app:async_details}.

\begin{figure}[!htbp]
  \centering
  \begin{subfigure}[b]{0.49\linewidth}
    \centering
    \includegraphics[width=\linewidth]{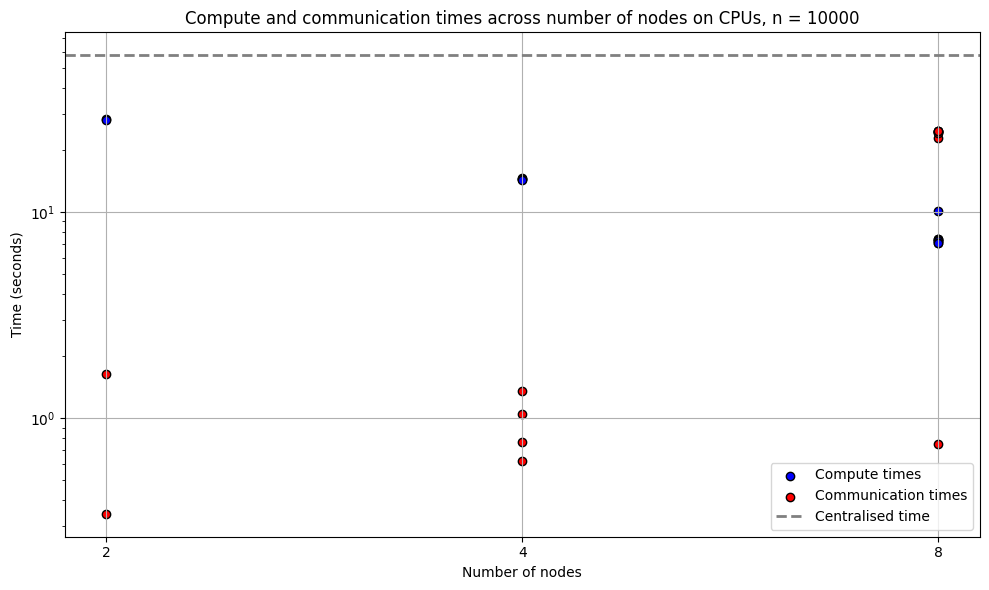}
  \end{subfigure}
  \hfill
  \begin{subfigure}[b]{0.49\linewidth}
    \centering
    \includegraphics[width=\linewidth]{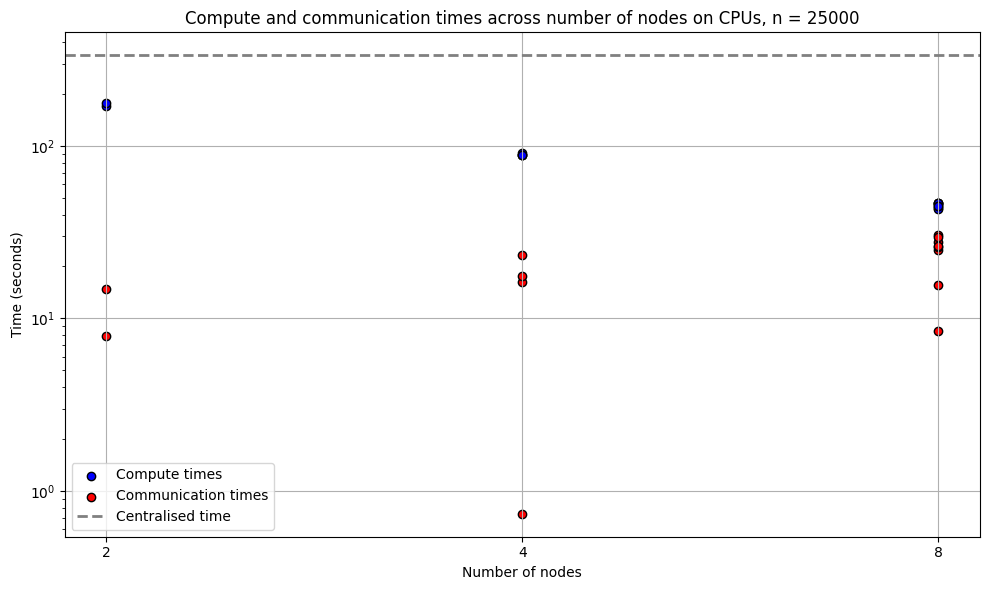}
  \end{subfigure}
    \caption{Synchronous CPU runs: per-node computation and communication
    times across multiple node counts for a fixed budget of 250 iterations,
    for $n = 10\,000$ (left) and $n = 25\,000$ (right).}
  \label{fig:cpu_node_times}
\end{figure}

On GPU, both centralized and synchronous federated variants exhibit stable iteration counts across this sweep; wall-clock time is instead governed by the compute/communication balance captured by the $\alpha$--$\beta$ model (Section~\ref{sec:comm-comp-model}). In particular, at moderate $n$ the exchange of global scaling vectors can dominate runtime, while at very large multitarget sizes federation can reduce per-device compute/memory pressure but remains communication-limited on our current cluster (cf. Figure~\ref{fig:cpu_node_times}).

On CPU, local matvecs are more expensive and communication is relatively cheaper, so strong scaling becomes visible. Figure~\ref{fig:cpu_node_times} shows clear reductions in per-node compute time and tangible wall-clock speedups with increasing $c$ for $n=10\,000$ and $n=25\,000$. Additional CPU convergence traces and per-node time distributions are deferred to the appendix.

Overall, these experiments support the main deployment guidance of this paper: use synchronous federation whenever global barriers are feasible (predictable convergence and well-modeled costs), and reserve asynchrony for settings where synchronization is impractical, with damping $\eta$ tuned for stability.

See~\ref{app:cpu_sync} and~\ref{app:cpu_async} for the corresponding CPU convergence trajectories and additional per-run variability plots.

\subsection{Embedding-based costs}
To complement synthetic sparsity/conditioning sweeps, we evaluate a more realistic OT cost derived from high-dimensional embeddings.
We sample $n$ vectors $x_i\in\mathbb{R}^d$ (representing embedding features) and define a symmetric cost
\[
C_{ij}=\|x_i-x_j\|_2^2,
\]
which corresponds to transport on a shared support in embedding space. We then run the same distributed Sinkhorn methods (All-to-All, Star, and the 2D baseline) and report convergence time and communication time. This experiment probes performance in a regime closer to common OT applications (e.g., matching distributions of learned features) while retaining dense-kernel structure.

\section{Privacy (optional measurement layer)}
Communicated Sinkhorn scalings can leak information about local data, especially under Star orchestration where a coordinator observes all uplink messages. As a lightweight baseline, we evaluate a mechanism that clips and perturbs \emph{log-domain} scaling vectors with Gaussian noise and tracks privacy loss via Rényi-DP composition across iterations. In our long-run setting with per-iteration releases, standard composition yields very loose bounds unless release frequency is reduced; clipping can also introduce a bias floor when the radius is too small. Because this layer is optional and not required for correctness of the distributed solvers, full mechanism details, accounting, and privacy--utility plots are deferred to~\ref{app:privacy}.

\section{Illustrative application: distributionally robust portfolio risk}
\label{sec:numt}

We briefly illustrate how Federated Sinkhorn can be used in a distributionally robust portfolio problem in the spirit of \cite{blanchet2019quantifying}. This example is illustrative only and uses synthetic return samples (no proprietary financial data), consistent with the Data Availability statement.

Let $\hat P$ denote the empirical distribution of historical asset returns and let $l(x)=w^\top x$ be the portfolio loss for weights $w\in\mathbb{R}^d$. The Blanchet--Murthy risk measure considers the worst expected loss over a Wasserstein ball of radius $\delta$ around $\hat P$: 
\begin{equation}
  \rho_{\text{worst}}(w)
  = \sup_{P:\, W_c(P,\hat P)\le\delta} \mathbb{E}_P[l(X)] ,
  \label{eq:bm_worst_case_loss}
\end{equation}
where $W_c$ is the Wasserstein distance with ground cost $c$.

Following \cite{blanchet2019quantifying}, this can be written in dual form as
\begin{equation}
  \rho_{\text{worst}}(w)
  = \inf_{\lambda\ge 0} \Bigl\{\lambda\delta
  + \mathbb{E}_{\hat P}\bigl[ h_\lambda(X)\bigr]\Bigr\},
  \qquad
  h_\lambda(x) = \sup_{x'}\bigl(l(x')-\lambda c(x,x')\bigr).
  \label{eq:bm_dual}
\end{equation}
Discretizing $\hat P$ on $n$ samples $\{x_i\}_{i=1}^n$ and restricting the supremum to a finite set $\{x'_j\}_{j=1}^n$ yields a discrete coupling $P\in\mathbb{R}_+^{n\times n}$ between ``historical'' and ``stress'' returns. As shown in~\ref{app:finance}, for fixed $\lambda$ the worst-case loss reduces to the entropic OT problem
\begin{mini}|l|
  {P\in\mathcal{U}}{
    \langle P, C_\lambda\rangle
    + \varepsilon_{\scriptscriptstyle\mathrm{OT}} \sum_{i,j} P_{ij}(\log P_{ij}-1)}{}{}
  \addConstraint{P\mathbf{1} = a,\quad P^\top\mathbf{1} = b,\quad P\ge 0,}
  \label{eq:finance_entropic_ot}
\end{mini}
with cost matrix $C_{\lambda,ij} = \lambda c(x_i,x'_j) - l(x'_j)$ and $(a,b)$ the source and target marginals. The Wasserstein radius constraint $\langle P,c\rangle = \delta$ is enforced by searching over $\lambda$ (e.g., via bisection), repeatedly solving \cref{eq:finance_entropic_ot} with Sinkhorn. Thus Blanchet--Murthy worst-case risk evaluation can be implemented as a sequence of entropic OT problems.

\subsection{Federated data layout}

In a multi-institution setting (e.g. several desks or banks), historical returns $\{x_i\}$ are horizontally partitioned across participants: client $j$ holds $\{x_i : i\in\mathcal{I}_j\}$ and the corresponding rows of the cost matrix $C_\lambda$. Forecast returns $\{x'_j\}$ and the portfolio weights $w$ may be shared globally (e.g. from a central risk team), so each client can assemble its local block $C_{\lambda,\mathcal{I}_j,:}$. Evaluating \cref{eq:finance_entropic_ot} then requires exactly the same communication pattern as our generic federated Sinkhorn algorithms: clients exchange only the scaling vectors $(u,v)$ (and a small number of global scalars such as $\langle P,c\rangle$), while raw return data never leave their host institution. Standard privacy tools, such as secure aggregation or DP noise on the exchanged scalings, can be layered on top.

\subsection{Toy portfolio experiment}

As a proof of concept we consider a toy portfolio of three technology stocks, with one-day historical returns $x\in\mathbb{R}^3$, analyst-predicted returns $x'\in\mathbb{R}^3$, and weights $w=(\tfrac{2}{5},\tfrac{1}{10},\tfrac{1}{2})$. We take a squared Euclidean ground cost $c(x_i,x'_j)=\|x_i-x'_j\|_2^2$, choose $\delta$ and $\lambda$ in a moderately conservative range, and set the entropic regularization to $\varepsilon_{\scriptscriptstyle\mathrm{OT}}=10^{-2}$. After shifting and normalizing $x$ and $x'$ to define marginals $(a,b)$ (details in~\ref{app:finance}), the federated Sinkhorn solver finds a coupling $P^\star$ that satisfies the Wasserstein constraint and yields a worst-case loss $\rho_{\text{worst}}(w)\approx -0.48$, i.e. a $48\%$ one‑day loss under the adverse model. This value is intentionally extreme due to the toy nature and parameter choices, but it demonstrates that the method recovers the Blanchet \& Murthy risk in a federated setting.

\begin{figure}[!htbp]
  \centering
  \begin{subfigure}[b]{0.49\linewidth}
    \centering
    \includegraphics[width=\linewidth]{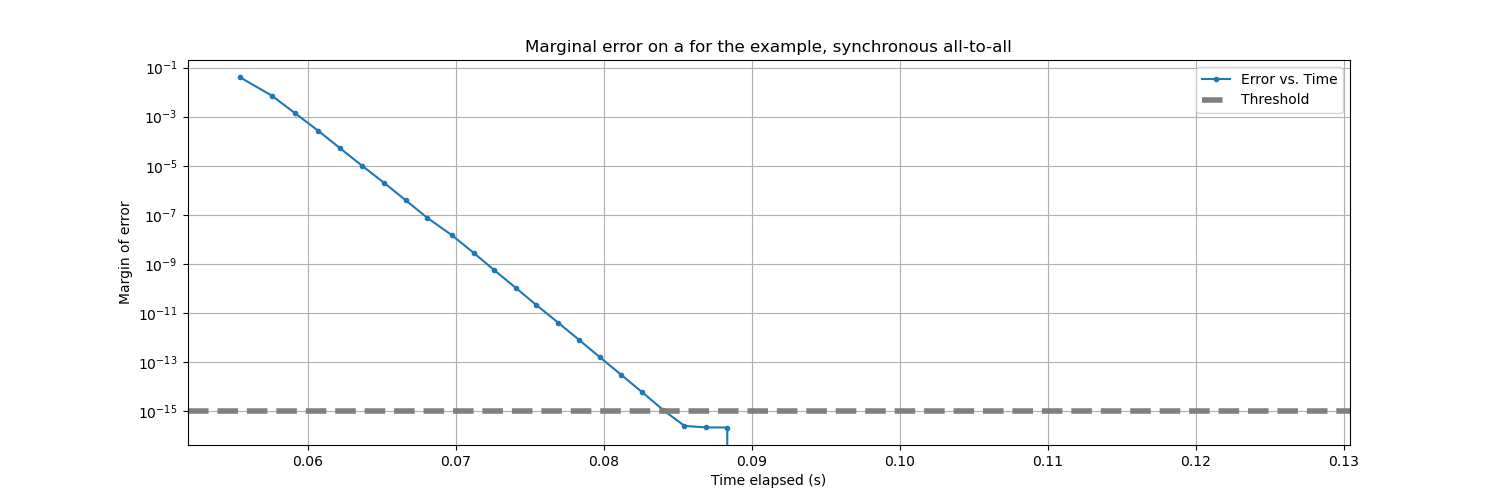}
    \caption{Sync. All-to-All}
    \label{fig:example_sync}
  \end{subfigure}
  \hfill
  \begin{subfigure}[b]{0.49\linewidth}
    \centering
    \includegraphics[width=\linewidth]{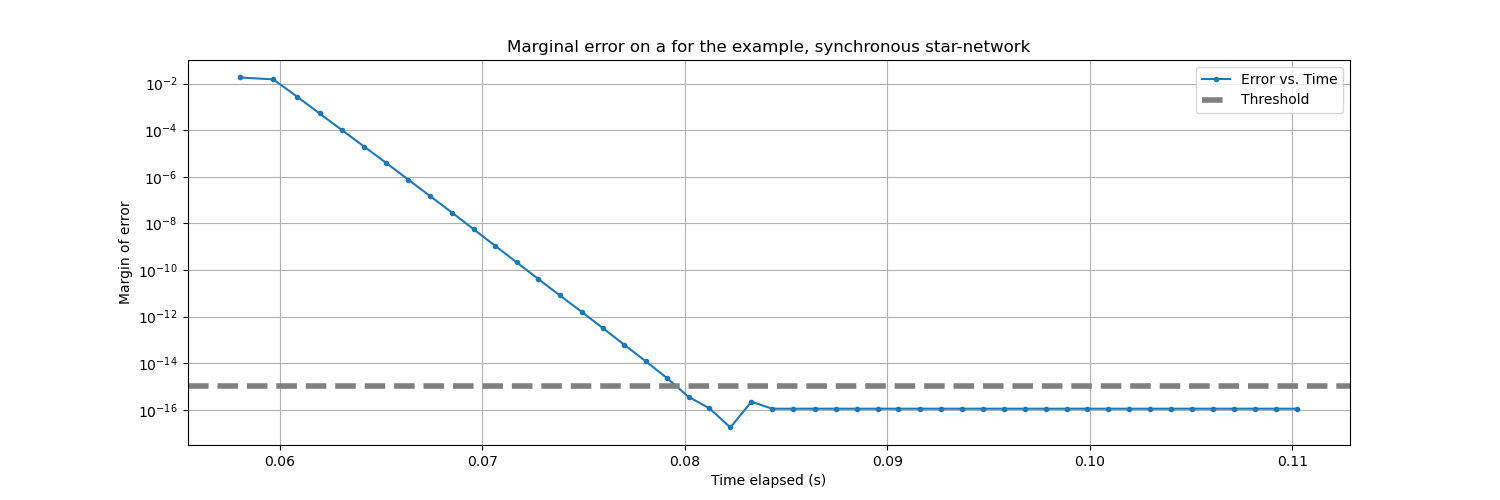}
    \caption{Sync. star-network}
    \label{fig:example_sync_star}
  \end{subfigure}
  \\
  \vspace{0.5em}
  \begin{subfigure}[b]{0.49\linewidth}
    \centering
    \includegraphics[width=\linewidth]{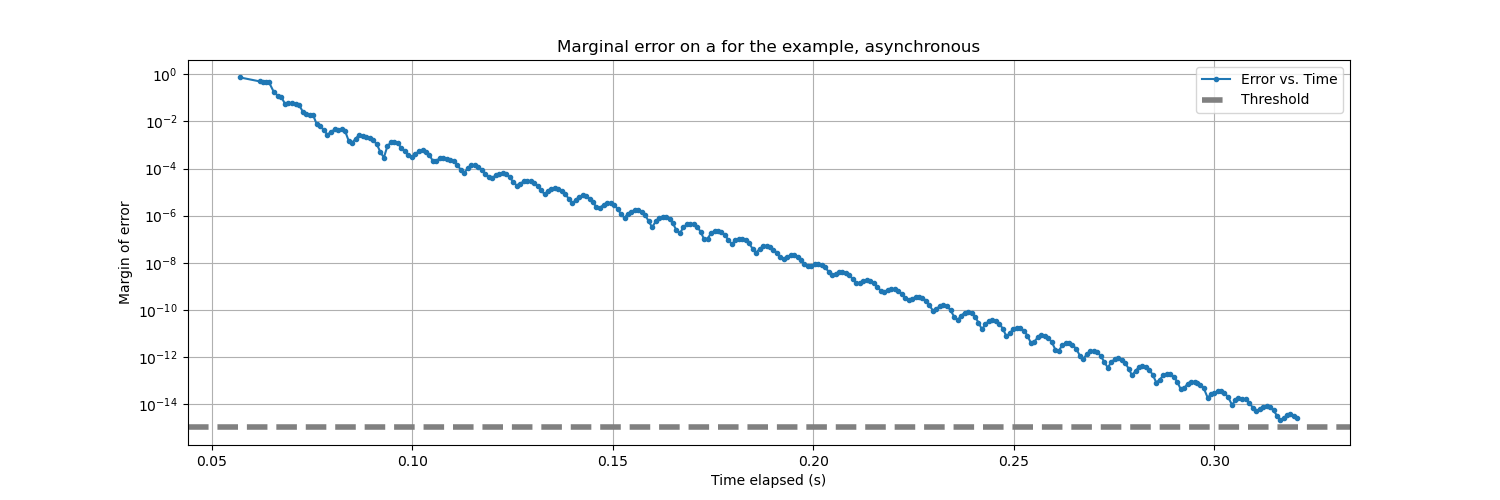}
    \caption{Async. All-to-All}
    \label{fig:example_async}
  \end{subfigure}
    \caption{Time of convergence for the three settings applied to this example. The Synchronous All-to-All's marginal error drops to 0 after a few iterations, probably due to rounding.}
    \label{fig:example_cvg}
\end{figure}

\begin{remark}
    The numerical values in this illustration are hand-constructed for clarity and are not drawn from market datasets; the goal is to demonstrate the end-to-end formulation and communication patterns rather than provide an empirical finance evaluation.
\end{remark}

We implemented this example with three ``offices'', each owning one row of the cost matrix. Figure~\ref{fig:example_cvg} reports the marginal error versus time for our three communication patterns. All variants converge within $0.5\,$s, with synchronous all‑to‑all being fastest and asynchronous slightly slower but still stable, illustrating that the federated algorithm can handle small portfolio problems with negligible overhead compared to a centralized solve.

\section{Discussion and Conclusion}
\label{sec:conclusion}

We studied entropic OT when the kernel operator is row-partitioned across $c$ workers and cannot be centralized. We presented exact synchronous federated Sinkhorn protocols for two orchestration patterns, All-to-All collectives and Star-Network (parameter-server), and derived per-iteration compute/communication/memory costs under an $\alpha$--$\beta$ latency--bandwidth model. Experiments on multi-node CPU/GPU systems validate the model and show that, for dense kernels, repeated exchange of global scaling state can dominate runtime as $c$ increases. We also considered an optional bounded-delay asynchronous schedule, which removes barriers but introduces variability under stale reads.

\noindent Practical guidance across the configurations:
\begin{itemize}
\item Use synchronous All-to-All when peer-to-peer collectives are feasible and predictable convergence is required.
\item Use Star when connectivity constraints preclude peer-to-peer links or when clients cannot store the kernel.
\item Provision the interconnect for repeated exchange of $\Theta(nk)$ scaling state per iteration, or reduce this traffic via compression/algorithmic restructuring (future work).
\end{itemize}

\newpage

\section*{CRediT authorship contribution statement}
\noindent \textbf{Jeremy Kulcsar:} Conceptualization; Methodology; Investigation; Software; Visualization; Writing -- original draft; Writing -- review \& editing.\\
\noindent \textbf{Vyacheslav Kungurtsev:} Supervision; Conceptualization; Methodology; Formal analysis; Writing -- original draft; Writing -- review \& editing.\\
\noindent \textbf{Georgios Korpas:} Supervision; Investigation; Formal analysis.\\
\noindent \textbf{Giulio Giaconi:} Writing -- review \& editing.\\
\noindent \textbf{William Shoosmith:} Writing -- review \& editing.

\section*{Declaration of competing interest}
The authors declare that they have no known competing financial interests or personal relationships that could have appeared to influence the work reported in this paper.

\section*{Data availability}
All experiments in this paper use synthetic data (including the toy portfolio example in~\ref{sec:numt}); no proprietary or personal datasets are used. Release of the reference implementation, synthetic data generators, and scripts is subject to internal open-source clearance and will be made publicly available in a permanent repository upon acceptance of the affiliated manuscript, and at the latest at the time of publication.

\section*{Funding}
This research did not receive any specific grant from funding agencies in the public, commercial, or not-for-profit sectors. Authors affiliated with HSBC Holdings Plc. conducted this work as part of their employment (salaried employees). Authors affiliated with Czech Technical University in Prague contributed under an institutional collaboration agreement between HSBC Holdings Plc. and Czech Technical University in Prague; no dedicated project funding was provided for this work.

\section*{Declaration of Generative AI and AI-assisted technologies in writing}
During the preparation of this work, the authors used OpenAI’s ChatGPT to improve language and readability. After using this tool/service, the authors reviewed and edited the content as needed and take full responsibility for the content of the published article.

\section*{Copyright disclaimer}
This paper was prepared for information purposes and is not a product of HSBC Bank Plc. or its affiliates. Neither HSBC Bank Plc. nor any of its affiliates make any explicit or implied representation or warranty and none of them accept any liability in connection with this paper, including, but not limited to, the completeness, accuracy, reliability of information contained herein and the potential legal, compliance, tax or accounting effects thereof. Copyright HSBC Group 2025.

\newpage

\section*{Glossary and abbreviations}

\begin{description}
  \item[OT] Optimal transport: the problem of finding a minimum-cost coupling (transport plan) between two probability distributions.

  \item[EOT / Entropic OT] Entropy-regularized optimal transport: OT with an entropic penalty that yields a strictly convex objective and enables fast fixed-point scaling methods.

  \item[Cost matrix $C$] Matrix of pairwise transport costs; $C_{ij}$ is the cost of transporting mass from source index $i$ to target index $j$.

  \item[Entropy regularization $\varepsilon_{\mathrm{OT}}$] Positive scalar controlling the strength of entropic smoothing in EOT; smaller values approximate unregularized OT but increase numerical stiffness.

  \item[Gibbs kernel $K$] Elementwise exponential kernel used in entropic OT, defined by $K = \exp(-C/\varepsilon_{\mathrm{OT}})$ (applied elementwise).

  \item[Coupling / transport plan $P$] Nonnegative matrix whose row and column sums match prescribed marginals; in entropic OT the optimal plan has the form $P^\star = \mathrm{diag}(u)\,K\,\mathrm{diag}(v)$.

  \item[Marginals $a,b$] Nonnegative vectors (with equal total mass) specifying the desired row sums ($a$) and column sums ($b$) of the coupling.

  \item[Scaling vectors $u,v$] Positive vectors that scale the kernel to satisfy marginal constraints; Sinkhorn updates alternate between updating $u$ and $v$.

  \item[Sinkhorn iterations / Sinkhorn--Knopp] Fixed-point matrix scaling procedure that alternates row and column normalizations (or their stabilized log-domain equivalents) to satisfy marginals in entropic OT.

  \item[Log-domain stabilization] Performing updates on $\log u$ and $\log v$ (often using log-sum-exp) to improve numerical stability when entries of $K$ are very small.

  \item[Row partition / client index sets $I_j$] Distributed layout where client $j$ owns a subset of rows (and corresponding scaling slice) indexed by $I_j$, typically with $|I_j|=m$ and $n=cm$.

  \item[Local matvecs] The dominant per-iteration operations: computing $(Kv)_{I_j}$ and $(K^\top u)_{I_j}$ on each client’s shard.

  \item[All-to-All (A2A) topology] Decentralized topology where all clients exchange information symmetrically (no central coordinator), typically via collective communication.

  \item[Star / parameter-server topology] Centralized topology where a server orchestrates communication (e.g., Bcast + point-to-point send/recv) and clients communicate only with the server.

  \item[MPI] Message Passing Interface: a standard programming model for distributed-memory parallelism and collective communication.

  \item[Allgather] MPI collective operation that gathers slices from all ranks and distributes the concatenated global vector to all ranks.

  \item[Broadcast / Bcast] MPI collective operation in which a designated root process sends the same buffer (e.g., a full vector or matrix) to all participating ranks.

  \item[send/recv] MPI point-to-point communication primitives (\texttt{Send}/\texttt{Recv}) where one rank transmits a message directly to another rank. In our Star topology, clients send scaling slices to the server via point-to-point send/recv, which is functionally a Gather pattern but not an MPI \texttt{Gather} collective.

  \item[Synchronous protocol] Execution model with global barriers/lockstep iterations; all clients use the same iteration index and exchange fresh scalings each round.

  \item[Asynchronous protocol] Execution model without global barriers; clients may proceed at different speeds and may use stale (delayed) information.

  \item[Non-blocking communication] MPI operations that initiate communication and return immediately, allowing computation to proceed while transfers are in flight.

  \item[\texttt{Isend}/\texttt{Irecv}] MPI non-blocking point-to-point send/receive primitives. They return a \emph{request handle} that can later be checked for completion or waited on.
    
  \item[\texttt{Test}] Non-blocking completion check for an outstanding MPI request (returns immediately indicating whether the operation has completed).
    
  \item[\texttt{Wait}] Blocking completion call for an outstanding MPI request (returns only once the operation has completed). In our asynchronous implementation, termination drains outstanding requests via \texttt{Wait} rather than enforcing a global barrier.
    
  \item[Request handle] An MPI object returned by non-blocking operations (\texttt{Isend}/\texttt{Irecv}) that represents an in-flight communication and is used with \texttt{Test}/\texttt{Wait}.
    
  \item[Slice cache / last-seen state] Local replica of remote scaling slices maintained by a client. In asynchronous federation, each client updates its cache opportunistically when new slices arrive and otherwise uses the most recent (possibly stale) values.
    
  \item[Staleness / delay $\tau$] The age of a cached remote slice, measured as the number of local logical iterations performed since the sender last updated that slice before the receiver uses it.

  \item[Bounded-delay asynchrony / staleness $\tau$] Assumption that any client’s view of communicated variables is at most $\tau$ iterations out of date.

  \item[Damping / under-relaxation $\eta$] Relaxation parameter $\eta\in(0,1]$ used to blend old and new updates to improve stability under staleness.

  \item[Latency--bandwidth ($\alpha$--$\beta$) model] Communication model where sending a message of size $S$ costs approximately $\alpha + \beta S$, with $\alpha$ capturing latency and $\beta$ inverse bandwidth.

  \item[Hilbert projective metric $d_H$] Metric on the positive orthant used in classical convergence analysis of Sinkhorn scaling and positive linear operators.

  \item[Contraction factor / Birkhoff contraction coefficient] Quantifies the contractivity of positive linear maps in projective metrics and governs linear convergence rates.

  \item[DP] Differential privacy: a formal guarantee limiting how much outputs can change when a single individual (or record) in the input dataset changes.

  \item[Gaussian mechanism] DP mechanism adding i.i.d.\ Gaussian noise calibrated to an $\ell_2$ sensitivity bound.

  \item[Clipping radius $S$] Bound used to clip communicated vectors (here, log-scalings) to control sensitivity before adding noise.

  \item[Noise scale $\sigma$] Standard deviation of the Gaussian noise added to clipped quantities.

  \item[RDP] R\'enyi differential privacy: a DP accounting framework parameterized by the R\'enyi order; composes additively across iterations and can be converted to $(\varepsilon,\delta)$-DP.

  \item[Privacy--utility trade-off] The empirical relationship between privacy strength (e.g., smaller $\varepsilon$) and solution accuracy (e.g., marginal error or objective value).
\end{description}

\newpage

\section*{Acknowledgements}

Computational resources were provided by the RCI computational cluster at Czech Technical University in Prague, which was used for the experiments reported in this paper.

\bibliographystyle{elsarticle-num}
\bibliography{refs.bib}

\newpage

\appendix

\section{Additional details on local steps}
\label{app:local_steps}

To decouple computation from communication, we allow each client to perform multiple local updates between global synchronizations. A local–step parameter $w\in\mathbb{N}$, analogous to the number of local steps in Local SGD~\citep{lin2019don}, controls this trade‑off:
\begin{itemize}
    \item $w=1$ yields fully synchronous Sinkhorn, with communication every iteration;
    \item $w>1$ yields local federated Sinkhorn, where each client performs $w$ updates using its current views of $(u,v)$ before participating in a collective that refreshes the global scalings (all‑to‑all or via the server).
\end{itemize}

\subsection{Synchronous effects}

\begin{figure}[!htbp]
  \centering
  \begin{subfigure}{0.48\linewidth}
    \centering
    \includegraphics[width=\linewidth]{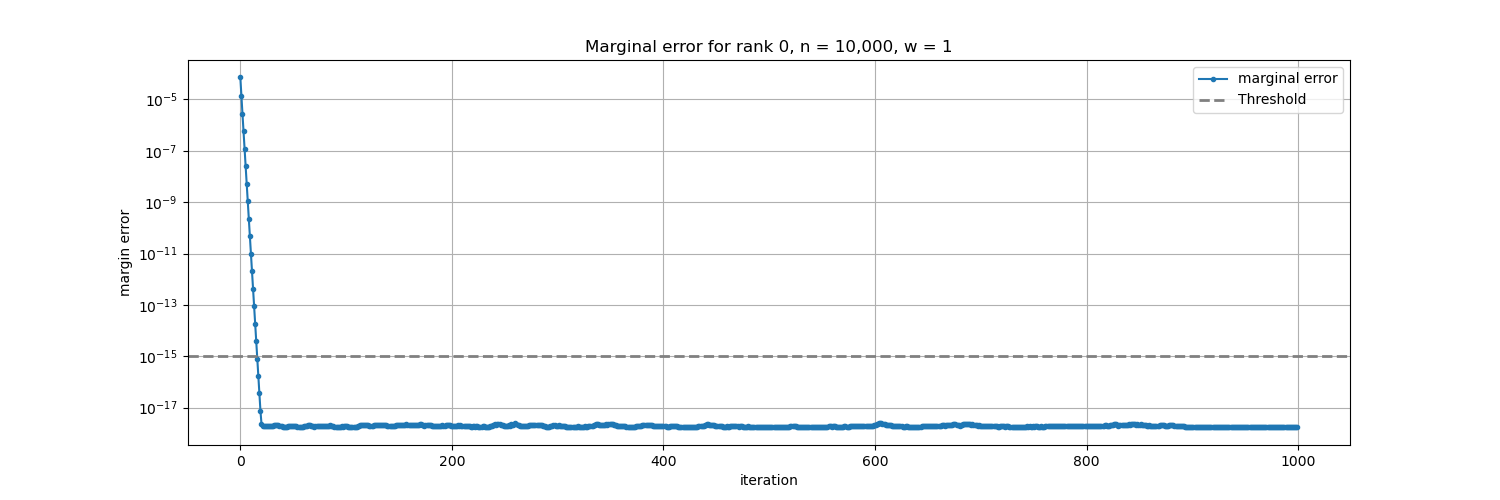}
    \caption{$w = 1$}
  \end{subfigure}\hfill
  \begin{subfigure}{0.48\linewidth}
    \centering
    \includegraphics[width=\linewidth]{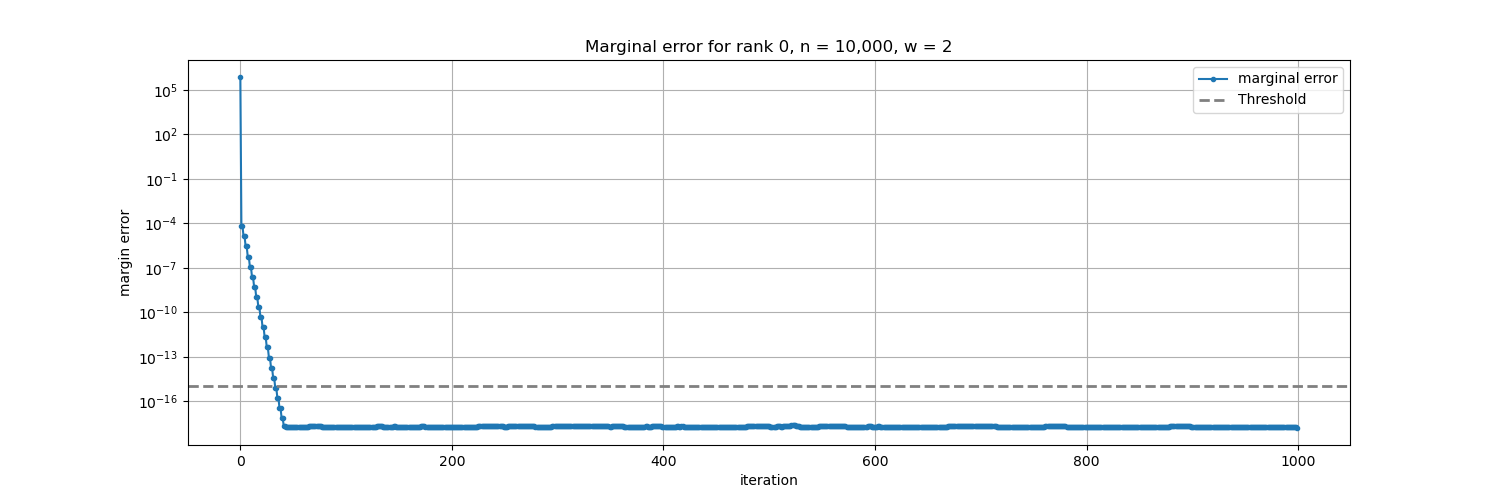}
    \caption{$w = 2$}
  \end{subfigure}
  \medskip
  \begin{subfigure}{0.48\linewidth}
    \centering
    \includegraphics[width=\linewidth]{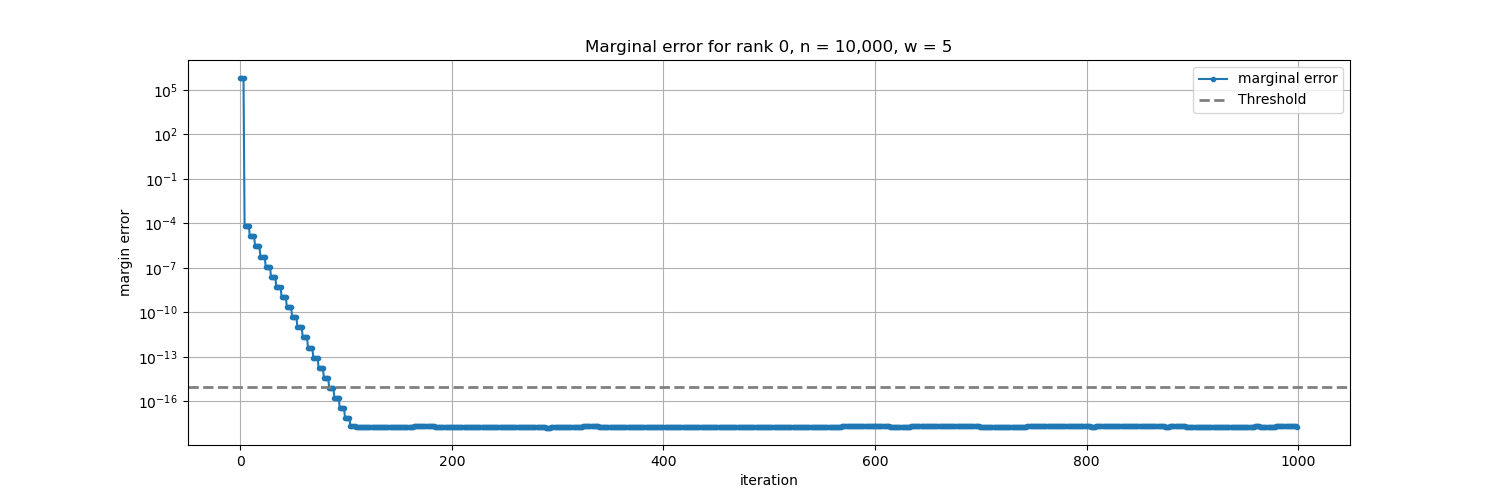}
    \caption{$w = 5$}
  \end{subfigure}\hfill
  \begin{subfigure}{0.48\linewidth}
    \centering
    \includegraphics[width=\linewidth]{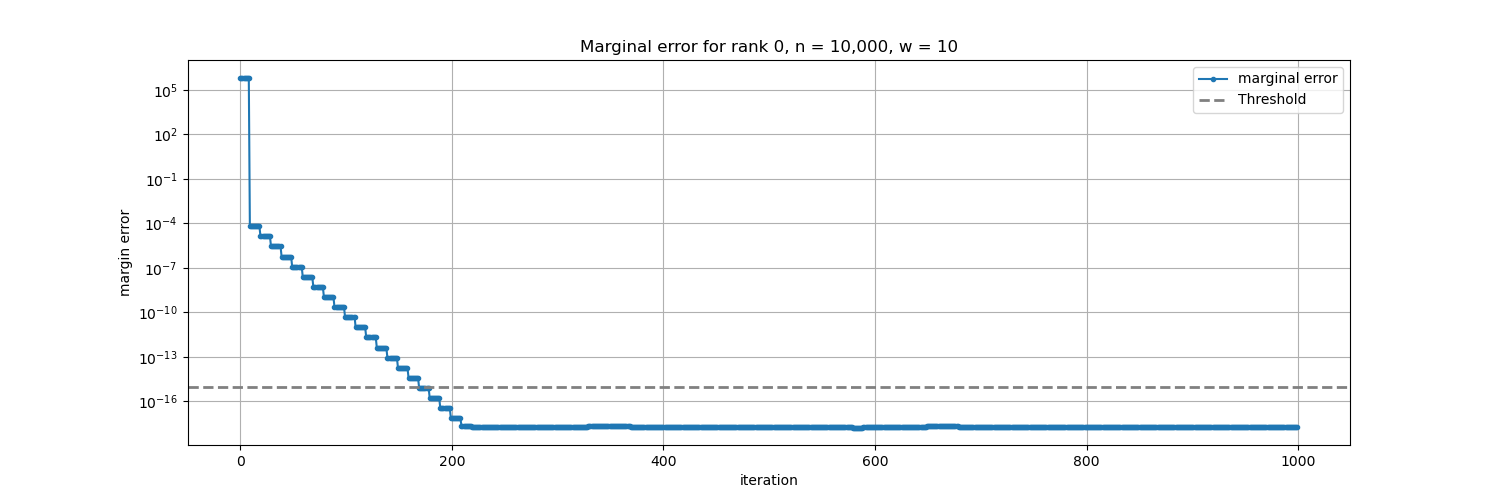}
    \caption{$w = 10$}
  \end{subfigure}
  \medskip
  \begin{subfigure}{0.48\linewidth}
    \centering
    \includegraphics[width=\linewidth]{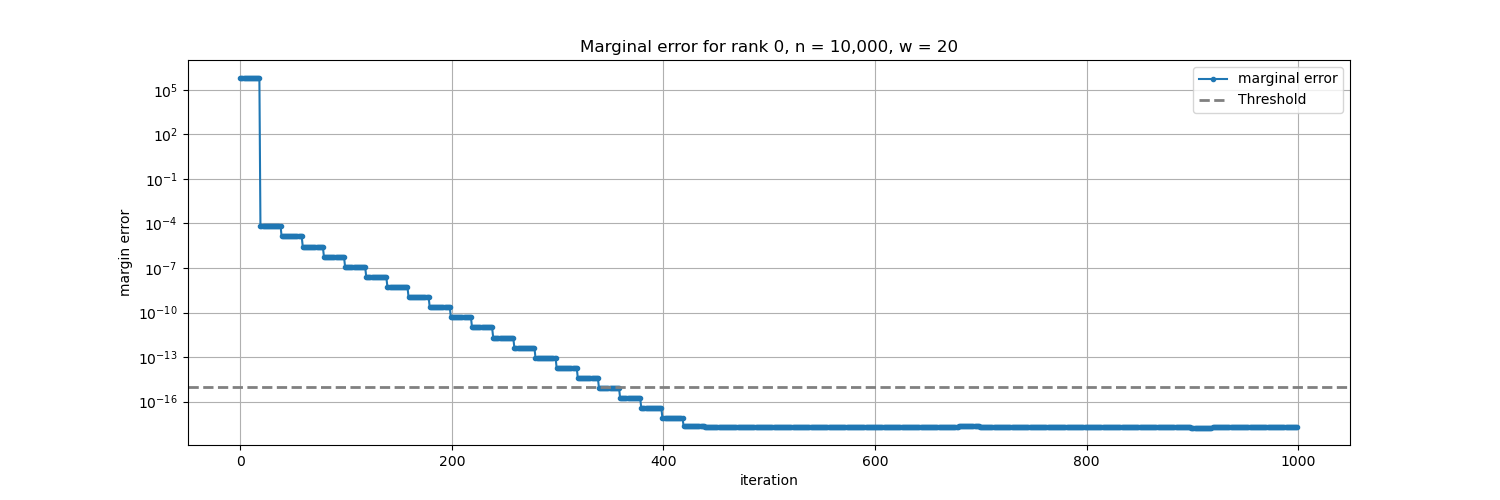}
    \caption{$w = 20$}
  \end{subfigure}\hfill
  \begin{subfigure}{0.48\linewidth}
    \centering
    \includegraphics[width=\linewidth]{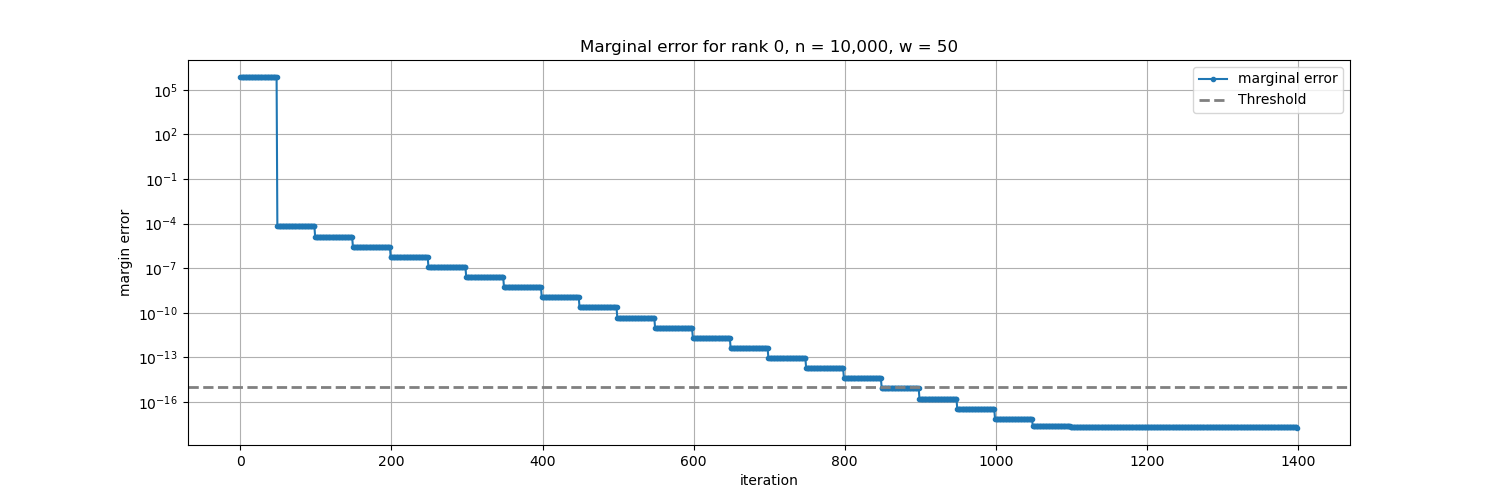}
    \caption{$w = 50$}
  \end{subfigure}
  \medskip
  \begin{subfigure}{0.48\linewidth}
    \centering
    \includegraphics[width=\linewidth]{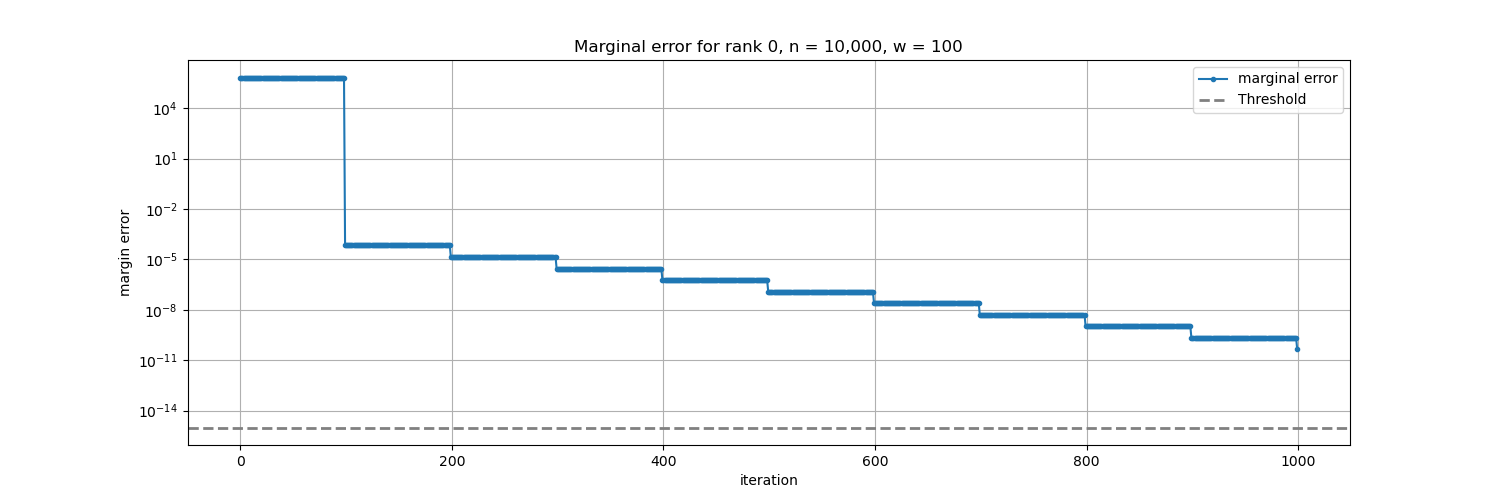}
    \caption{$w = 100$}
  \end{subfigure}
  \caption{Effect of performing $w$ local iterations between broadcasts on the marginal error for the synchronous federated setting. Each panel shows a different value of $w$.}
  \label{fig:bcast_sync}
\end{figure}

Increasing $w$ reduces collective frequency but exacerbates staleness of $u$ and $v$ within a round. Empirically we find $w>1$ harms convergence speed (Figure~\ref{fig:bcast_sync}), matching the model in Section~\ref{sec:comm-comp-model}: as $w$ grows, compute/communication overlap improves but the contraction per unit time degrades due to older scaling vectors.

Across the settings in Figures~~\ref{fig:bcast_sync}-~\ref{fig:bcast_async}, increasing the local-step parameter $w>1$ consistently increases time-to-tolerance, indicating that reduced synchronization frequency does not compensate for the loss of fresh global coupling in our implementation. This observation is empirical and may shift with different kernels, problem regimes, or collective implementations (e.g., different MPI algorithms and network characteristics), so we do not claim that $w>1$ is universally suboptimal.

\subsection{Asynchronous effects}

\begin{figure}[!htbp]
  \centering
  \begin{subfigure}{0.48\linewidth}
    \centering
    \includegraphics[width=\linewidth]{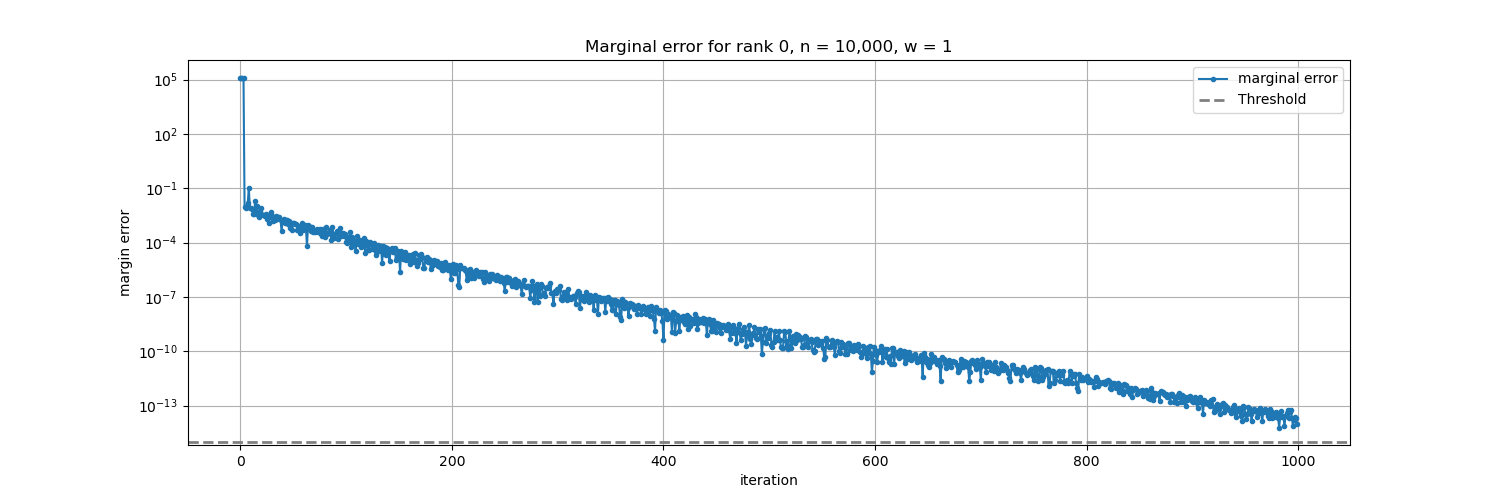}
    \caption{$w = 1$}
  \end{subfigure}\hfill
  \begin{subfigure}{0.48\linewidth}
    \centering
    \includegraphics[width=\linewidth]{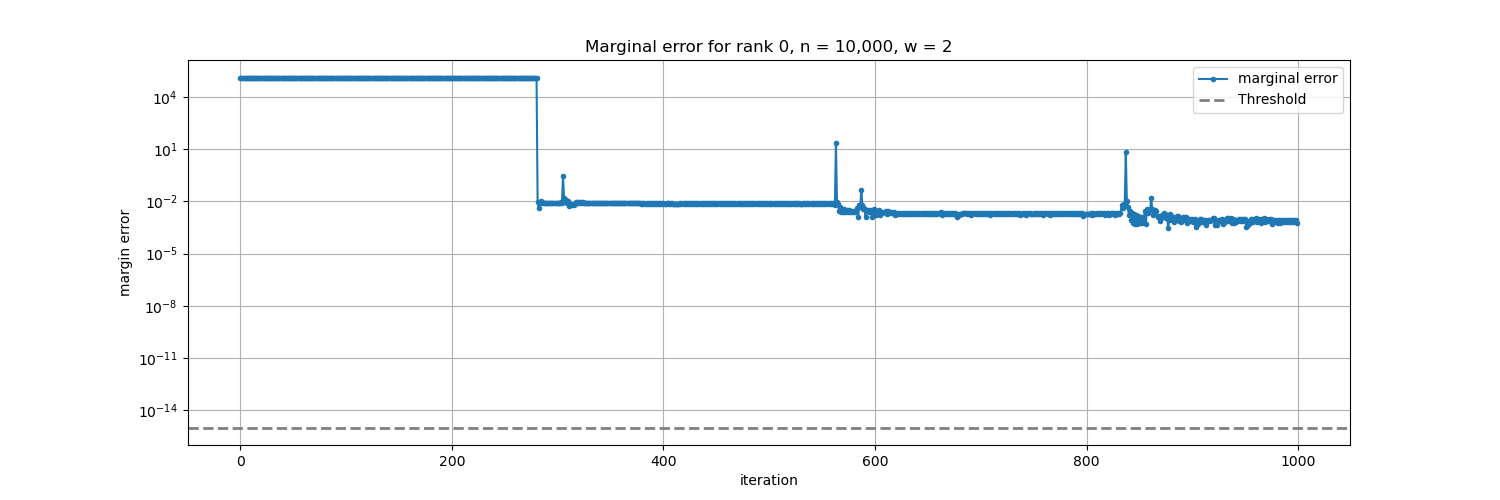}
    \caption{$w = 2$}
  \end{subfigure}
  \medskip
  \begin{subfigure}{0.48\linewidth}
    \centering
    \includegraphics[width=\linewidth]{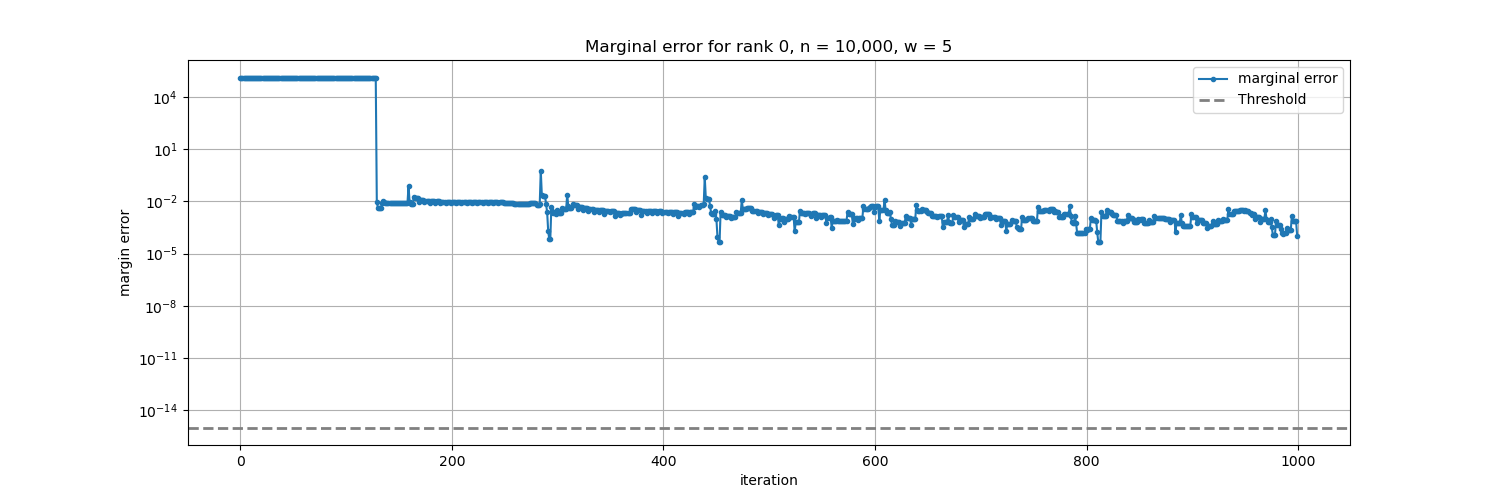}
    \caption{$w = 5$}
  \end{subfigure}\hfill
  \begin{subfigure}{0.48\linewidth}
    \centering
    \includegraphics[width=\linewidth]{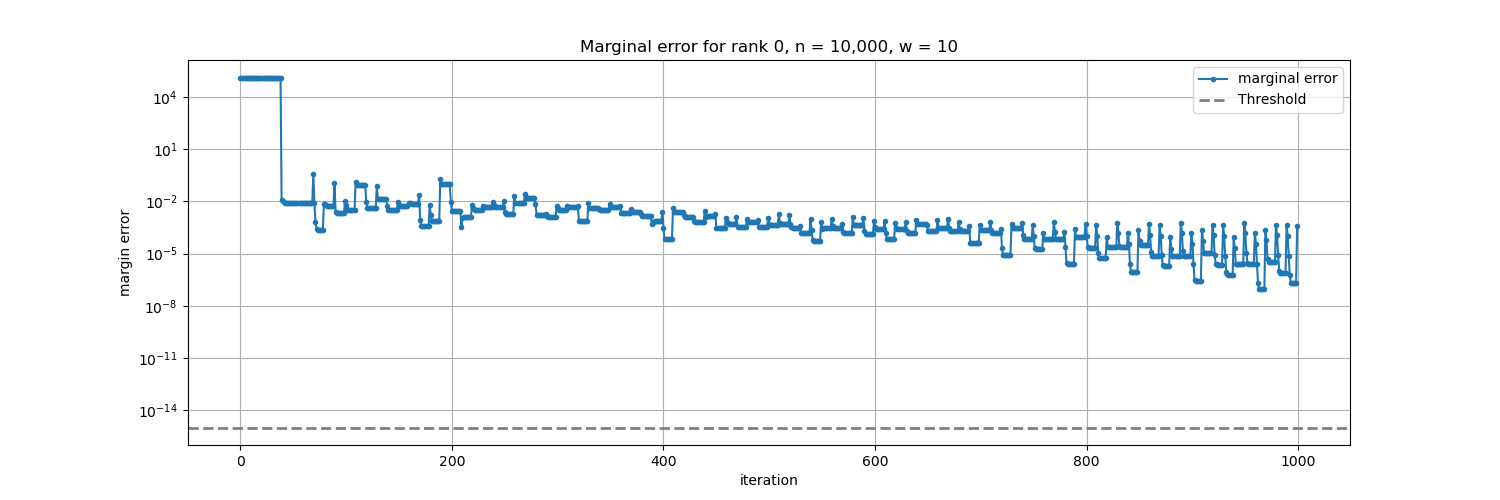}
    \caption{$w = 10$}
  \end{subfigure}
  \medskip
  \begin{subfigure}{0.48\linewidth}
    \centering
    \includegraphics[width=\linewidth]{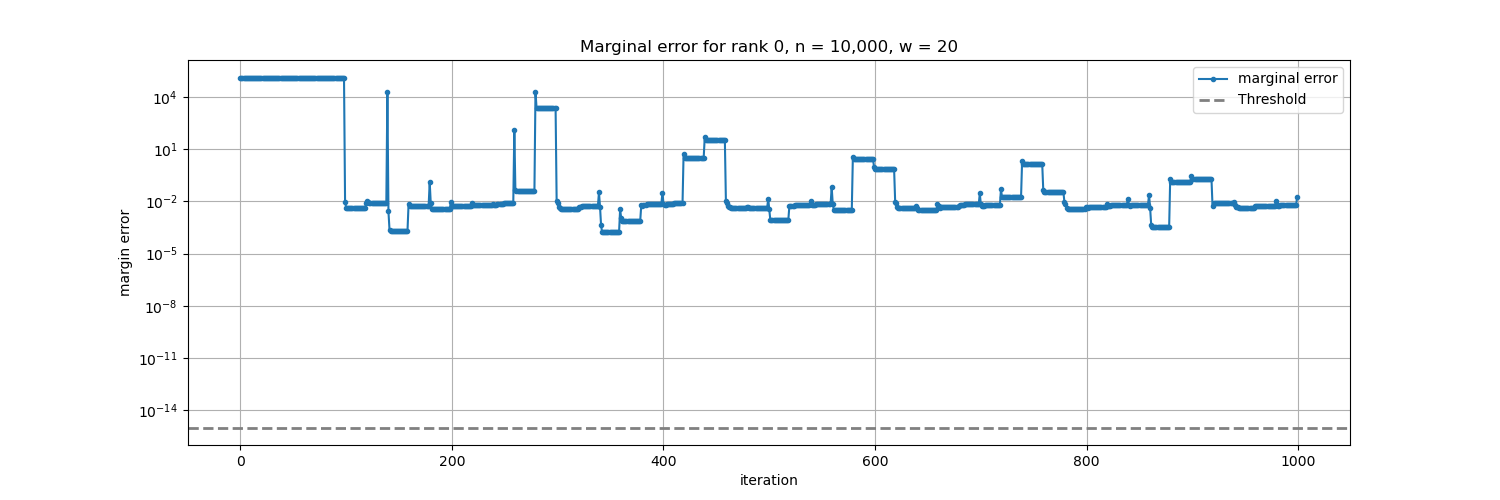}
    \caption{$w = 20$}
  \end{subfigure}\hfill
  \begin{subfigure}{0.48\linewidth}
    \centering
    \includegraphics[width=\linewidth]{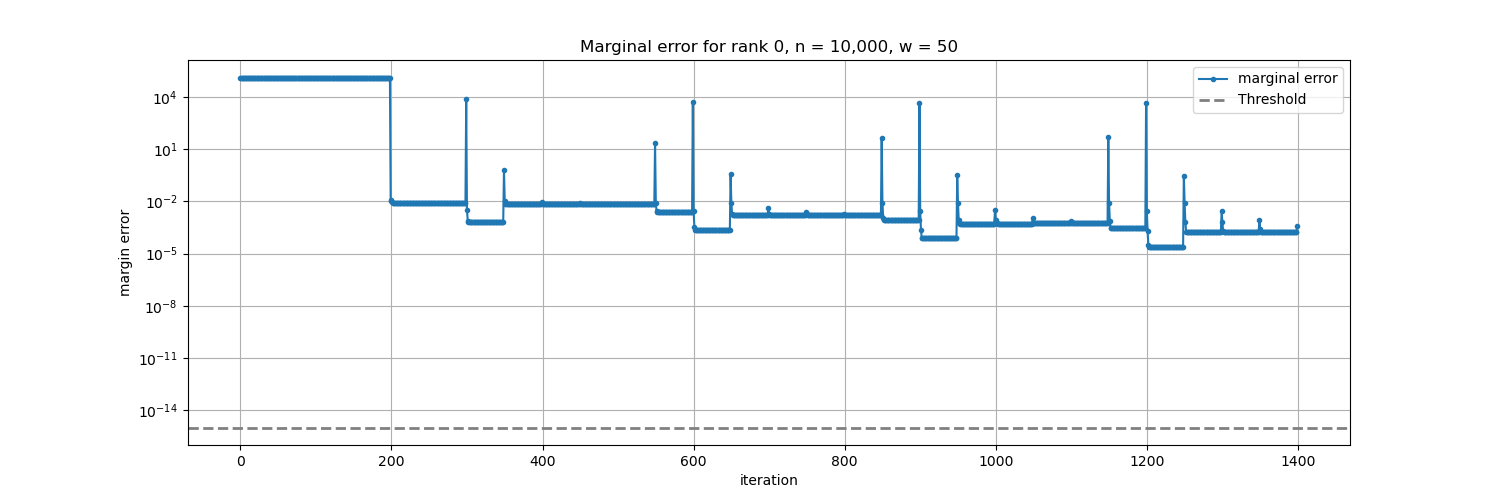}
    \caption{$w = 50$}
  \end{subfigure}
  \medskip
  \begin{subfigure}{0.48\linewidth}
    \centering
    \includegraphics[width=\linewidth]{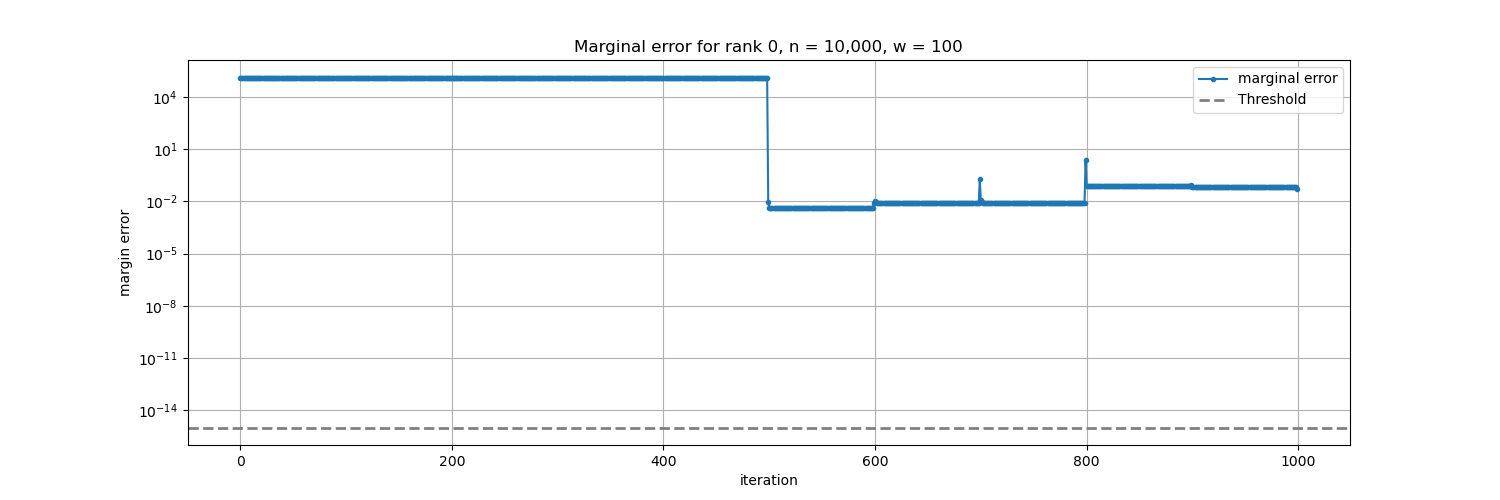}
    \caption{$w = 100$}
  \end{subfigure}
  \caption{Effect of performing $w$ local iterations between broadcasts on the marginal error for the asynchronous federated setting. Each panel shows a different value of $w$.}
  \label{fig:bcast_async}
\end{figure}

Empirically we find $w>1$ harms convergence speed as well (Figure~\ref{fig:bcast_async}), which only worsens the instability of the asynchronous approach.

\begin{figure}[!htbp]
    \centering
    \includegraphics[width=0.6\linewidth]{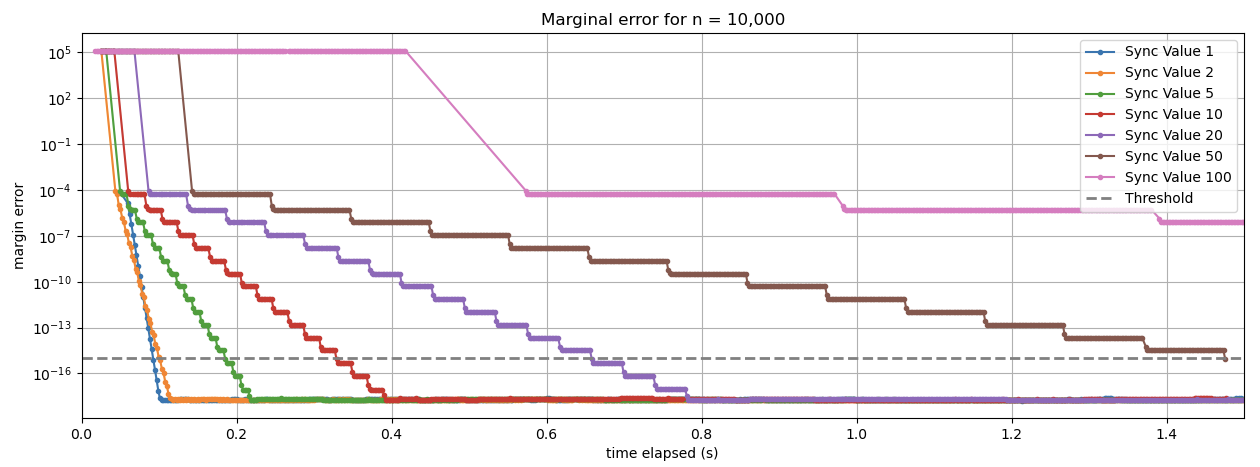}
    \caption{Effect of local iterations before broadcast for different values of $w$ for synchronous federated setting, against time elapsed.}
    \label{fig:bcast_time}
\end{figure}

Moreover, by plotting the marginal error against elapsed time for different values of $w$ on Figures~\ref{fig:bcast_time}, we observe that it worsens the convergence not only in terms of iterations, but also in terms of convergence time in seconds.

\section{Convergence Properties}
\label{sec:con}

We assume: 
(A1) strictly positive kernel entries $K_{ij}>0$; 
(A2) $a,b\in\mathbb{R}^n_{++}$ with $\mathbf{1}^\top a=\mathbf{1}^\top b$; 
(A3) in the asynchronous setting there exists $\tau<\infty$ such that any read of a remote block of $(u,v)$ is at most $\tau$ logical updates stale, and every client-owned block of $(u,v)$ is updated infinitely often (no permanent failures/stragglers).

\begin{proposition}[Synchronous federated iterates match centralized Sinkhorn]
\label{prop:sync}
Under (A1)--(A2), Algorithm~\ref{alg:fedsinksyncall} (All-to-All) and Algorithm~\ref{alg:fedsinksyncstar} (Star-Network) generate exactly the same sequence of global iterates $(u^t,v^t)$ as the centralized Sinkhorn--Knopp iteration, provided they are initialized identically. Consequently, all classical convergence results for Sinkhorn with strictly positive kernels apply directly: $(u^t,v^t)$ converges linearly (in the Hilbert projective metric) to the unique fixed point $(u^\star,v^\star)$, and $P^\star=\mathrm{diag}(u^\star)K\mathrm{diag}(v^\star)$ solves \cref{eq:otprob}.
\end{proposition}

\begin{proposition}[Asynchronous All-to-All convergence under bounded delay]
\label{prop:async}
Under (A1)--(A3), Algorithm~\ref{alg:fedsinkasyncall} converges (in the Hilbert projective metric) to the Sinkhorn scaling solution, i.e., to a fixed point $(u^\star,v^\star)$ that is unique up to the standard rescaling ambiguity $(u^\star,v^\star)\sim(c^{-1}u^\star,cv^\star)$, and yields the unique transport plan
$P^\star=\mathrm{diag}(u^\star)K\mathrm{diag}(v^\star)$ solving \cref{eq:otprob}. This guarantee holds for the standard (undamped) updates $\eta=1$; we use $\eta<1$ in practice as an empirical stabilizer under staleness (see Section~\ref{subsec:async_exp}), without requiring it for the convergence guarantee.
\end{proposition}

\noindent The following section,~\ref{app:convergence}, provides full proof of those propositions.

\section{Convergence proofs}
\label{app:convergence}

We collect here the proofs of Propositions~\ref{prop:sync} and~\ref{prop:async}. We recall assumptions (A1)--(A3) from~\ref{sec:con}.

\subsection{Preliminaries}

Let $\mathbb{R}^n_{++}$ denote the set of strictly positive vectors in $\mathbb{R}^n$. For $x,y \in \mathbb{R}^n_{++}$, the Hilbert projective metric is
\[
  d_{\mathrm{H}}(x,y)
  = \log \left(
    \max_i \frac{x_i}{y_i}
  \right)
  - \log \left(
    \min_i \frac{x_i}{y_i}
  \right).
\]
If $K \in \mathbb{R}^{n \times n}_{++}$ is a matrix with strictly positive entries, the map $x \mapsto Kx$ is a strict contraction in $d_{\mathrm{H}}$\footnote{See e.g.\ Theorem~2.2 in Franklin and Lorenz \cite{franklin1989scaling}, or Chapter~3 of Berman and Plemmons \cite{berman1994nonnegative}.}. Moreover, if $K$ is bistochasticable, then the iterative matrix scaling algorithm of Sinkhorn and Knopp converges linearly in $d_{\mathrm{H}}$ to the unique pair of scaling vectors $(u^\star, v^\star)$ that make $\mathrm{diag}(u^\star)K\mathrm{diag}(v^\star)$ doubly stochastic \cite{sinkhorn1967concerning,franklin1989scaling}.

In our setting, with strictly positive $a,b \in \mathbb{R}^n_{++}$, the classical Sinkhorn--Knopp iteration can be written as the fixed-point map $T : \mathbb{R}^n_{++} \times \mathbb{R}^n_{++} \to \mathbb{R}^n_{++} \times \mathbb{R}^n_{++}$,
\begin{equation}
  T(u,v)
  = \Bigl(
      a \oslash (Kv),\;
      b \oslash (K^\top (a \oslash (Kv)))
    \Bigr),
  \label{eq:T-operator}
\end{equation}
where $\oslash$ denotes elementwise division. Under (A1)--(A2), $T$ is a contraction in the product Hilbert metric and admits a unique fixed point $(u^\star,v^\star)$, and the centralized iteration $(u^{t+1},v^{t+1})=T(u^t,v^t)$ converges linearly to $(u^\star,v^\star)$ (see, e.g., \cite{peyre2019computational}).

\subsection{Proof of Proposition~\ref{prop:sync} (synchronous case)}

We show that both synchronous federated variants (All-to-All and star) generate the same global sequence of iterates as the centralized Sinkhorn iteration, and therefore inherit its linear convergence.

\subsubsection{All-to-All}

Consider Algorithm~\ref{alg:fedsinksyncall} with $w=1$ and suppose that all clients are initialized with the same global vectors $u^0,v^0$. Let $(u_{\text{cent}}^t,v_{\text{cent}}^t)$ denote the iterates produced by the centralized Sinkhorn algorithm, and $(u^t,v^t)$ the concatenation of all local slices in the federated run. We prove by induction on $t$ that $(u^t,v^t)=(u_{\text{cent}}^t,v_{\text{cent}}^t)$ for all $t$.

For $t=0$ this holds by construction. Assume it holds for some $t\ge 0$. At iteration $t+1$, the centralized algorithm computes $q = K v_{\text{cent}}^t$ and updates $u_{\text{cent}}^{t+1} = a \oslash q$. In the federated algorithm, each client $j$ holds the same $v^t$ (by the previous \textbf{Allgather}) and computes its local block $q_{\mathcal{I}_j} = K_{\mathcal{I}_j,:} v^t$. Stacking these blocks yields exactly $q = Kv^t = K v_{\text{cent}}^t$. Each client then updates $u_{\mathcal{I}_j}^{t+1} = a_j \oslash q_{\mathcal{I}_j}$. Concatenating the slices gives $u^{t+1}=a \oslash q = u_{\text{cent}}^{t+1}$. The subsequent \textbf{Allgather} ensures all clients hold this global $u^{t+1}$.

The same argument applies to the column update. Given $u^{t+1}$, each client computes $r_{\mathcal{I}_j} = (K_{:,\mathcal{I}_j})^\top u^{t+1}$, so that stacking yields $r = K^\top u^{t+1}$. The updates $v_{\mathcal{I}_j}^{t+1} = b_j \oslash r_{\mathcal{I}_j}$ concatenate to $v^{t+1} = b \oslash r = v_{\text{cent}}^{t+1}$. Thus, by induction, $(u^t,v^t)=(u_{\text{cent}}^t,v_{\text{cent}}^t)$ for all $t$.

\subsubsection{Star-Network}

In the star variant, the server explicitly stores the full $K$ and global $(u,v)$ and performs the matvecs $q=Kv$ and $r=K^\top u$ centrally. Clients only rescale their local marginals. By inspection of Algorithm~\ref{alg:fedsinksyncstar}, the server's iterates $(u_{\text{srv}}^t,v_{\text{srv}}^t)$ follow exactly the same update rule as the centralized Sinkhorn algorithm, with clients simply acting as remote workers that compute the elementwise divisions $a_j \oslash q_{\mathcal{I}_j}$ and $b_j \oslash r_{\mathcal{I}_j}$. Thus $(u_{\text{srv}}^t,v_{\text{srv}}^t)=(u_{\text{cent}}^t,v_{\text{cent}}^t)$ for all~$t$.

\subsubsection{Convergence}

By the equivalence above, both synchronous federated schemes generate the same global iterates as centralized Sinkhorn. Under (A1)--(A2), the centralized iteration converges linearly in the Hilbert metric to the unique fixed point $(u^\star,v^\star)$ of \cref{eq:T-operator} \cite{sinkhorn1967concerning,franklin1989scaling,peyre2019computational}. Hence the same holds for the federated iterates, which proves Proposition~\ref{prop:sync}.
\qed

\subsection{Proof of Proposition~\ref{prop:async} (asynchronous case)}

We analyze the asynchronous All-to-All scheme under the bounded-delay and fairness assumption (A3). The key point is that the asynchronous implementation performs
block updates (client-owned slices) of a contractive fixed-point map.

\paragraph{A contractive map with the Sinkhorn fixed point}
Define the (Jacobi-form) scaling map $F:\mathbb{R}^n_{++}\times\mathbb{R}^n_{++}\to
\mathbb{R}^n_{++}\times\mathbb{R}^n_{++}$ by
\begin{equation}
  F(u,v)\;:=\;\Bigl(a \oslash (Kv),\; b \oslash (K^\top u)\Bigr).
  \label{eq:F-jacobi}
\end{equation}
A pair $(u^\star,v^\star)$ is a fixed point of $F$ if and only if it satisfies the Sinkhorn scaling equations $u^\star = a\oslash(Kv^\star)$ and $v^\star = b\oslash(K^\top u^\star)$, and therefore yields the unique entropic-OT plan $P^\star=\mathrm{diag}(u^\star)K\mathrm{diag}(v^\star)$ (the scalings are unique up to $(u^\star,v^\star)\sim(c^{-1}u^\star,cv^\star)$, while $P^\star$ is unique).

Equip $\mathcal{X}=\mathbb{R}^{2n}_{++}$ with the product Hilbert metric
\[
d\big((u,v),(u',v')\big)\;:=\;\max\!\big\{d_{\mathrm H}(u,u'),\ d_{\mathrm H}(v,v')\big\}.
\]
Under (A1)--(A2), the positive linear maps $x\mapsto Kx$ and $x\mapsto K^\top x$ are strict contractions in $d_{\mathrm H}$ (Birkhoff contraction; see Franklin--Lorenz
\cite{franklin1989scaling} or Berman--Plemmons \cite{berman1994nonnegative}). Moreover, for fixed $a\in\mathbb{R}^n_{++}$, the map $x\mapsto a\oslash x$ is an
\emph{isometry} in $d_{\mathrm H}$:
$d_{\mathrm H}(a\oslash x, a\oslash y)=d_{\mathrm H}(x,y)$.
Therefore there exists $\gamma\in(0,1)$ (depending on $K$) such that
\[
d_{\mathrm H}\!\big(a\oslash(Kv),\, a\oslash(Kv')\big)\le \gamma\, d_{\mathrm H}(v,v'),
\qquad
d_{\mathrm H}\!\big(b\oslash(K^\top u),\, b\oslash(K^\top u')\big)\le \gamma\, d_{\mathrm H}(u,u').
\]
Taking a maximum over the two blocks shows that $F$ is $\gamma$-contractive in $d$:
\begin{equation}
d\big(F(u,v),F(u',v')\big)\;\le\;\gamma\, d\big((u,v),(u',v')\big).
\label{eq:F-contraction}
\end{equation}

\paragraph{Asynchronous block-coordinate model}
Let $x^t=(u^t,v^t)$ denote the global state after $t$ logical block-updates. Blocks correspond to client-owned slices of $u$ and $v$. At each logical time $t$, some block is updated using a delayed read $\hat x^t$ whose components satisfy $0\le t-s_i(t)\le\tau$ (bounded delay), while other blocks are left unchanged; and every block is updated infinitely often (fairness), per (A3). With $\eta=1$, the update applied by Algorithm~\ref{alg:fedsinkasyncall} is exactly a totally asynchronous block update of the contractive map $F$ in \cref{eq:F-jacobi}.

\paragraph{Convergence}
Since $F$ is a contraction in the complete metric space induced by the product Hilbert metric (modulo the standard projective scaling equivalence), the totally asynchronous
iteration theorem of Bertsekas--Tsitsiklis \cite[Ch.~6]{bertsekas1989parallel} applies. Hence the asynchronous iterates converge geometrically (in $d$) to the fixed-point
class of $(u^\star,v^\star)$, and therefore to the unique transport plan $P^\star$. This proves Proposition~\ref{prop:async}.
\qed

\paragraph{Remark (practical damping)}
For $\eta<1$, Algorithm~\ref{alg:fedsinkasyncall} performs an under-relaxed update in the \emph{value domain}, which empirically improves stability under staleness (Section~\ref{subsec:async_exp}). A formal contraction proof for this arithmetic under-relaxation in the Hilbert metric is not required for Proposition~\ref{prop:async} as stated (which covers $\eta=1$).

\subsection{Tree-based collective approximations}
\label{app:collectives}

For back-of-the-envelope reasoning on a well-provisioned cluster with a tree-based collective implementation, a common leading-order approximation is
\begin{align}
T_{\mathrm{AG}}(n,c)
&\approx
\alpha \log_2 c \;+\; \beta\,\frac{c-1}{c}\,n,
\label{eq:collectives-closedform_AG}
\\
T_{\mathrm{BC}}(n,c)
&\approx
\alpha \log_2 c \;+\; \beta\,n,
\label{eq:collectives-closedform_SC}
\\
T_{\mathrm{SR}}(n,c)
&\approx
\alpha \log_2 c \;+\; \beta\,n,
\label{eq:collectives-closedform_GA}
\end{align}
up to constant factors that depend on the MPI/NCCL implementation and routing. In the main text we instead treat $(\alpha_{\cdot}(c),\beta_{\cdot}(c))$ as measured parameters for the target system.

\section{Additional details for Synchronous All-to-All federation}
\label{app:sync_details}

\subsection{CPU marginal-error-versus-time convergence}
\label{app:cpu_sync}

\begin{figure}[!htbp]
  \centering
  \begin{subfigure}[b]{0.49\linewidth}
    \centering
    \includegraphics[width=\linewidth]{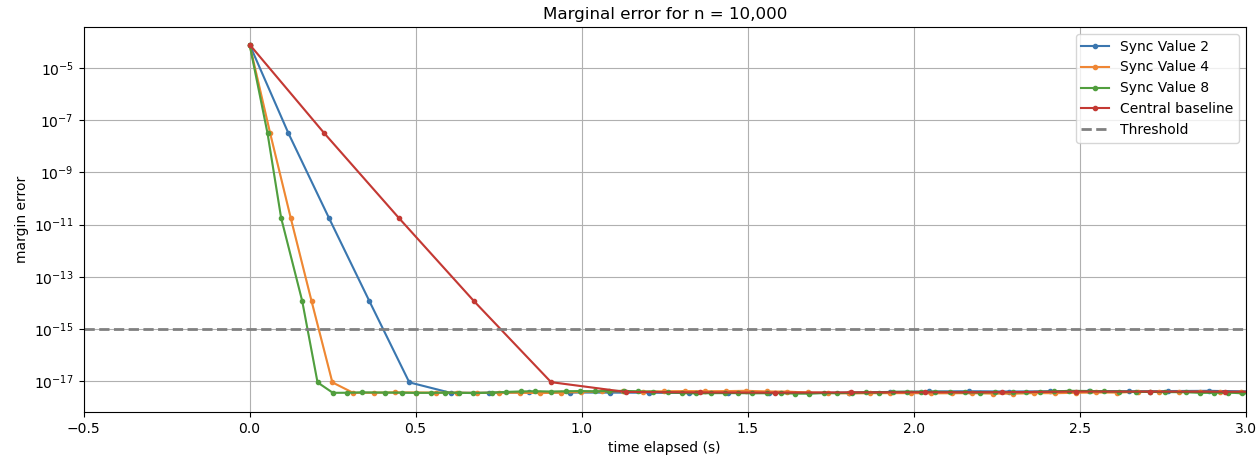}
    \caption{$n = 10{,}000$.}
    \label{fig:sync_cpu_cvg_offset}
  \end{subfigure}
  \hfill
  \begin{subfigure}[b]{0.49\linewidth}
    \centering
    \includegraphics[width=\linewidth]{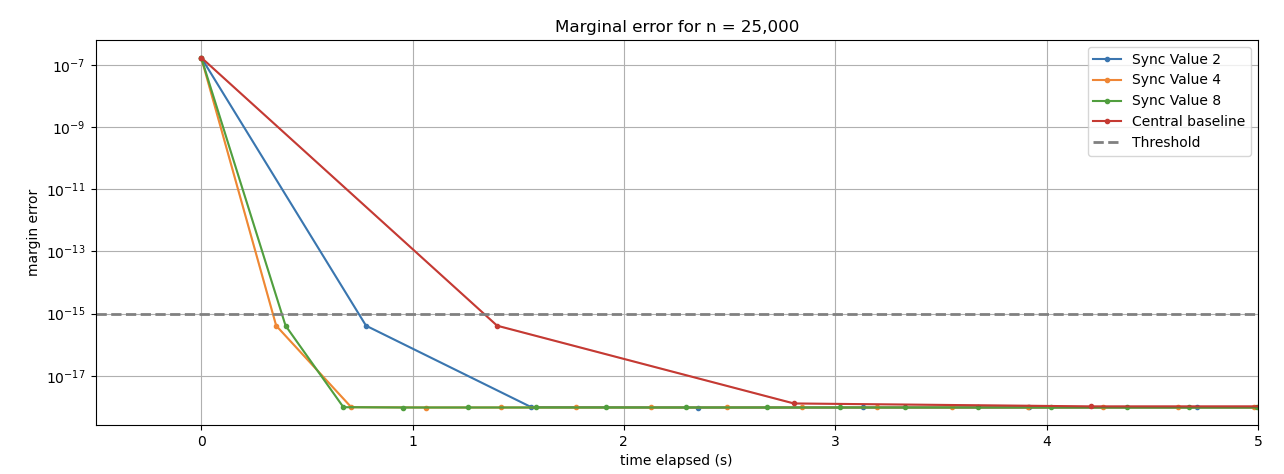}
    \caption{$n = 25{,}000$.}
    \label{fig:sync_cpu_cvg_25k}
  \end{subfigure}
  \caption{Synchronous CPU convergence: marginal error on $a$ versus elapsed time for different node counts and problem sizes. Starting points were equalized to isolate the steady-state convergence behavior.}
  \label{fig:sync_cpu_cvg_all}
\end{figure}

\subsection{Synchronous Star-Network Pseudocode}
\label{app:sync_start_netowkr_pseudo}

\begin{algorithm}[!htbp]
\DontPrintSemicolon
\caption{Synchronous Federated Sinkhorn (Star-Network)}
\label{alg:fedsinksyncstar}
\small
\KwIn{Client $j$ holds $a_{j},b_{j}$; \textbf{Server} holds $C$ (or $K$)}
\KwOut{$u,v$ (and optionally $P=\mathrm{diag}(u)K\mathrm{diag}(v)$) at the server}
\textbf{Server:} initialize $u\gets\mathbf{1}$, $v\gets\mathbf{1}$\;
\For{$t=1,2,\dots,\texttt{IterMax}$}{
  \tcp{Update $u$ (enforce row marginal $a$)}
  \textbf{Client $j \to$ Server:} send $v_{\mathcal{I}_j}$\;
  \textbf{Server:} assemble $v$; compute $q\gets K v$;\; \textbf{Bcast} $q$ to all clients\;
  \textbf{Client $j$:} $u_{\mathcal{I}_j}\gets a_{j}\oslash q_{\mathcal{I}_j}$;\; \textbf{Client $j \to$ Server:} send $u_{\mathcal{I}_j}$\;
  \tcp{Update $v$ (enforce column marginal $b$)}
  \textbf{Server:} assemble $u$; compute $r\gets K^\top u$;\; \textbf{Bcast} $r$ to all clients\;
  \textbf{Client $j$:} $v_{\mathcal{I}_j}\gets b_{j}\oslash r_{\mathcal{I}_j}$\;
}
\end{algorithm}

\section{Additional details for Asynchronous All-to-All federation}
\label{app:async_details}

\subsection{Asynchronous All-to-All pseudocode}
\label{app:async_alltoall_pseudo}

\begin{algorithm}[!htbp]
\DontPrintSemicolon
\caption{Asynchronous Federated Sinkhorn (All-to-All)}
\label{alg:fedsinkasyncall}
\small
\KwIn{Client $j$ holds $a_j,b_j$ and local kernel shards $K_{\mathcal{I}_j,:}$ and $K_{:,\mathcal{I}_j}$; relaxation $\eta\in(0,1]$; sync period $w$}
\KwOut{Local views of global scalings $u,v$ (eventually consistent after draining in-flight messages)}
\textbf{Each client $j$:} initialize local slice $u_{\mathcal{I}_j}\gets\mathbf{1}$, $v_{\mathcal{I}_j}\gets\mathbf{1}$ and cached global vectors $u,v$\;
\For{$t=1,2,\dots,\texttt{IterMax}$}{
  \tcp{Use the most recent cached global scalings (may be stale)}
  \textbf{Client $j$:} compute $q_{\mathcal{I}_j}\gets K_{\mathcal{I}_j,:}v$;\;
  \textbf{Client $j$:} $\hat u_{\mathcal{I}_j}\gets a_j\oslash q_{\mathcal{I}_j}$;\;
  \textbf{Client $j$:} $u_{\mathcal{I}_j}\gets (1-\eta)u_{\mathcal{I}_j}+\eta\,\hat u_{\mathcal{I}_j}$;\;
  \textbf{Client $j$:} compute $r_{\mathcal{I}_j}\gets (K_{:,\mathcal{I}_j})^\top u$;\;
  \textbf{Client $j$:} $\hat v_{\mathcal{I}_j}\gets b_j\oslash r_{\mathcal{I}_j}$;\;
  \textbf{Client $j$:} $v_{\mathcal{I}_j}\gets (1-\eta)v_{\mathcal{I}_j}+\eta\,\hat v_{\mathcal{I}_j}$;\;

  \If(\tcp*[f]{Periodic non-blocking exchange}){$t \bmod w = 0$}{
    \textbf{Client $j$:} post non-blocking \textbf{Isend} of $u_{\mathcal{I}_j}$ and $v_{\mathcal{I}_j}$ to all peers\;
    \textbf{Client $j$:} post non-blocking \textbf{Irecv} buffers for $u_{\mathcal{I}_k},v_{\mathcal{I}_k}$ from all peers $k\neq j$\;
  }
  \tcp{Progress communication opportunistically}
  \textbf{Client $j$:} \textbf{Test} for completed receives; for each completed message, update the cached global slices\;
}
\tcp{Termination: drain outstanding requests (no global barrier required)}
\textbf{Client $j$:} \textbf{Wait} for outstanding send/recv requests; apply any final received slices\;
\end{algorithm}

\subsection{Robustness across node counts and step sizes}

\begin{figure}[!htbp]
  \centering
  \begin{subfigure}[b]{0.49\linewidth}
    \centering
    \includegraphics[width=\linewidth]{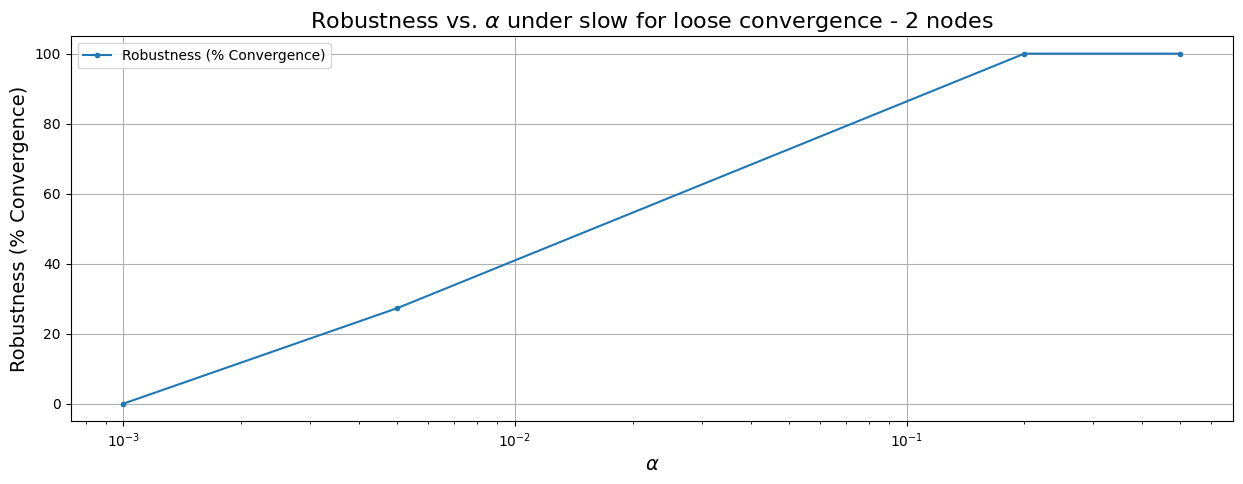}
    \caption{2 nodes.}
    \label{fig:robust_2}
  \end{subfigure}
  \hfill
  \begin{subfigure}[b]{0.49\linewidth}
    \centering
    \includegraphics[width=\linewidth]{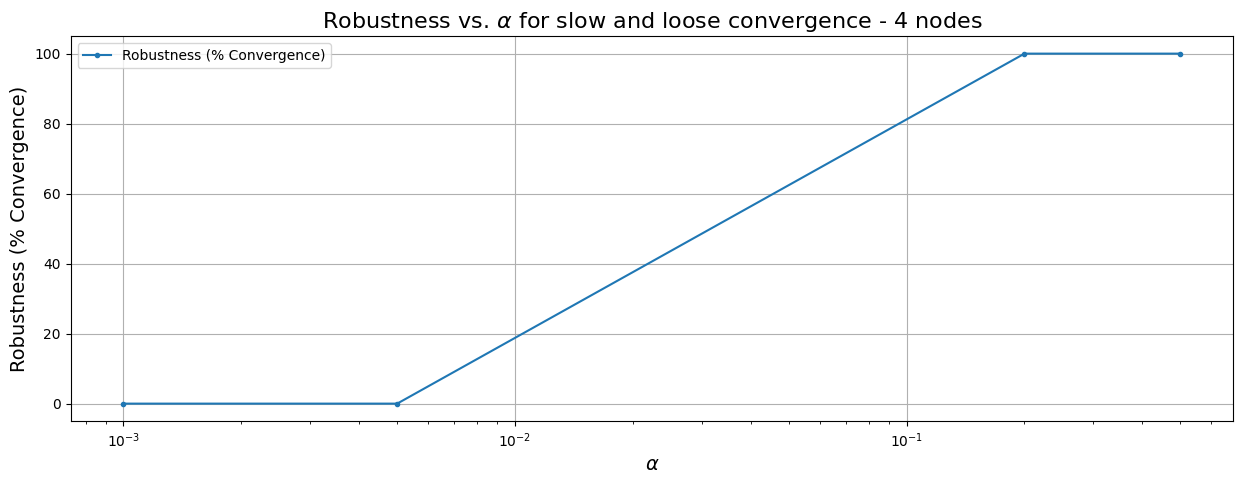}
    \caption{4 nodes.}
    \label{fig:robust_4}
  \end{subfigure}
  \\
  \vspace{0.5em}
  \begin{subfigure}[b]{0.49\linewidth}
    \centering
    \includegraphics[width=\linewidth]{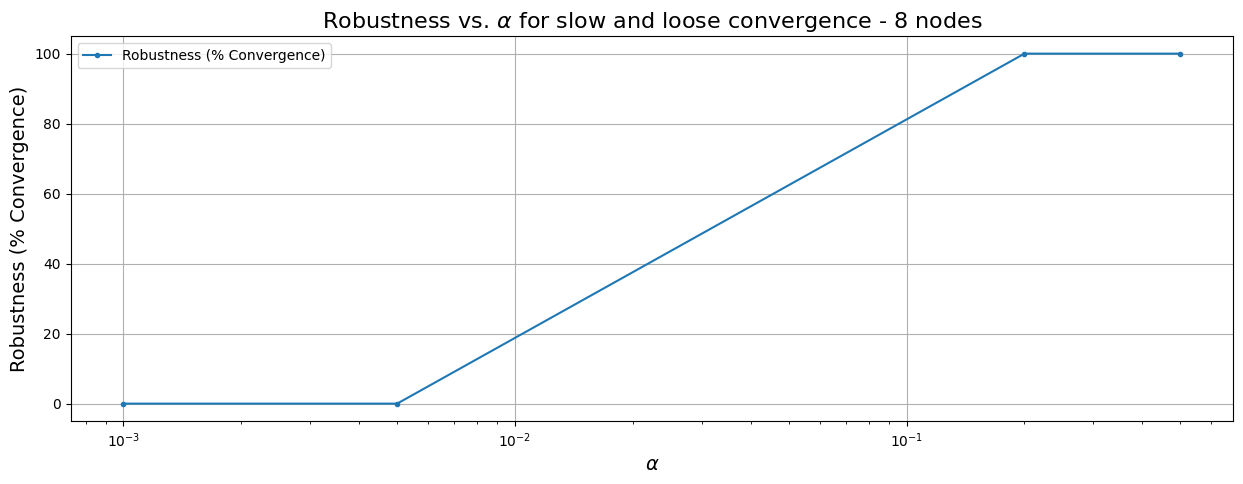}
    \caption{8 nodes.}
    \label{fig:robust_8}
  \end{subfigure}
  \caption{Percentage of simulations that converged within the experiment as a function of the step size $\eta$ for different node counts.}
  \label{fig:robust}
\end{figure}

We performed a robustness experiment with randomized inputs. For each configuration, we generated independent problems with $n = 10\,000$ and recorded:
(i) the percentage of runs that reached a marginal error below a ``loose'' tolerance of $10^{-5}$ or a ``tight'' tolerance of $10^{-12}$;
(ii) the average time to convergence; and
(iii) the fraction of runs that timed out or diverged (no convergence within $3000$ iterations).
Table~\ref{tab:summary_stability_all} summarizes the results for $c = 2,4,8$ nodes, keeping only the best performing step size $\eta$ for each asynchronous setting, and Figure~\ref{fig:robust} shows the convergence probability as a function of~$\eta$.

\begin{table}[!htbp]
  \centering
  \scriptsize
  \setlength{\tabcolsep}{3pt}
  \renewcommand{\arraystretch}{0.9}
  \begin{tabular}{cllrrrrrrrr}
    \toprule
    Nodes & Scheme & Limit &
    \multicolumn{2}{c}{Time (s)} &
    \multicolumn{2}{c}{Conv. (\%)} &
    \multicolumn{2}{c}{Timeout (\%)} &
    \multicolumn{2}{c}{Divergence (\%)} \\
    & & &
    fast & slow & fast & slow & fast & slow & fast & slow \\
    \cmidrule(lr){1-3}\cmidrule(lr){4-5}\cmidrule(lr){6-7}\cmidrule(lr){8-9}\cmidrule(lr){10-11}
    2 & Sync All-to-All & loose & 0.77 & 1.01 & 96.77 & 100.0 & 3.22 & 0.0 & 0.0 & 0.0 \\
    2 & Sync All-to-All & tight & 1.10 & 1.38 & 97.87 & 100.0 & 2.13 & 0.0 & 0.0 & 0.0 \\
    2 & Sync Star       & loose & 1.00 & 0.66 & 95.83 & 100.0 & 4.17 & 0.0 & 0.0 & 0.0 \\
    2 & Sync Star       & tight & 1.01 & 1.69 & 100.0 & 100.0 & 0.0 & 0.0 & 0.0 & 0.0 \\
    2 & Async ($\eta=0.5$) & loose & 0.77 & 1.01 & 96.77 & 100.0 & 3.23 & 0.0 & 0.0 & 0.0 \\
    2 & Async ($\eta=0.5$) & tight & 1.10 & 1.38 & 97.87 & 100.0 & 2.13 & 0.0 & 0.0 & 0.0 \\
    \midrule
    4 & Sync All-to-All & loose & 9.71 & 3.36 & 66.67 & 100.0 & 33.33 & 0.0 & 0.0 & 0.0 \\
    4 & Sync All-to-All & tight & 4.58 & 4.89 & 80.00 & 100.0 & 20.00 & 0.0 & 0.0 & 0.0 \\
    4 & Sync Star       & loose & 12.86 & 15.51 & 66.67 & 100.0 & 33.33 & 0.0 & 0.0 & 0.0 \\
    4 & Sync Star       & tight & 16.79 & 16.31 & 66.67 & 100.0 & 33.33 & 0.0 & 0.0 & 0.0 \\
    4 & Async ($\eta=0.5$) & loose & 4.48 & 11.21 & 66.67 & 100.0 & 33.33 & 0.0 & 0.0 & 0.0 \\
    4 & Async ($\eta=0.5$) & tight & 6.20 & 45.58 & 66.67 & 66.67 & 33.33 & 0.0 & 0.0 & 33.33 \\
    \midrule
    8 & Sync All-to-All & loose & 1.19 & 1.94 & 100.0 & 100.0 & 0.0 & 0.0 & 0.0 & 0.0 \\
    8 & Sync All-to-All & tight & 1.79 & 1.36 & 100.0 & 100.0 & 0.0 & 0.0 & 0.0 & 0.0 \\
    8 & Sync Star       & loose & 0.48 & 0.49 & 100.0 & 100.0 & 0.0 & 0.0 & 0.0 & 0.0 \\
    8 & Sync Star       & tight & 0.95 & 0.95 & 100.0 & 100.0 & 0.0 & 0.0 & 0.0 & 0.0 \\
    8 & Async ($\eta=0.5$) & loose & 2.97 & 6.96 & 80.0 & 100.0 & 20.0 & 0.0 & 0.0 & 0.0 \\
    8 & Async ($\eta=0.5$) & tight & 4.39 & 16.75 & 80.0 & 80.0 & 20.0 & 0.0 & 0.0 & 20.0 \\
    \bottomrule
  \end{tabular}
  \caption{Robustness of synchronous and asynchronous federated Sinkhorn across node counts, tolerances, and time limits. Only the asynchronous run with the best-performing step size $\eta$ is shown for each configuration. For 2 nodes, the async schedule degenerates to sync under our communication implementation, yielding identical timings.}
  \label{tab:summary_stability_all}
\end{table}

Figure~\ref{fig:robust} shows the convergence robustness for different values of $\eta$, ranging from $0.001$ to $0.5$, under the slow/loose criterion.

\subsection{Additional asynchronous trajectories with varying step sizes}

\begin{figure}[!htbp]
  \centering
  \begin{subfigure}[b]{0.48\textwidth}
    \centering
    \includegraphics[width=\linewidth]{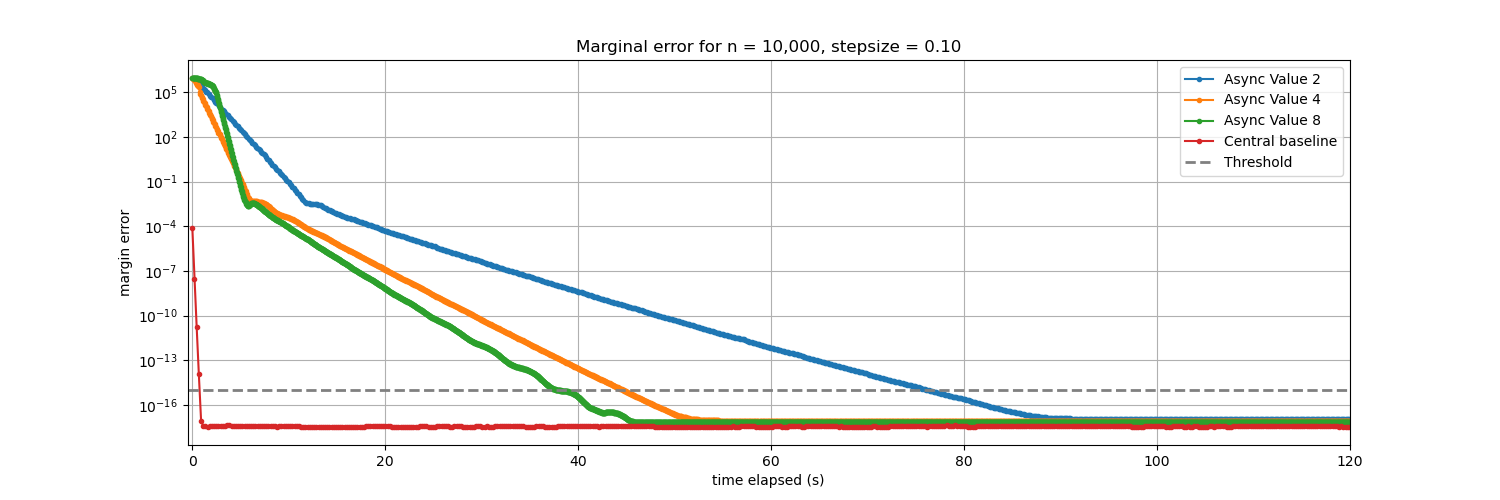}
    \includegraphics[width=\linewidth]{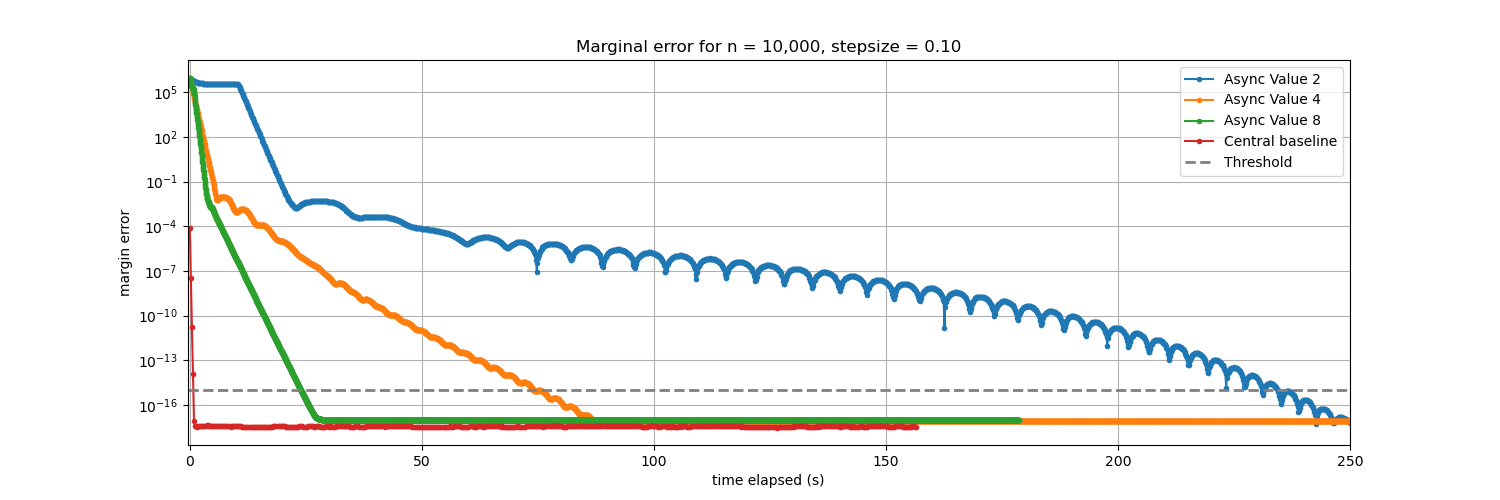}
    \caption{Two runs at $\eta = 0.1$.}
    \label{fig:times_and_error_cpu_async_stepsize10}
  \end{subfigure}
  \hfill
  \begin{subfigure}[b]{0.48\textwidth}
    \centering
    \includegraphics[width=\linewidth]{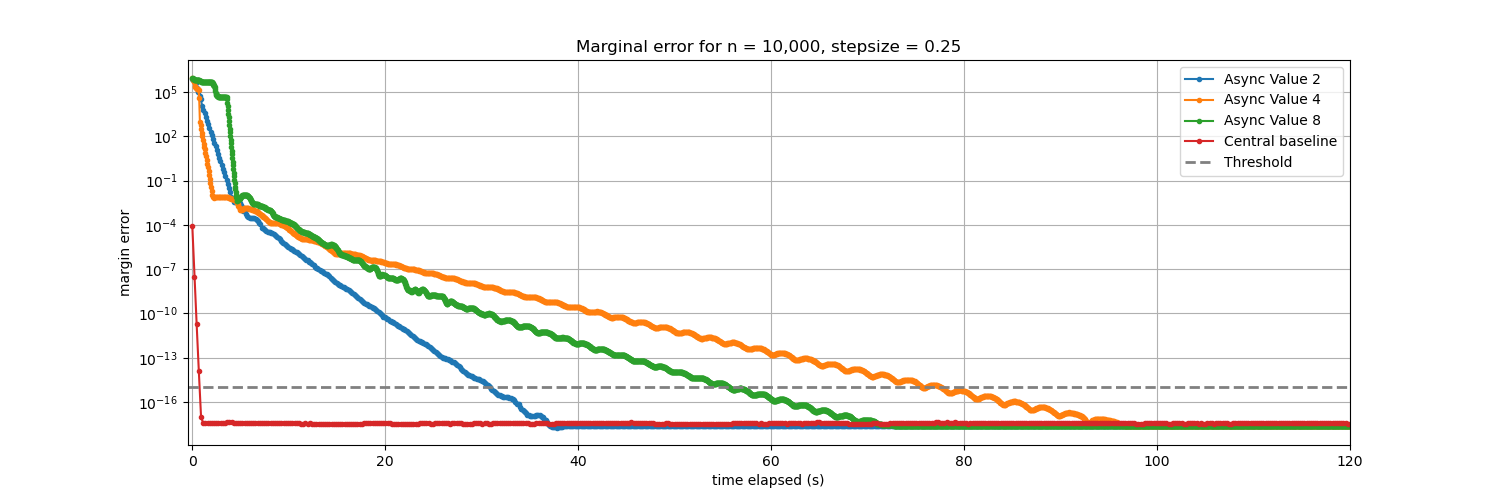}
    \includegraphics[width=\linewidth]{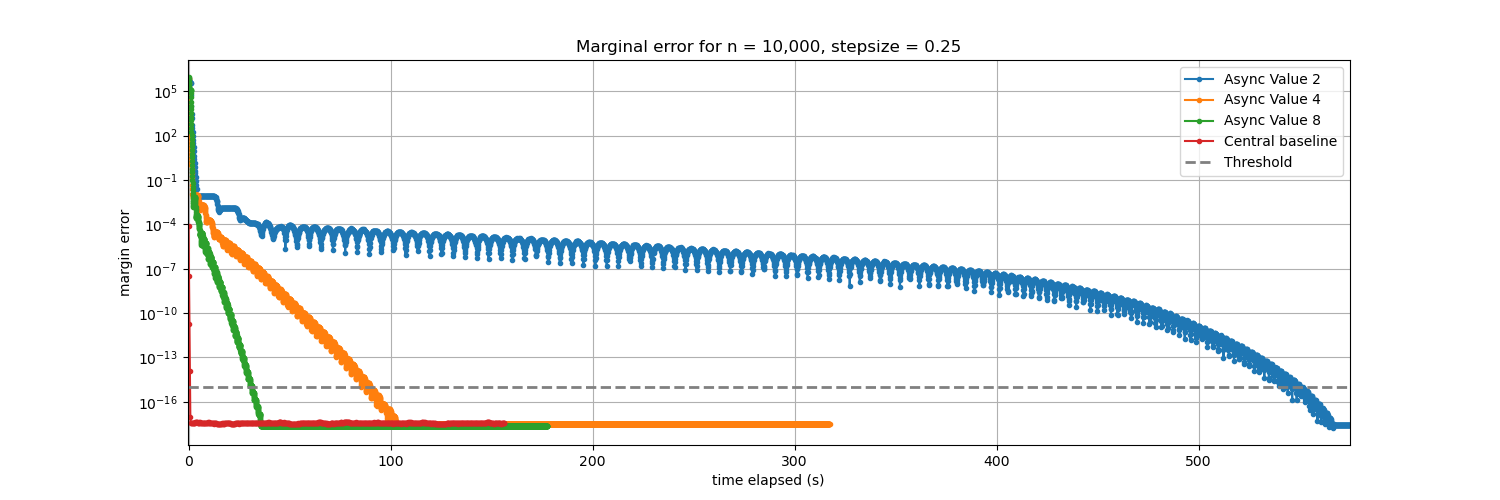}
    \caption{Two runs at $\eta = 0.25$.}
    \label{fig:times_and_error_cpu_async_stepsize25}
  \end{subfigure}
  \caption{Additional asynchronous trajectories for smaller step sizes
  $\eta \in \{0.1, 0.25\}$ with identical initial conditions.}
  \label{fig:async_cpu_appendix}
\end{figure}

\subsection{Computation and communication times}

\begin{figure}[!htbp]
    \centering
    \includegraphics[width=0.5\linewidth]{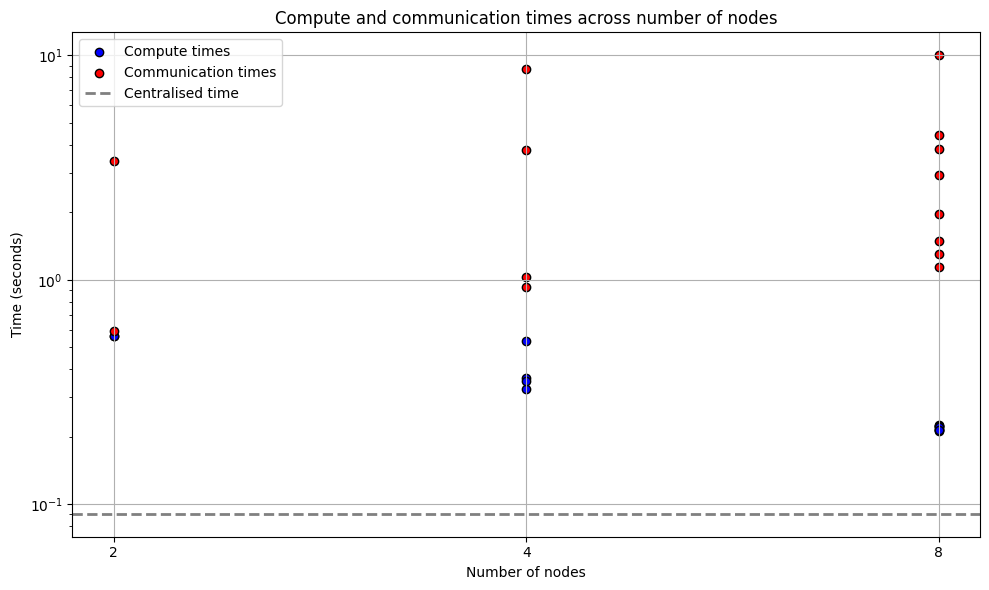}
    \caption{Execution times for different numbers of nodes at $n = 10\,000$ and $T = 250$ iterations for the asynchronous All-to-All scheme. Experiments were performed on a cluster, allowing up to 8 local MPI processes.}
    \label{fig:async_node_times}
\end{figure}

Figure~\ref{fig:async_node_times} reports, for each node and each configuration, the time spent in local matvecs and the time spent in communication. Compute times stay in a similar range across $c \in \{2,4,8\}$ and remain above the centralized baseline, while communication dominates the overall runtime and becomes more variable as the number of nodes increases.

\subsection{Delay distributions}
\label{app:delay_distributions}

\begin{figure}[!htbp]
    \centering
    \includegraphics[width=0.6\linewidth]{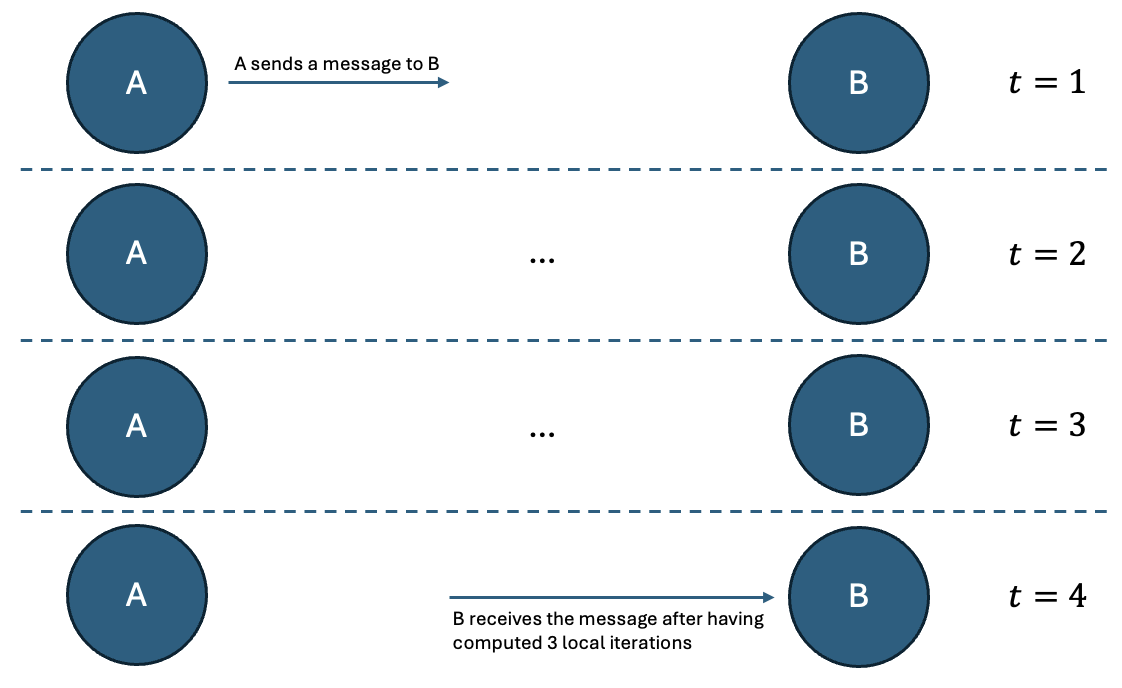}
    \caption{Illustration of how the delays are counted. In this example, node B has the time to perform 3 local iterations before receiving node A's message, making it 3 iterations old. We use $\tau$ to denote the message's age.}
    \label{fig:delay_count_illustration}
\end{figure}

A key source of variability in the asynchronous scheme is the age of the
information each node uses. When node $A$ publishes an updated slice and
node $B$ only receives it after performing several local iterations, we say that
the message is $\tau$ iterations stale at $B$ (cf.\ Figure~\ref{fig:delay_count_illustration} for an illustration of how $\tau$ is counted).

\begin{figure}[!htbp]
    \centering
    \includegraphics[width=0.5\linewidth]{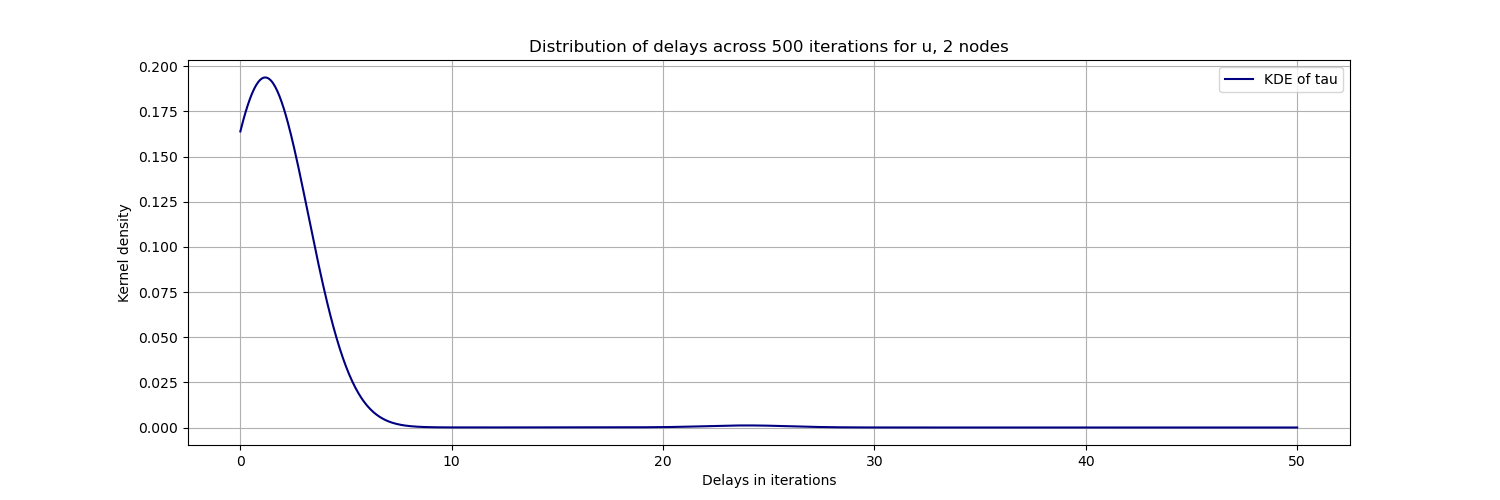}\hfill
    \includegraphics[width=0.5\linewidth]{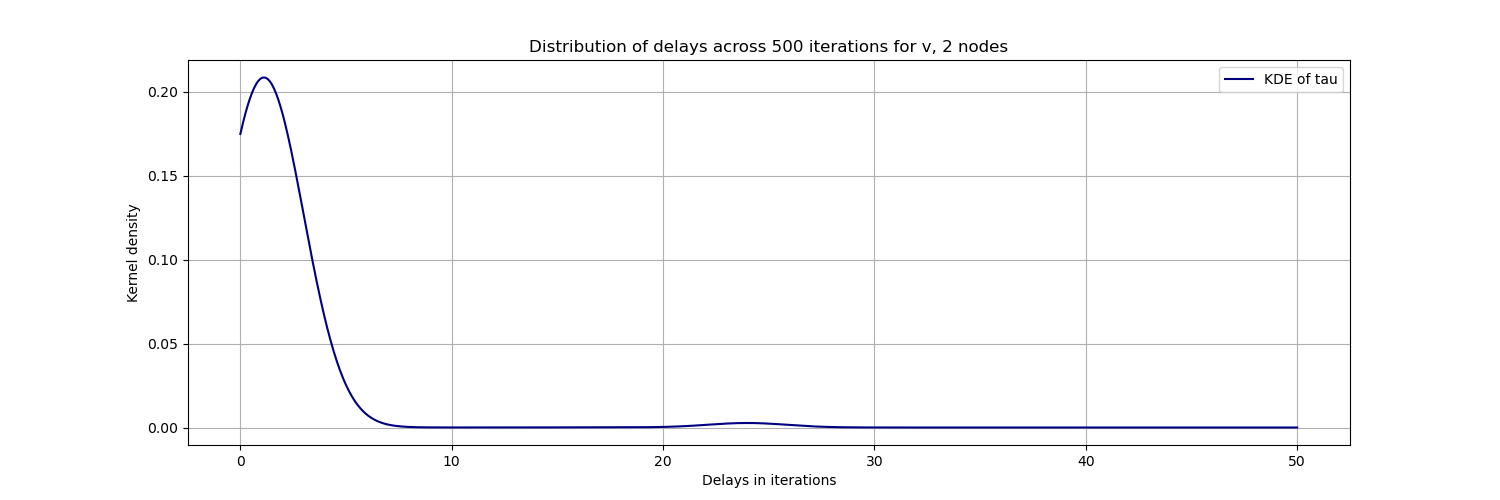}

    \vspace{1em}

    \includegraphics[width=0.5\linewidth]{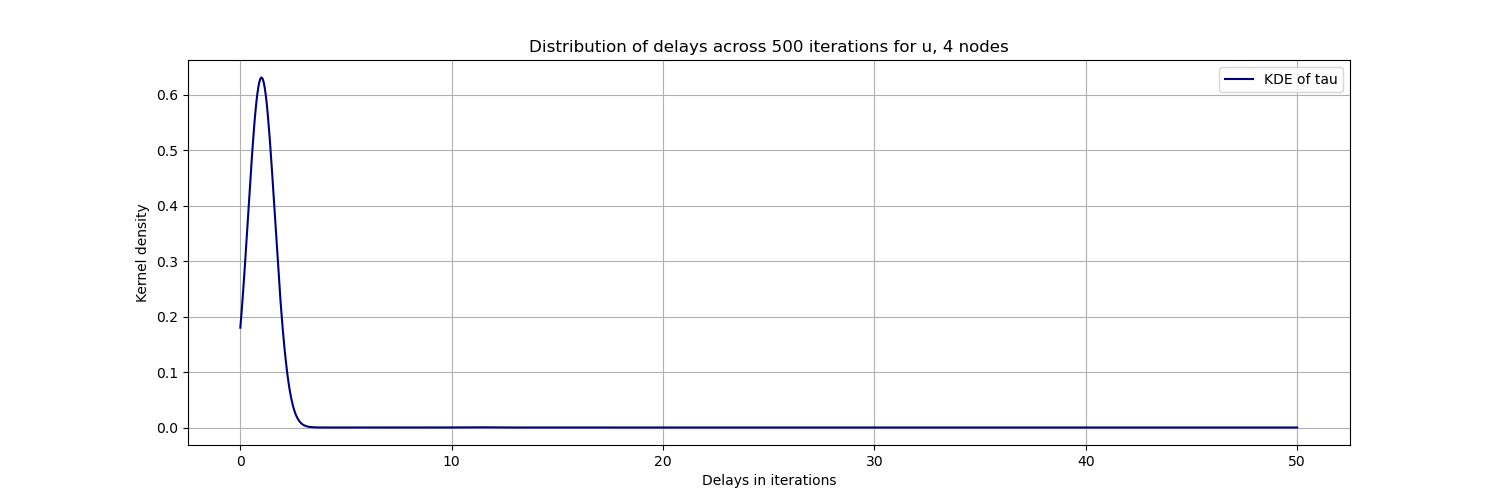}\hfill
    \includegraphics[width=0.5\linewidth]{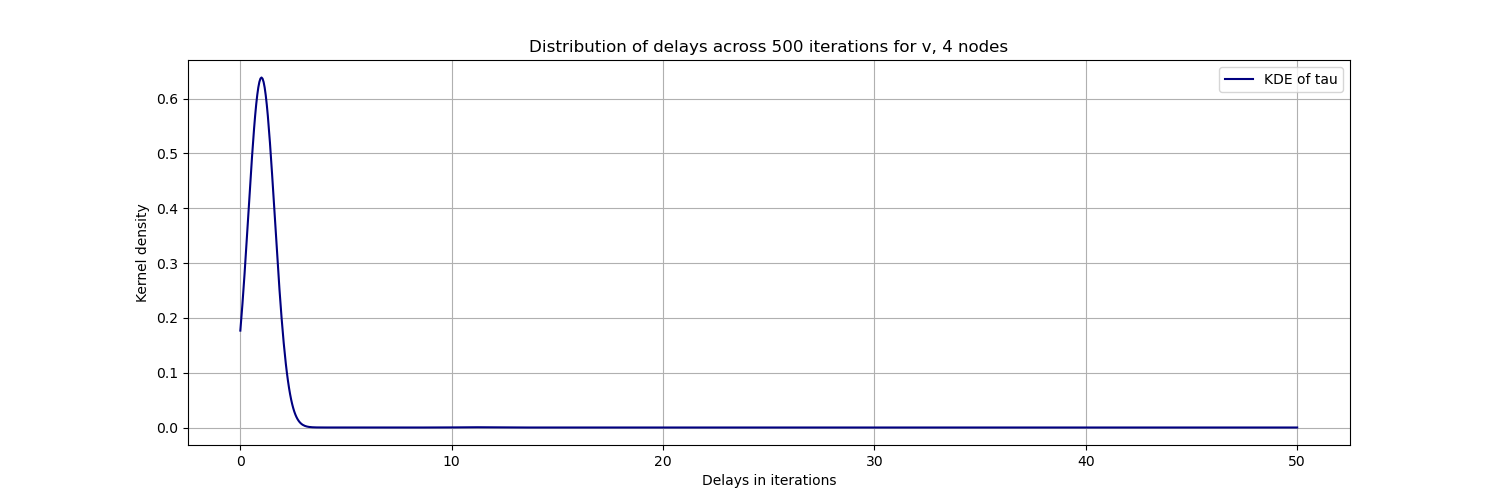}

    \vspace{1em}

    \includegraphics[width=0.5\linewidth]{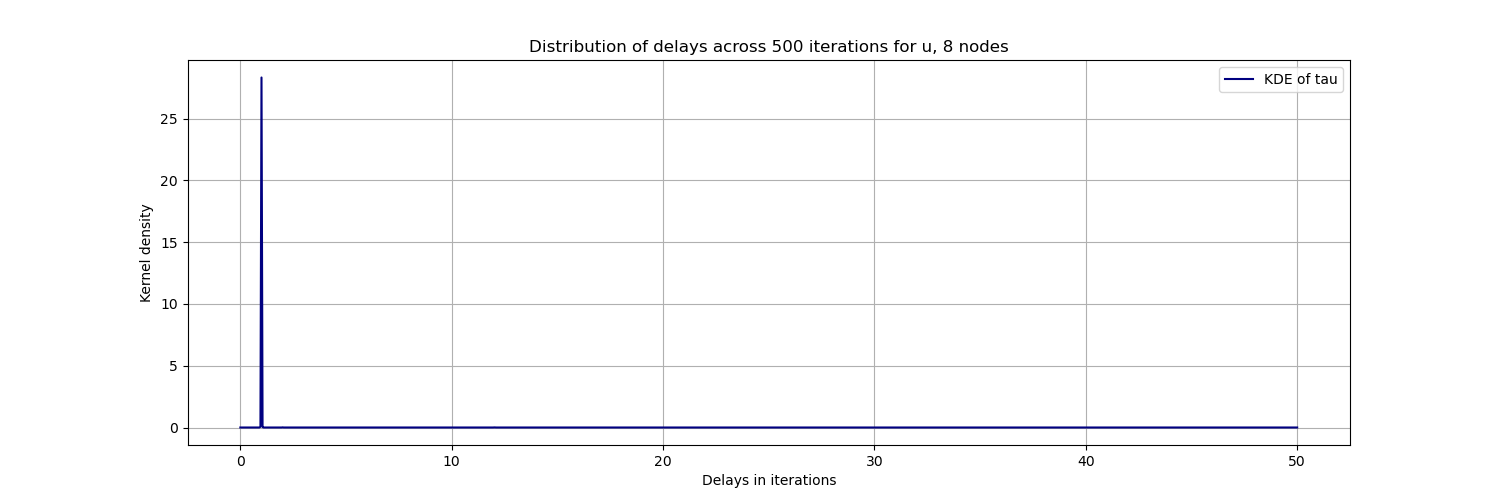}\hfill
    \includegraphics[width=0.5\linewidth]{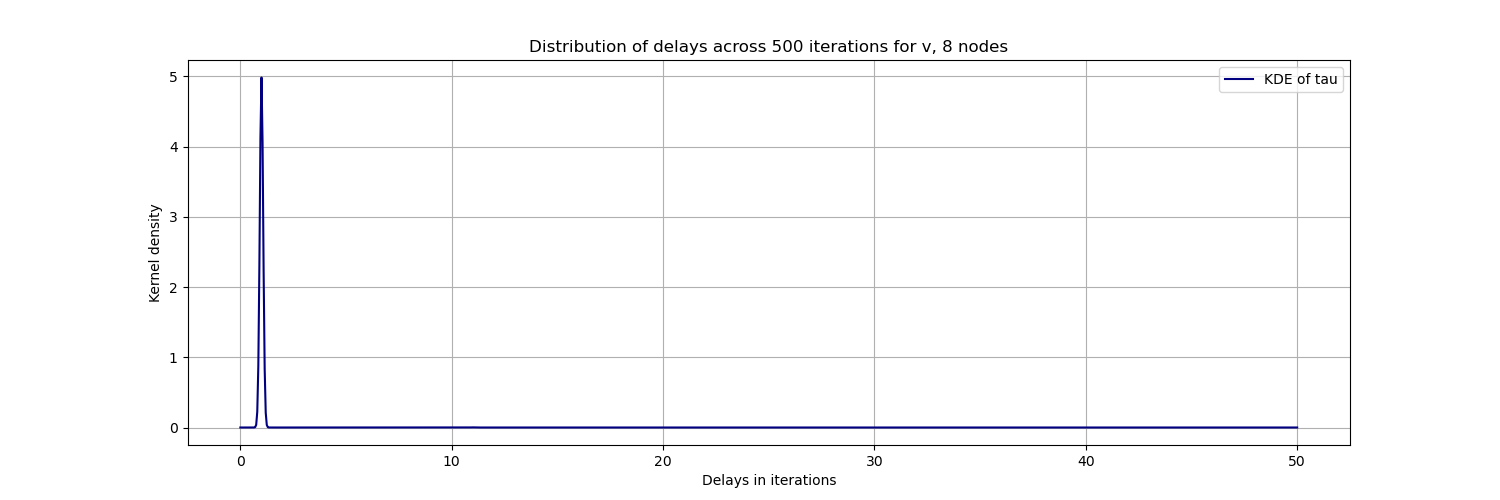}

    \caption{Kernel density estimates of the delay $\tau$ (in iterations) for $T=500$ across different settings for 1000 simulations, for $u$ (left) and $v$ (right) messages. The density is plotted for $\tau \in \{1,\dots,50\}$.}
    \label{fig:async_delays_to_50}
\end{figure}

\begin{figure}[!htbp]
    \centering
    \includegraphics[width=0.5\linewidth]{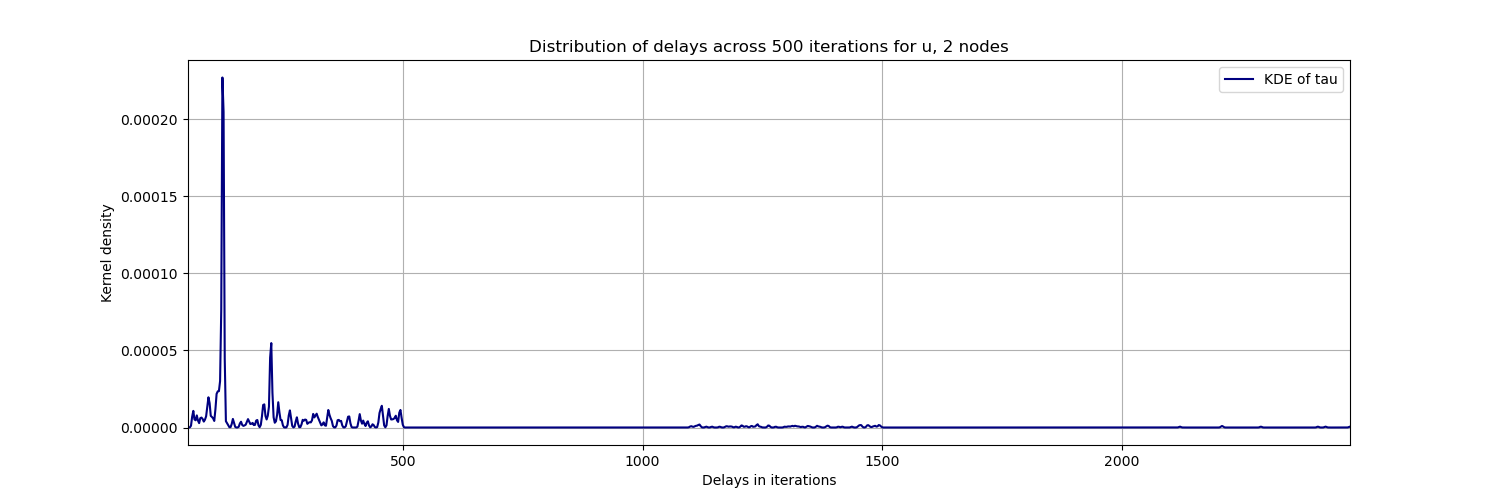}\hfill
    \includegraphics[width=0.5\linewidth]{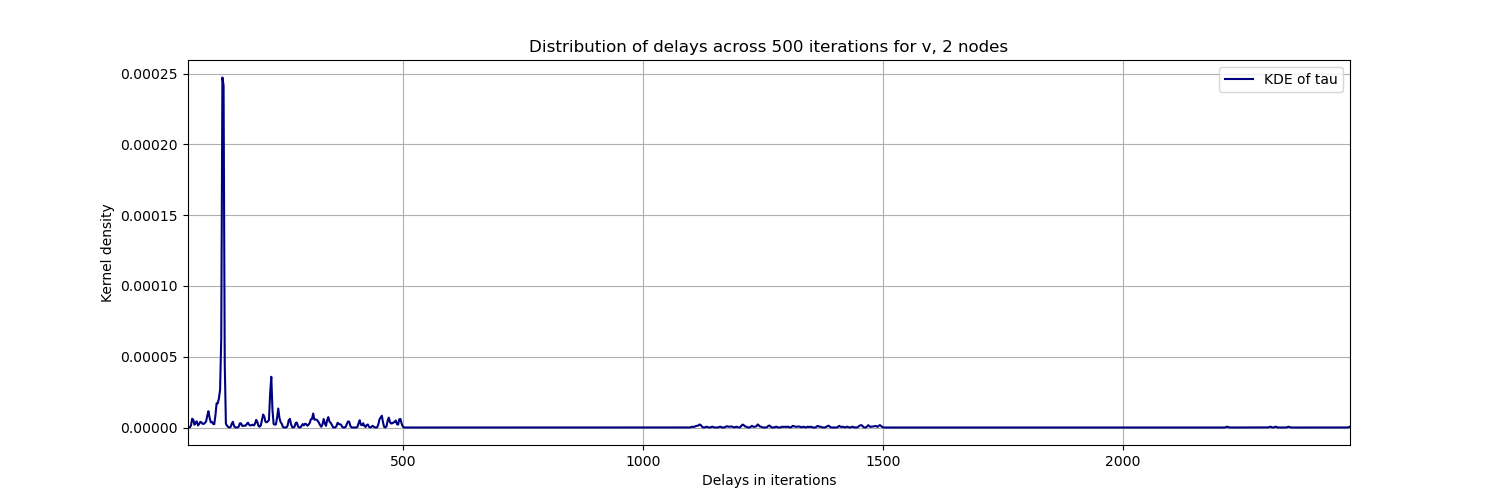}

    \vspace{1em}

    \includegraphics[width=0.5\linewidth]{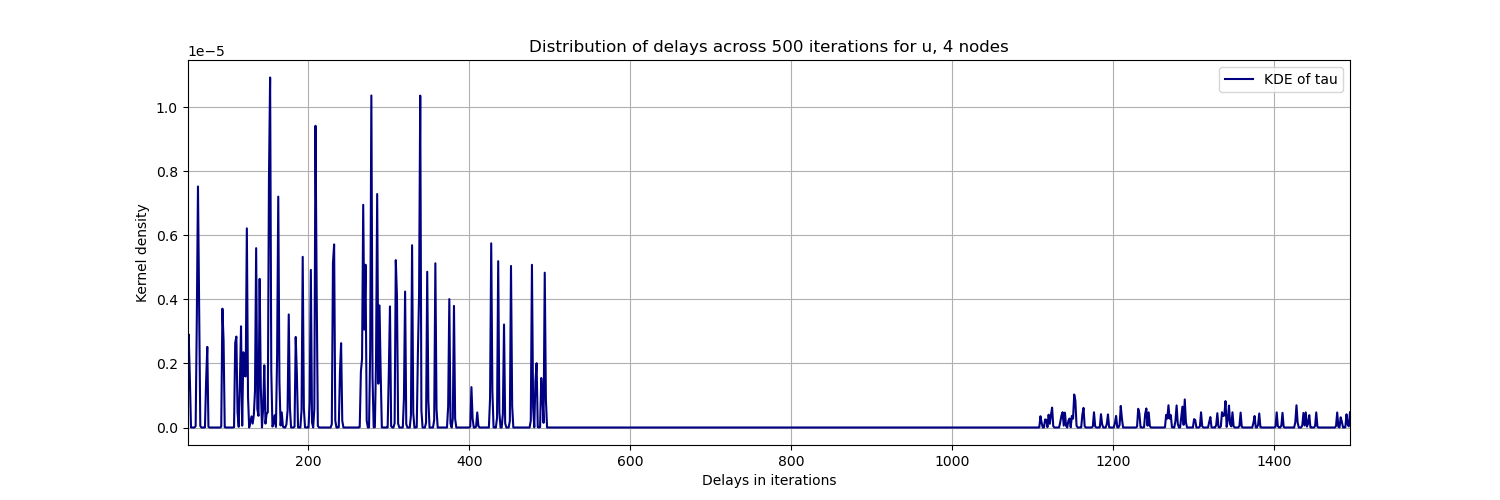}\hfill
    \includegraphics[width=0.5\linewidth]{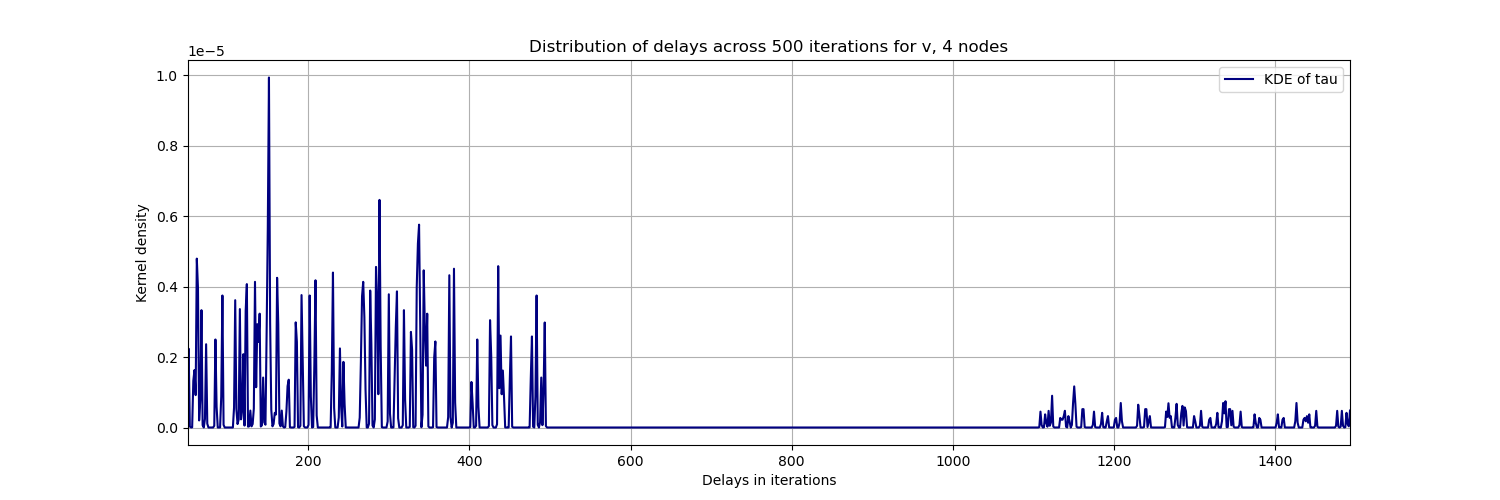}

    \vspace{1em}

    \includegraphics[width=0.5\linewidth]{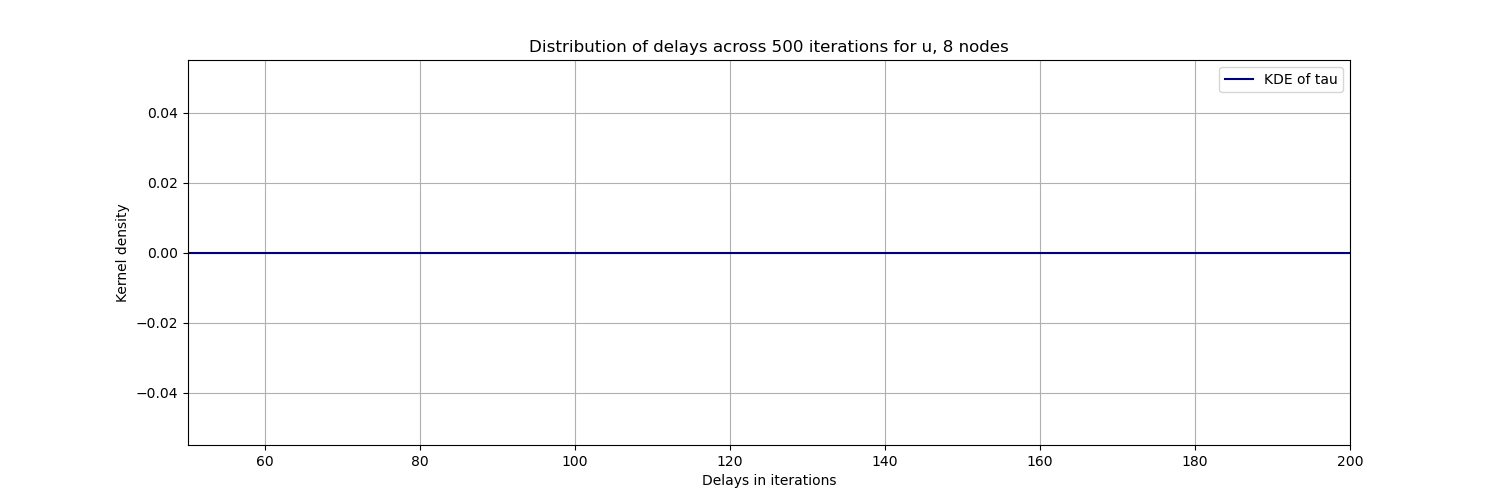}\hfill
    \includegraphics[width=0.5\linewidth]{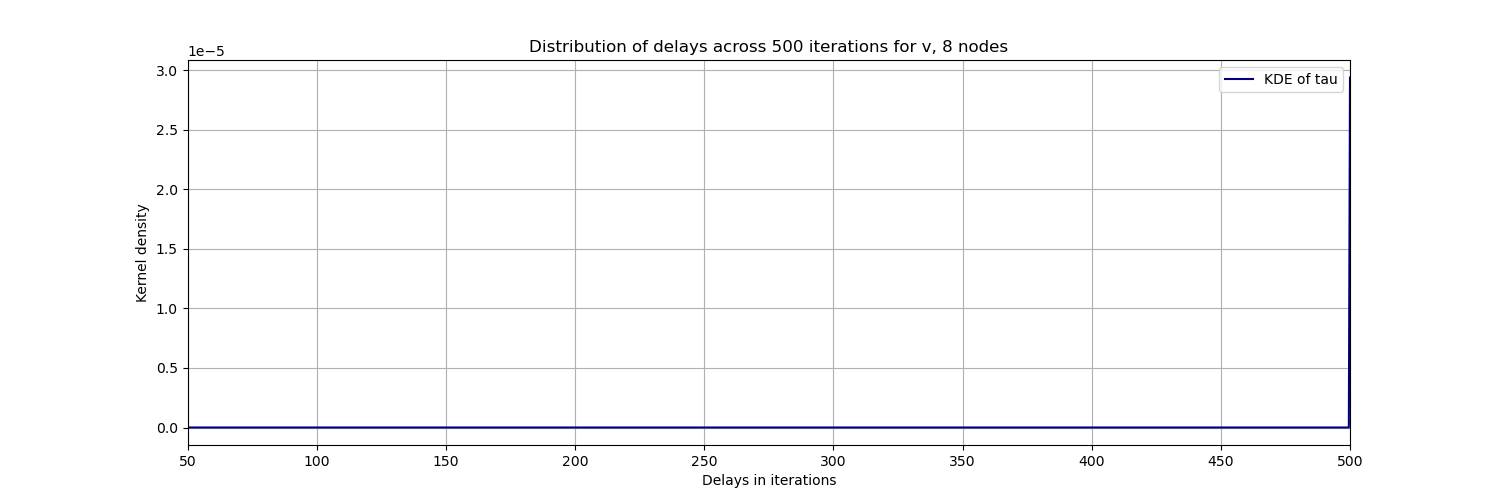}
    \caption{KDE of the delay $\tau$ for $T = 500$ across different settings for 1000 simulations, for $u$ (left) and $v$ (right) messages. The density is plotted for $\tau > 50$. Note the different vertical scale compared to Figure~\ref{fig:async_delays_to_50}.}
    \label{fig:async_delays_from_50}
\end{figure}

To quantify these delays, we ran the asynchronous federated Sinkhorn
algorithm $1000$ times on a problem with $n = 10\,000$ and a fixed iteration
budget $T = 500$, for $c \in \{2,4,8\}$ nodes and a single OT problem (no
multitarget vectorization). For each run, and for each ordered pair of nodes,
we recorded the delay $\tau$ between the sender's update and the receiver's
next use of that slice. Figures~\ref{fig:async_delays_to_50}--\ref{fig:async_delays_from_50}
and Table~\ref{table:delays_num_nodes} summarize the resulting distributions.

The average delay is close to one iteration in all cases, but we observe heavy
tails: the maximum delay $\tau^{\max}$ drops from $2477$ for 2 nodes to $500$
for 8 nodes. These long-tail events, which depend on transient network and
scheduling effects, contribute to the non-deterministic convergence behavior
of the asynchronous scheme: local updates based on very stale $(u,v)$ can
temporarily drive the iterates away from the fixed point, increasing both
variance and time to convergence.

\begin{table}[!htbp]
    \centering
    \begin{tabular}{lcccc}
    \toprule
    Number of nodes & $\tau^{\max}$ & $\tau^{\min}$ & $\tau^{\mathrm{mean}}$ & $\tau^{\mathrm{std}}$ \\
    \midrule
    2 & 2477 & 1 & 2.35 & 24.89 \\
    4 & 1494 & 1 & 1.18 & 10.53 \\
    8 & 500  & 1 & 1.01 & 0.72 \\
    \bottomrule
    \end{tabular}
    \caption{Statistics on delays for different numbers of nodes. The delays were averaged over $u$ and $v$, across 1000 simulations.}
    \label{table:delays_num_nodes}
\end{table}

The distributions in Figures~\ref{fig:async_delays_to_50}--\ref{fig:async_delays_from_50} show that most messages are used after one or a few local iterations, but rare long delays do occur, especially for small $c$. Table~\ref{table:delays_num_nodes} confirms that the mean delay stays close to one iteration, while the maximum delay shrinks as the number of nodes increases.

\subsection{Additional CPU asynchronous trajectories}
\label{app:cpu_async}

\begin{figure}[!htbp]
  \centering
  \begin{subfigure}[b]{0.49\linewidth}
    \centering
    \includegraphics[width=\linewidth]{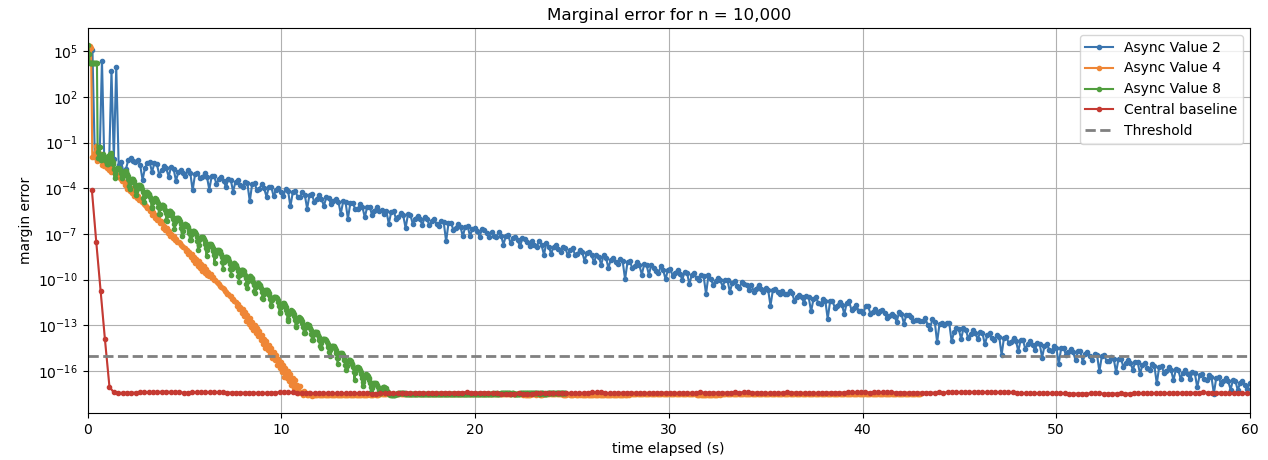}
    \caption{Example 1, $n = 10{,}000$.}
    \label{fig:async_cpu_cvg_10k}
  \end{subfigure}
  \hfill
  \begin{subfigure}[b]{0.49\linewidth}
    \centering
    \includegraphics[width=\linewidth]{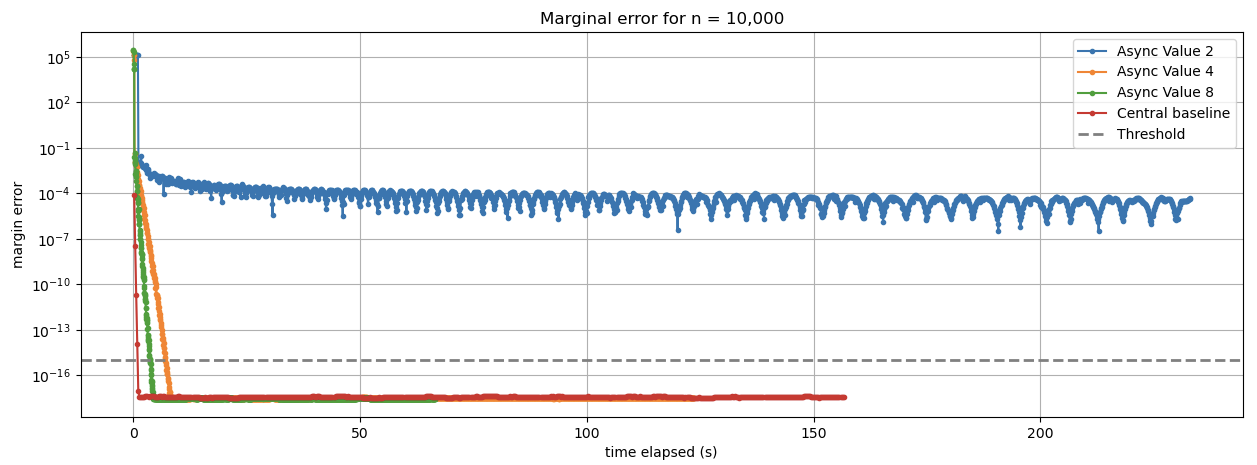}
    \caption{Example 2, $n = 10{,}000$.}
    \label{fig:async_cpu_cvg_10k_3}
  \end{subfigure}
  \\
  \vspace{0.5em}
  \begin{subfigure}[b]{0.49\linewidth}
    \centering
    \includegraphics[width=\linewidth]{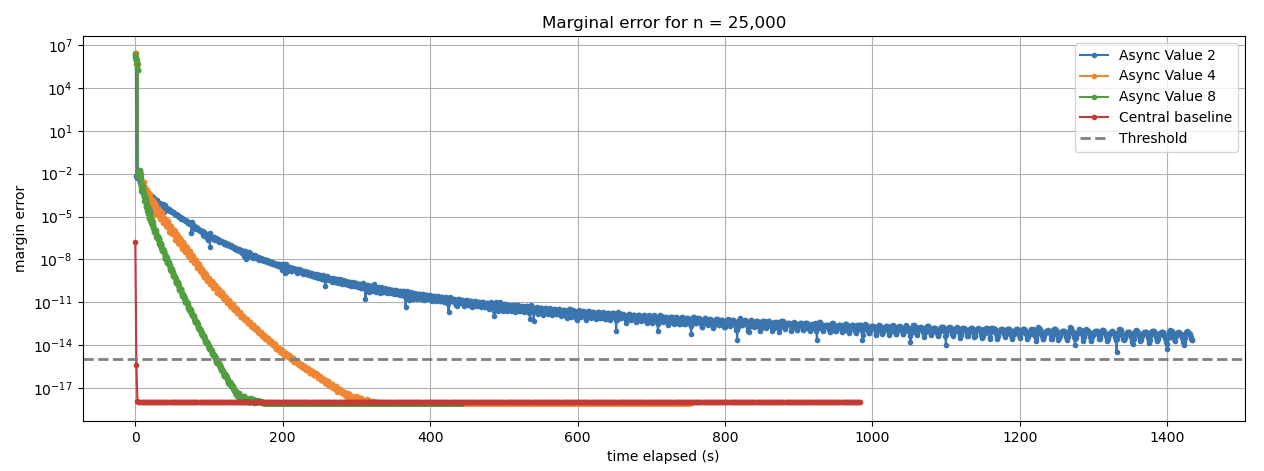}
    \caption{Example 3, $n = 25{,}000$.}
    \label{fig:async_cpu_cvg_25k}
  \end{subfigure}
  \hfill
  \begin{subfigure}[b]{0.49\linewidth}
    \centering
    \includegraphics[width=\linewidth]{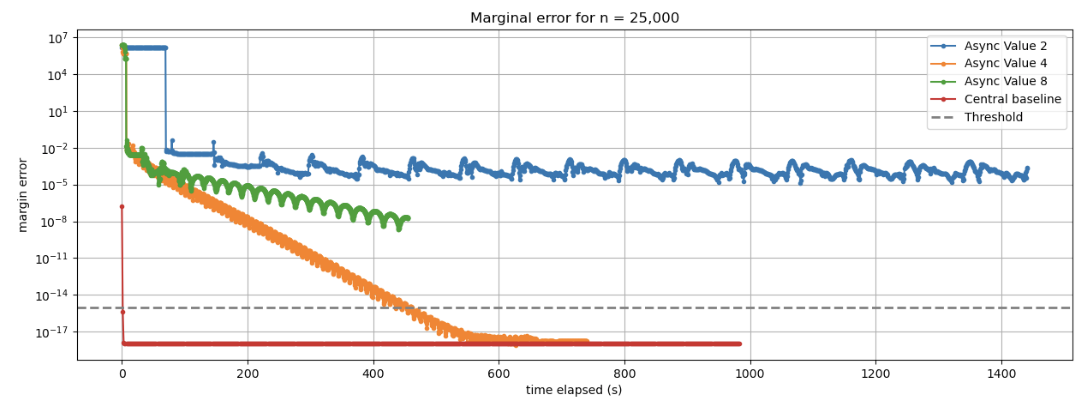}
    \caption{Example 4, $n = 25{,}000$.}
    \label{fig:async_cpu_cvg_25k_2}
  \end{subfigure}
  \caption{Asynchronous CPU convergence: marginal error on $a$ against elapsed time
  for several representative runs. The panels illustrate strong run-to-run
  variability and occasional very slow convergence, particularly for small
  node counts.}
  \label{fig:async_cpu_cvg_all}
\end{figure}

\subsection{Per-node CPU computation and communication times}
\label{appendix:cpu_dists}

\begin{figure}[!htbp]
  \centering
  \begin{subfigure}[b]{0.32\linewidth}
    \centering
    \includegraphics[width=\linewidth]{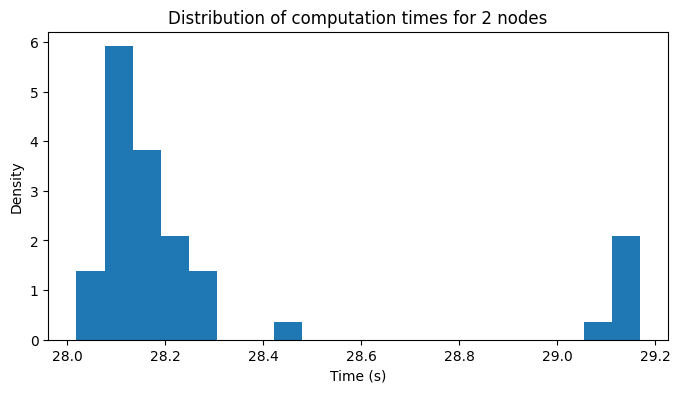}
    \caption{Compute, 2 nodes.}
    \label{fig:cpu_dist_comp_2}
  \end{subfigure}
  \hfill
  \begin{subfigure}[b]{0.32\linewidth}
    \centering
    \includegraphics[width=\linewidth]{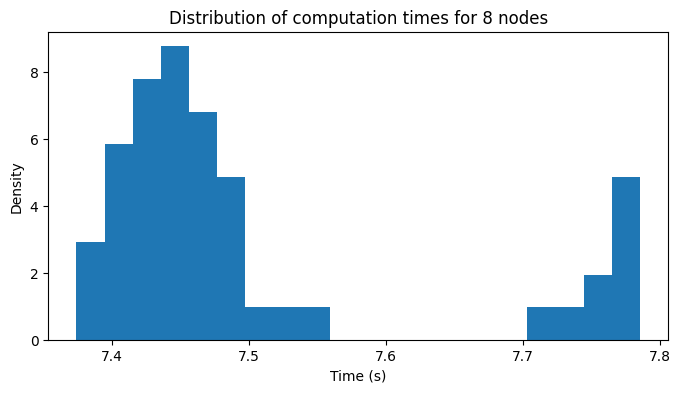}
    \caption{Compute, 4 nodes.}
    \label{fig:cpu_dist_comp_4}
  \end{subfigure}
  \hfill
  \begin{subfigure}[b]{0.32\linewidth}
    \centering
    \includegraphics[width=\linewidth]{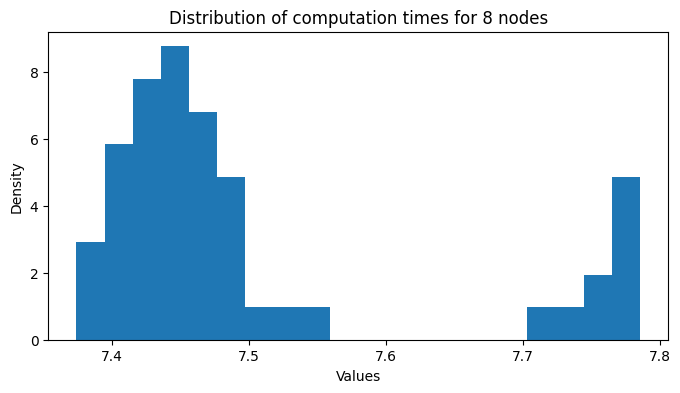}
    \caption{Compute, 8 nodes.}
    \label{fig:cpu_dist_comp_8}
  \end{subfigure}
  \\
  \vspace{0.5em}
  \begin{subfigure}[b]{0.32\linewidth}
    \centering
    \includegraphics[width=\linewidth]{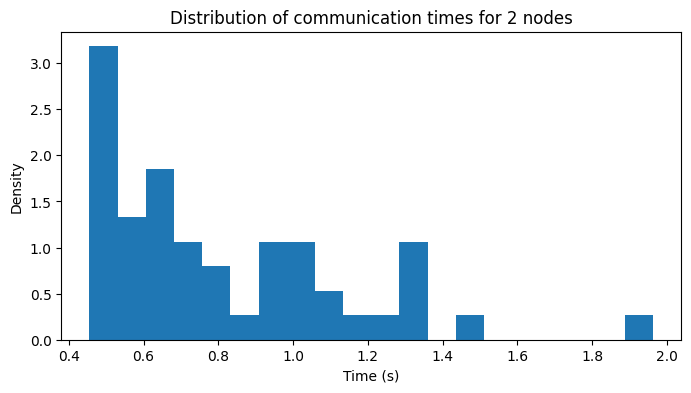}
    \caption{Comm., 2 nodes.}
    \label{fig:cpu_dist_comm_2}
  \end{subfigure}
  \hfill
  \begin{subfigure}[b]{0.32\linewidth}
    \centering
    \includegraphics[width=\linewidth]{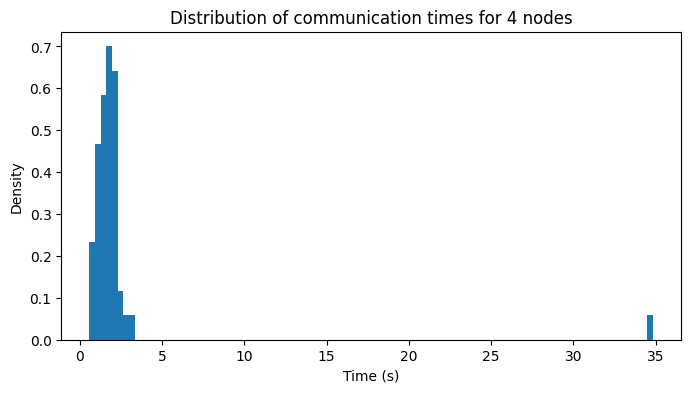}
    \caption{Comm., 4 nodes.}
    \label{fig:cpu_dist_comm_4}
  \end{subfigure}
  \hfill
  \begin{subfigure}[b]{0.32\linewidth}
    \centering
    \includegraphics[width=\linewidth]{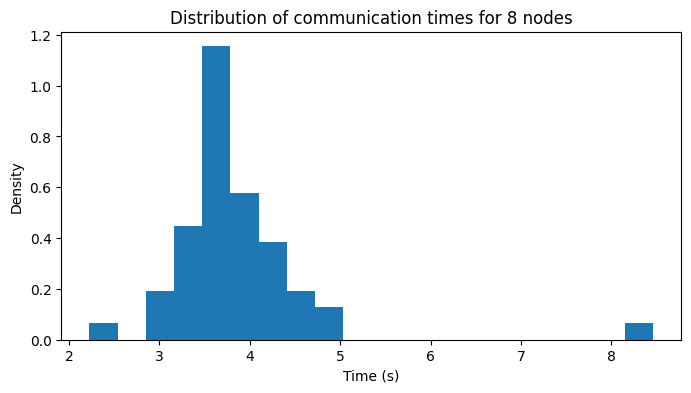}
    \caption{Comm., 8 nodes.}
    \label{fig:cpu_dist_comm_8}
  \end{subfigure}
  \caption{Distribution of per-node computation times (top row) and communication
  times (bottom row) for synchronous CPU runs with 2, 4, and 8 nodes.}
  \label{fig:cpu_dist_times}
\end{figure}

\section{Additional details on implementation}
\label{app:impl-details}

\subsection{Software stack}
The reference implementation is written in Python with \texttt{mpi4py} and PyTorch. \texttt{mpi4py} exposes the standard MPI programming model to Python and provides the blocking and non‑blocking primitives (\texttt{Allgather}, \texttt{Isend}, \texttt{Irecv}, etc.) that underpin the synchronous and asynchronous communication patterns used in our algorithms. PyTorch serves solely as a high‑performance tensor library: the Sinkhorn updates are expressed in terms of batched matrix–vector products and
elementwise operations on GPU tensors.

\subsection{MPI and GPU communication}
MPI and GPU‑native message passing are not always directly compatible. MPI implementations guarantee native support for host (CPU) buffers, while direct GPU‑to‑GPU transfers require either CUDA‑aware MPI extensions or explicit staging of data through host memory. In our setting only the scaling vectors and small intermediate quantities are communicated; the cost matrix blocks remain local to each client. Thus, any overhead from GPU–CPU–GPU transfers affects a low‑dimensional part of the state and is included in the timing measurements reported in Section~\ref{sec:experiments}.

\subsection{Numeric precision}
By default we use double precision (\texttt{float64}) for the marginals, scalings, and intermediate vectors. For the $\varepsilon_{\scriptscriptstyle\mathrm{OT}}$‑sensitivity experiment we additionally employ higher‑precision arithmetic with 50 decimal digits in order to eliminate rounding‑error effects when comparing convergence behavior across different regularization levels.

\subsection{Synchronous vs. asynchronous modes}
Conceptually, the synchronous federation enforces a global barrier after each round of updates: every process completes its local matvecs, participates in a blocking collective to exchange the relevant slices of $(u,v)$, and only then advances to the next iteration. In contrast, the asynchronous all‑to‑all scheme lets each rank initiate non‑blocking sends/receives and immediately continue with local computation, so updates can be applied using slightly stale versions of the global scalings. This distinction aligns with the theoretical model in~\ref{sec:con}: synchronous variants realize the exact centralized fixed‑point iteration, while the asynchronous variant trades determinism for better overlap between communication and computation.

\subsection{Interconnection details.}
The nodes are connected by an InfiniBand fabric (nominal 100~Gb/s link speed), and we use an OpenMPI-based MPI stack
(\texttt{mpi4py/3.1.5-gompi-2023b}; PyTorch~2.2.1 with CUDA~12.4). Unless otherwise stated, we store histograms and scaling vectors in double precision (\texttt{float64}). (If CUDA-aware MPI or NCCL collectives are available, they can reduce constant factors in communication time; our topology-aware complexity models remain unchanged.)

\section{Additional details on Experiments}
\label{app:exp_details}

\subsection{HPC baseline: 2D block distribution with row/column \texttt{Allreduce}}

To contextualize our All-to-All and Star designs against a strong HPC-style baseline, we also implement a standard \emph{2D block} distribution of the kernel $K$ over a $p_r \times p_c$ process grid.
Each rank $(i,j)$ stores a block $K_{ij}\in\mathbb{R}^{n_i\times n_j}$ and maintains local scalings $u_i$ (row block) and $v_j$ (column block).
Each Sinkhorn iteration computes
\[
q_i = \sum_{j=1}^{p_c} K_{ij} v_j
\quad\text{via a row-wise } \texttt{Allreduce},
\qquad
r_j = \sum_{i=1}^{p_r} K_{ij}^\top u_i
\quad\text{via a column-wise } \texttt{Allreduce}.
\]
This baseline replaces global scaling-vector gathers with two structured \texttt{Allreduce} operations per iteration, reducing per-rank communication volume from $\Theta(n)$ (global gather) to $\Theta(n/p_r + n/p_c)$.

\begin{table}[!htbp]
\centering
\caption{Timing summary (10 repeats per setting). For each run we take the maximum across ranks as a wall-clock proxy; we report median [p25,p75] across repeats.}
\begin{tabular}{llcc}
\toprule
Method & $c$ & Total time (s) & Comm.\ fraction \\
\midrule
All-to-All (sync) & 2 & 1.171 [1.085,1.195] & 0.939 [0.930,0.943] \\
All-to-All (sync) & 4 & 2.582 [2.476,2.674] & 0.970 [0.969,0.971] \\
All-to-All (sync) & 8 & 6.673 [6.538,6.728] & 0.987 [0.987,0.987] \\
Star (sync) & 2 & 5.538 [5.443,5.719] & 0.617 [0.606,0.639] \\
Star (sync) & 4 & 6.022 [5.928,6.177] & 0.830 [0.827,0.833] \\
Star (sync) & 8 & 9.122 [9.085,9.203] & 0.936 [0.935,0.937] \\
HPC 2D baseline (sync) & 2 & 0.680 [0.676,0.684] & 0.920 [0.915,0.921] \\
HPC 2D baseline (sync) & 4 & 0.410 [0.408,0.412] & 0.859 [0.856,0.860] \\
HPC 2D baseline (sync) & 8 & 0.307 [0.306,0.310] & 0.811 [0.810,0.814] \\
\bottomrule
\end{tabular}
\label{tab:timing_summary_all}
\end{table}

Table~\ref{tab:timing_summary_all} reports end-to-end runtime measurements for synchronous Sinkhorn on an embedding-derived dense cost matrix ($n=10{,}000$), comparing the All-to-All and Star federated topologies against the HPC 2D block baseline, for $c\in\{2,4,8\}$.
All methods use the same OT instance and numerical settings (same $\varepsilon_{\mathrm{OT}}$, precision, and iteration budget / stopping rule); only the communication pattern (and, for the 2D baseline, the kernel layout) differs.
Each timing point is the median of 10 independent runs; for each run we take the maximum over ranks as a wall-clock proxy and report $[p25,p75]$ across repeats.

The All-to-All topology shows negative scaling with node count: total time grows from 1.17\,s ($c{=}2$) to 6.67\,s ($c{=}8$), and the communication fraction increases to $\approx 0.99$, indicating that global scaling-vector exchange dominates at higher $c$.
The Star topology is slower for this dense-kernel setting (5.54--9.12\,s) and also becomes increasingly communication-bound as $c$ grows (comm.\ fraction 0.62$\rightarrow$0.94), reflecting the cost of repeated server broadcasts and client-to-server uplinks.
In contrast, the 2D block-distributed baseline improves with $c$ (0.68$\rightarrow$0.31\,s) and achieves a lower communication fraction (0.92$\rightarrow$0.81) by reducing per-rank communication volume to $\Theta(n/p_r + n/p_c)$.
Crucially, this baseline assumes a kernel layout that supports both row and column access and therefore serves as an optimistic upper bound relative to federated deployments where a strict row partition (and restricted transpose access) is the primary constraint.

Taken together, these results clarify the trade-off space: the 2D block baseline represents the communication-efficient extreme under full data mobility, while the federated topologies trade increased communication for feasibility under realistic deployment constraints.

\section{Additional details on communication models}
\label{app:com_details}

\subsection{$\alpha$--$\beta$ calibration table}

\begin{table}[!htbp]
\centering
\caption{Calibrated latency--bandwidth parameters ($T(B)=\alpha+\beta B$) for the communication primitives used by our implementations (float64, $k=1$). We report $\alpha$ in $\mu$s and effective bandwidth $1/\beta$ in GB/s.}
\begin{tabular}{llrrrr}
\toprule
Primitive & $c$ & $\alpha$ ($\mu$s) & $\beta$ (s/byte) & BW (GB/s) & $R^2$ \\
\midrule
Allgather (A2A) & 2 & 481.45 & $1.0047{\times}10^{-9}$ & 0.995 & 0.993 \\
Allgather (A2A) & 4 & 801.35 & $8.7851{\times}10^{-10}$ & 1.138 & 0.968 \\
Allgather (A2A) & 8 & 1322.50 & $9.4662{\times}10^{-10}$ & 1.056 & 0.936 \\
\midrule
Star broadcast (BC) & 2 & 19.63 & $2.8493{\times}10^{-10}$ & 3.510 & 0.995 \\
Star broadcast (BC) & 4 & 21.94 & $4.3798{\times}10^{-10}$ & 2.283 & 0.998 \\
Star broadcast (BC) & 8 & 28.26 & $7.5910{\times}10^{-10}$ & 1.317 & 0.997 \\
\midrule
Star uplink (SR) & 2 & 19.67 & $1.0994{\times}10^{-10}$ & 9.096 & 0.998 \\
Star uplink (SR) & 4 & 28.06 & $1.2795{\times}10^{-10}$ & 7.816 & 0.994 \\
Star uplink (SR) & 8 & 45.27 & $1.6485{\times}10^{-10}$ & 6.066 & 0.999 \\
\bottomrule
\end{tabular}
\label{tab:alpha_beta_calib}
\end{table}

\section{Privacy Layer and Evaluation}
\label{app:privacy}

\subsection{Threat model}

We assume an honest--but--curious adversary that can observe all messages exchanged during federation: the server in the Star-Network topology and the peer-to-peer traffic in the All-to-All topology. We protect the communicated log-domain Sinkhorn scaling vectors with differential privacy (DP), and we report privacy for communicated messages under sequential composition across iterations. We track cumulative privacy loss using R\'enyi differential privacy (RDP) and convert the final bound to $(\varepsilon_{\scriptscriptstyle\mathrm{DP}},\delta)$ at report time~\citep{mironov2017renyi}. Secure aggregation~\citep{bonawitz2017practical} is complementary and could be combined with our mechanism; we do not implement it in this paper.

\begin{table}[t]
\centering
\caption{What different parties can observe under each topology (honest-but-curious model).}
\begin{tabular}{p{2.2cm}p{5.0cm}p{5.0cm}}
\toprule
Topology & Who observes messages? & Typical leakage surface (without protections) \\
\midrule
All-to-All & Every participant receives (broadcast/gather) all clients' communicated scalings each iteration. & Any single compromised participant can observe all communicated scalings; scalings may encode information about local marginals/kernels across iterations. \\
Star & The server observes all clients' uplink messages each iteration; clients typically receive only server broadcasts. & The server is a single point that can observe all scalings (and any aggregated quantities it computes). Clients do not necessarily observe other clients' scalings unless the server rebroadcasts them. \\
\bottomrule
\end{tabular}
\label{tab:threatmodel_topology}
\end{table}

\paragraph{Interpretation.}
The key distinction is whether the “aggregator view” is held by a single server (Star) or effectively by every participant (All-to-All).
Our DP mechanism applies to the communicated log-scalings in either topology; secure aggregation can further reduce server visibility in Star settings, while peer-to-peer secure aggregation is substantially harder operationally in All-to-All deployments.

We report differential privacy for the communicated messages (the client log-scaling vectors) under the standard neighboring-dataset definition record-level: two datasets are neighbors if they differ in one record. Our mechanism clips each communicated log-scaling to radius $S$, which yields a conservative global $\ell_2$-sensitivity bound $\Delta \le 2S$ for each release (difference between any two clipped outputs). This bound is generally loose in long Sinkhorn runs, and our results should be interpreted as a measurement baseline showing how privacy degrades under per-iteration communication without amplification or reduced release frequency.

\subsection{Gaussian noise on log‑domain scalings}

We instantiate DP directly on the Sinkhorn scaling vectors. Let $u^t, v^t \in \mathbb{R}^n_{+}$ denote the standard multiplicative Sinkhorn scalings at iteration $t$:
\begin{align*}
u^{t+1} &= a \oslash (K v^t),\\
v^{t+1} &= b \oslash (K^\top u^{t+1}).
\end{align*}
and define stabilized log-variables $l_u^t=\log u^t$ and $l_v^t=\log v^t$~\citep{schmitzer2019stabilized,peyre2019computational}.

\subsubsection{Differential Privacy Mechanism}

After each local update, each client clips and perturbs the log-scalings and communicates only the noised values:
\[
\tilde l^t_u
= \mathrm{clip}_{S_u}\!\bigl(l^t_u\bigr) + \mathcal{N}(0,\sigma_u^2 I),
\qquad
\tilde l^t_v
= \mathrm{clip}_{S_v}\!\bigl(l^t_v\bigr) + \mathcal{N}(0,\sigma_v^2 I),
\]
where $\mathrm{clip}_S(\cdot)$ denotes \emph{row-wise} $\ell_2$ clipping applied independently across the multitarget dimension. Concretely, if $L\in\mathbb{R}^{m\times N}$ is a communicated log-scaling slice, then for each row $i$ we set
\[
(\mathrm{clip}_S(L))_{i,:} \;=\; L_{i,:}\cdot \min\Bigl\{1,\; \frac{S}{\lVert L_{i,:}\rVert_2}\Bigr\}.
\]
In the reference implementation we log convergence using the current scalings held by the algorithm (which include DP perturbations when enabled). By DP post-processing invariance, exponentiation and all downstream computations preserve privacy~\citep{dwork2014algorithmic}. We include the Gaussian-only case as an optimization ablation and the clip+Gaussian case as the formal DP instantiation, following the DP-Sinkhorn setup of~\citep{cao2021don}.

\subsubsection{Instantiations used in experiments}

We evaluate two instantiations:
\begin{itemize}
\item \textbf{Gaussian-only (ablation).} Add log-domain Gaussian noise without relying on a sensitivity bound (no formal DP guarantee); this isolates the optimization effect of log-domain noise.
\item \textbf{Clipping+Gaussian (DP).} Clip the log-scalings and add Gaussian noise, yielding a standard Gaussian mechanism with per-message $\ell_2$ sensitivity bounded by $2S_u$ and $2S_v$, respectively.
\end{itemize}

\begin{remark}
We treat $(S,\sigma)$ as mechanism hyperparameters and empirically report both utility (marginal error) and the resulting RDP-based privacy bound. We do not propose a deployment policy for selecting $(S,\sigma)$ under a fixed privacy budget.
\end{remark}

\subsubsection{Practical choices}

Unless otherwise stated we set $S_u=S_v=:S$ and $\sigma_u=\sigma_v=:\sigma$, and sweep $S\in\{1,4,8,12,16,20\}$ and $\sigma\in\{10^{-10},10^{-5},10^{-3},10^{-2},10^{-1},0.5,1\}$. We do not employ subsampling, adaptive clipping, or mean-centering of log-scalings in the experiments reported.

\subsection{RDP accountant and $(\varepsilon_{\scriptscriptstyle\mathrm{DP}},\delta)$ conversion}

For a Gaussian mechanism that releases a vector with $\ell_2$ sensitivity $\Delta$ and noise scale $\sigma$, the release satisfies $(p,\varepsilon_{\scriptscriptstyle\mathrm{RDP}})$-RDP with
\[
\varepsilon_{\scriptscriptstyle\mathrm{RDP}}(p) = p \frac{\Delta^2}{2\sigma^2}, \qquad p>1,
\]
and RDP composes additively across releases~\citep{mironov2017renyi}. In our clip+Gaussian mechanism, per-message $\ell_2$ clipping to radius $S$ yields the conservative sensitivity bound $\Delta \le 2S$ for each log-scaling release. We therefore accumulate $\varepsilon_{\scriptscriptstyle\mathrm{RDP}}^{\mathrm{tot}}(p)$ over all communicated releases and convert to $(\varepsilon_{\scriptscriptstyle\mathrm{DP}},\delta)$ by
\[
\varepsilon_{\scriptscriptstyle\mathrm{DP}}(\delta)
= \min_{p \in \mathcal{A}}
\left\{
\varepsilon_{\scriptscriptstyle\mathrm{RDP}}^{\mathrm{tot}}(p)
+ \frac{\log(1/\delta)}{p - 1}
\right\},
\]
with $\delta=10^{-5}$ and a standard grid of orders $\mathcal{A}$ (see Appendix~D for the exact set). We apply this accounting only to the clip+Gaussian mechanism; the Gaussian-only ablation is not interpreted as a formal DP guarantee.

\subsection{Experimental protocol}

We consider a synthetic entropic OT problem with $n=5000$ and cost matrix $C \in \mathbb{R}^{n \times n}$. We run distributed Sinkhorn for $T=1000$ iterations with $\varepsilon_{\scriptscriptstyle\mathrm{OT}}=0.01$ and $c \in \{2,4,8\}$ workers under both the All-to-All and star topologies. Unless otherwise stated we set $S_u=S_v=S$ and $\sigma_u=\sigma_v=\sigma$.

At iteration $t$ we compute the marginal violation using the current scalings held by the algorithm
(which include DP perturbations when enabled),
\[
\mathrm{err}^{(t)}\ :=\ \left\| P^{(t)}\mathbf{1}-a\right\|_2,\qquad P^{(t)}=\mathrm{diag}(u^{(t)})K\mathrm{diag}(v^{(t)}).
\]
and report: (i) $\mathrm{err}^{(t)}$ vs.\ iteration, (ii) a noise floor defined as the mean of $\mathrm{err}^{(t)}$ over the last 100 iterations, and (iii) the final RDP-based $(\varepsilon_{\scriptscriptstyle\mathrm{DP}},\delta)$ for clip+Gaussian runs with $\delta=10^{-5}$. \\
We sweep $\sigma \in \{10^{-10},10^{-5},10^{-3},10^{-2},10^{-1},0.5,1\}$ for the Gaussian-only ablation, and additionally sweep $S \in \{1,4,8,12,16,20\}$ for the clip+Gaussian mechanism. Unless otherwise stated we show rank~0 and the synchronous All-to-All topology in the main text; other ranks and the star topology are deferred to~\ref{app:privacy-extra}.

\begin{remark}
All privacy experiments use synthetic data; the reported $(\varepsilon_{\scriptscriptstyle\mathrm{DP}},\delta)$ values should be interpreted per configuration (as if chosen a priori), not jointly over the full hyperparameter sweep.
\end{remark}

\subsection{Results}
\label{sec:results}

We treat differential privacy as a measurement layer on communicated log-scalings, not as a claim of strong end-to-end privacy for long Sinkhorn runs. In particular, when composing per-iteration releases over many iterations, standard RDP accounting can yield very loose bounds ($\varepsilon_{\scriptscriptstyle\mathrm{DP}}\gg1$ in our sweeps), so the main purpose here is to characterize the privacy–utility tradeoff of simple clipping/noise baselines and to motivate directions needed for practically meaningful budgets.

\subsubsection{Effect of log-domain Gaussian noise}

\begin{figure}[!htbp]
  \begin{subfigure}[b]{0.49\linewidth}
    \centering
    \includegraphics[width=\linewidth]{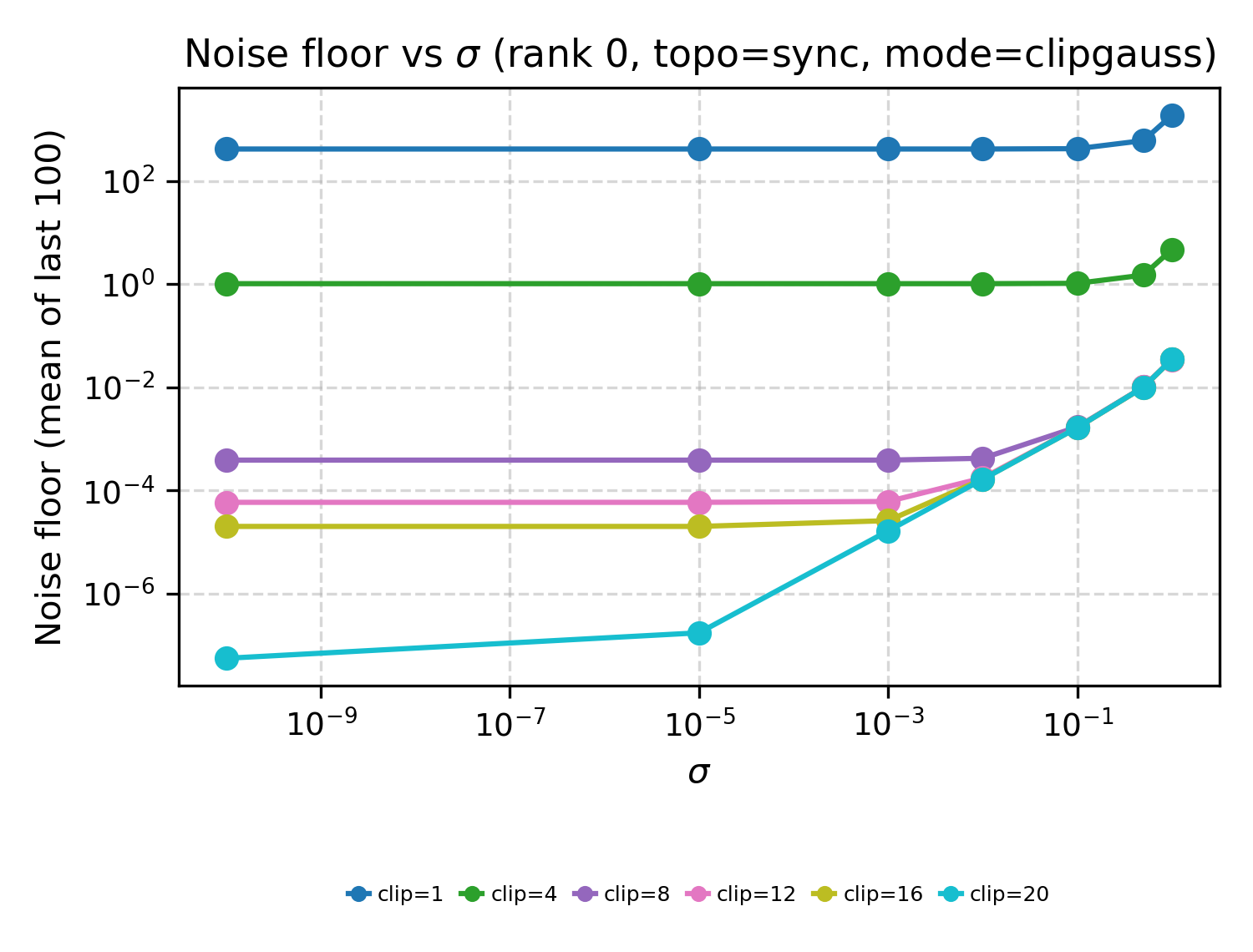}
    \caption{Clip+Gaussian mechanism.}
    \label{fig:noisefloor_sigma_clip}
  \end{subfigure}
  \hfill
  \centering
  \begin{subfigure}[b]{0.49\linewidth}
    \centering
    \includegraphics[width=\linewidth]{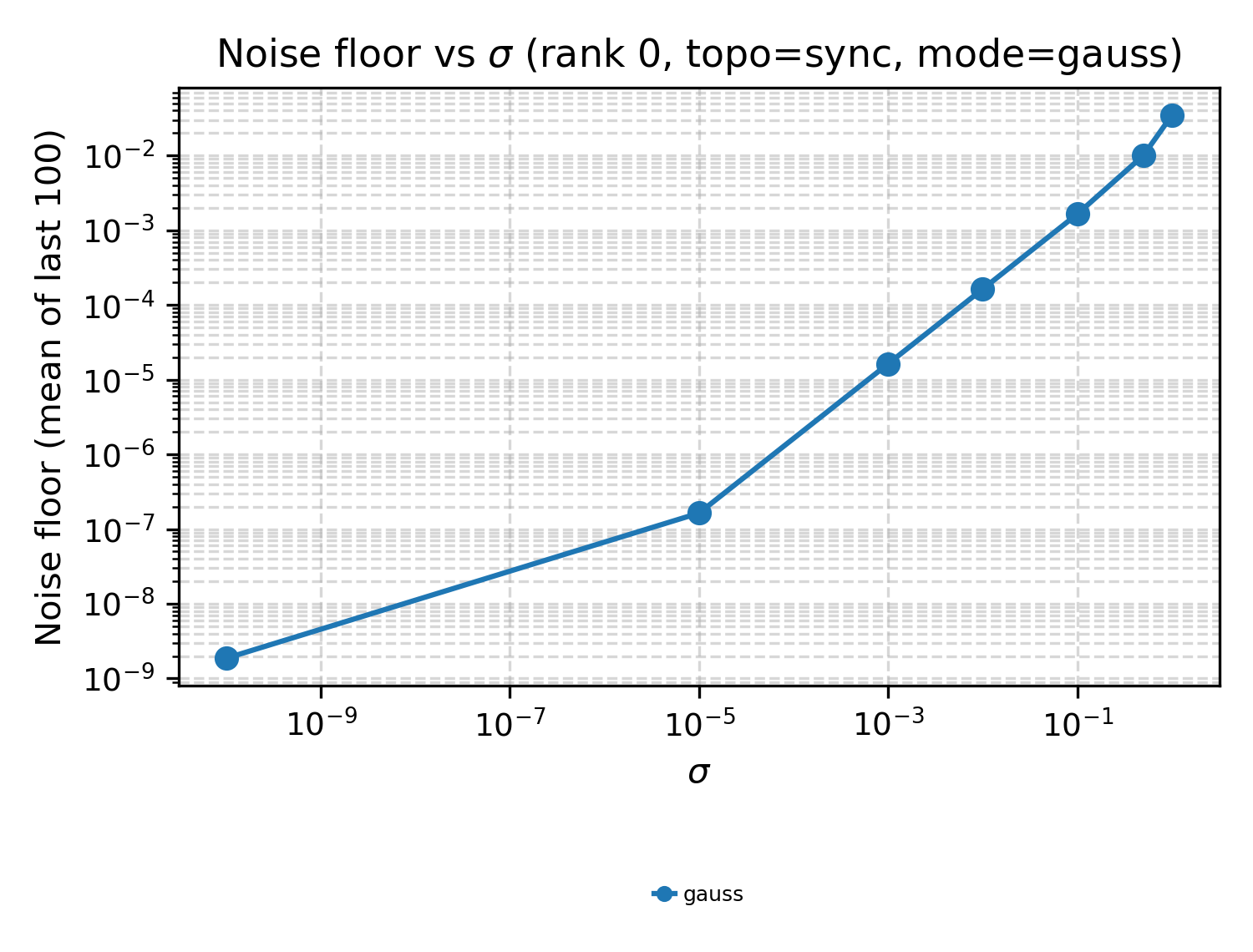}
    \caption{Gaussian-only ablation (no clipping).}
    \label{fig:noisefloor_sigma_gauss}
  \end{subfigure}
  \caption{Noise floor (mean marginal error over the last 100 iterations) as a function of log-domain noise scale $\sigma$ on the synchronous All-to-All topology (rank~0). Each curve in panel~(a) corresponds to a different
  clipping radius $S$; panel~(b) uses no clipping and is shown for robustness
  comparison only (no formal DP guarantee).}
  \label{fig:noisefloor_sigma}
\end{figure}

In the Gaussian-only ablation (no clipping), the marginal-error noise floor is essentially unchanged for $\sigma \le 10^{-3}$ and increases only once $\sigma \gtrsim 10^{-2}$ (Figure~\ref{fig:noisefloor_sigma}). We include these runs as an optimization ablation, without clipping they do not yield a finite global sensitivity bound and are therefore not interpreted as a formal DP guarantee.

\subsubsection{Impact of clipping on optimization}

With clipping+Gaussian, optimization quality is primarily controlled by the clipping radius (Figure~\ref{fig:noisefloor_sigma}). For small radii ($S \le 4$) the error floor is dominated by clipping bias and is nearly insensitive to $\sigma$, while for larger radii ($S \ge 8$) clipping bias becomes small and the noise floor increases smoothly as $\sigma$ grows. Full trajectories (including Star-Network topology and additional ranks) are reported in~\ref{app:privacy-extra}.

\subsubsection{Utility--privacy tradeoff}

\begin{figure}[!htbp]
  \centering
  \includegraphics[width=0.6\linewidth]{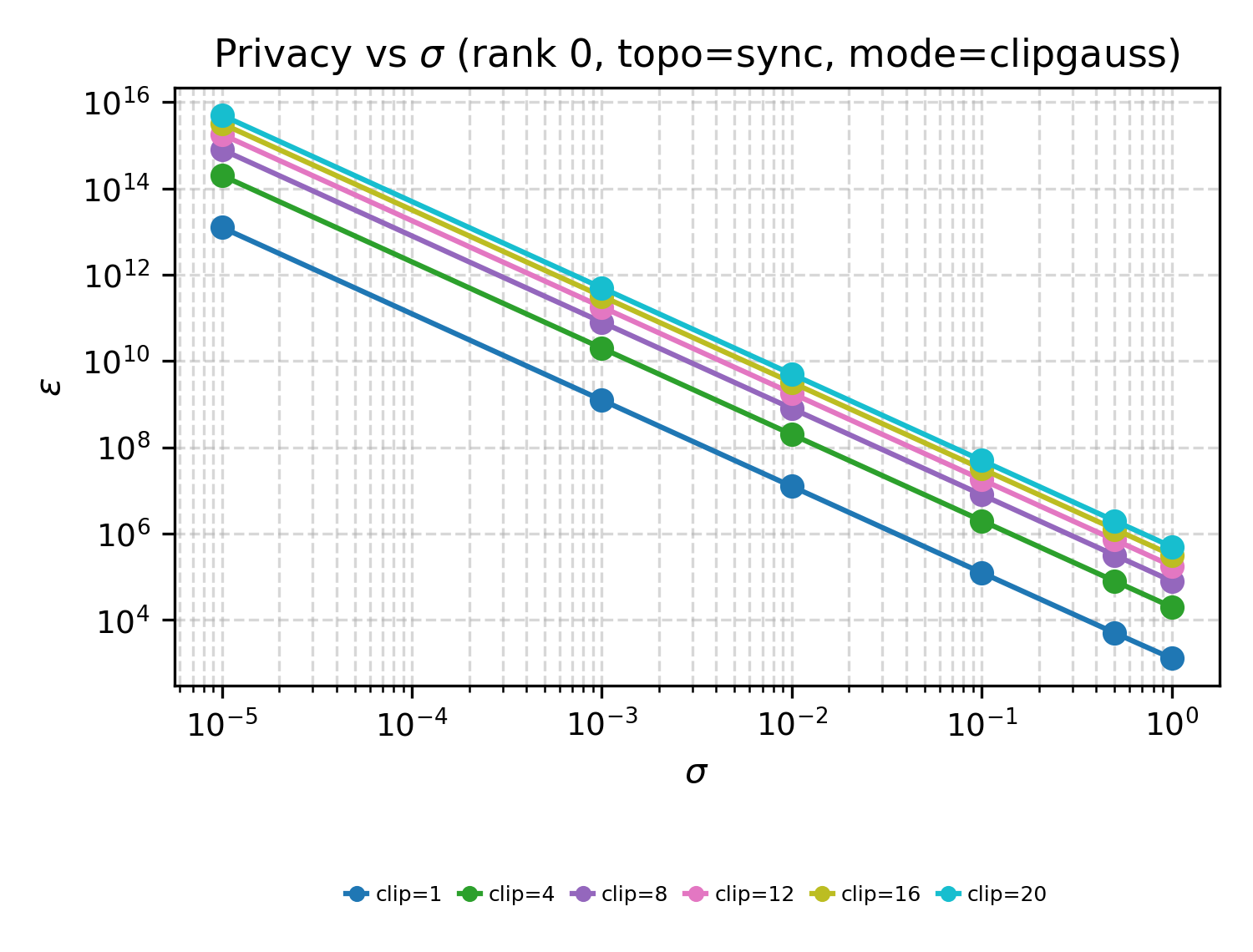}
  \caption{RDP-based privacy parameter
  $\varepsilon_{\scriptscriptstyle\mathrm{DP}}$ (for $\delta=10^{-5}$) as a
  function of log-domain noise scale $\sigma$ for the clip+Gaussian mechanism
  on the synchronous All-to-All topology (rank~0). Each curve corresponds to a
  different clipping radius $S$. Larger $\sigma$ improves the formal privacy
  bound but, as shown in Figure~\ref{fig:noisefloor_sigma}, also increases the
  optimization error.}
  \label{fig:eps_sigma_clip}
\end{figure}

Figure~\ref{fig:eps_sigma_clip} reports the RDP-based $\varepsilon_{\scriptscriptstyle\mathrm{DP}}$ (with $\delta=10^{-5}$) for the clip+Gaussian mechanism. Under non-subsampled, per-iteration releases over $T=1000$, the resulting bounds are extremely loose ($\varepsilon_{\scriptscriptstyle\mathrm{DP}}\gg 1$ across the sweep), and reducing $\varepsilon_{\scriptscriptstyle\mathrm{DP}}$ by shrinking $S$ incurs substantial clipping bias (Figure~\ref{fig:noisefloor_sigma}). Achieving practically tight budgets would require fewer releases (fewer iterations and/or subsampling), privacy amplification, or alternative mechanisms such as privacy-preserving aggregation.

\section{Additional privacy plots}
\label{app:privacy-extra}

\subsection{Full clip sweep: error vs iteration for each clip value (sync)}

\begin{figure}[!htbp]
  \centering
  \begin{subfigure}[b]{0.32\linewidth}
    \centering
    \includegraphics[width=\linewidth]{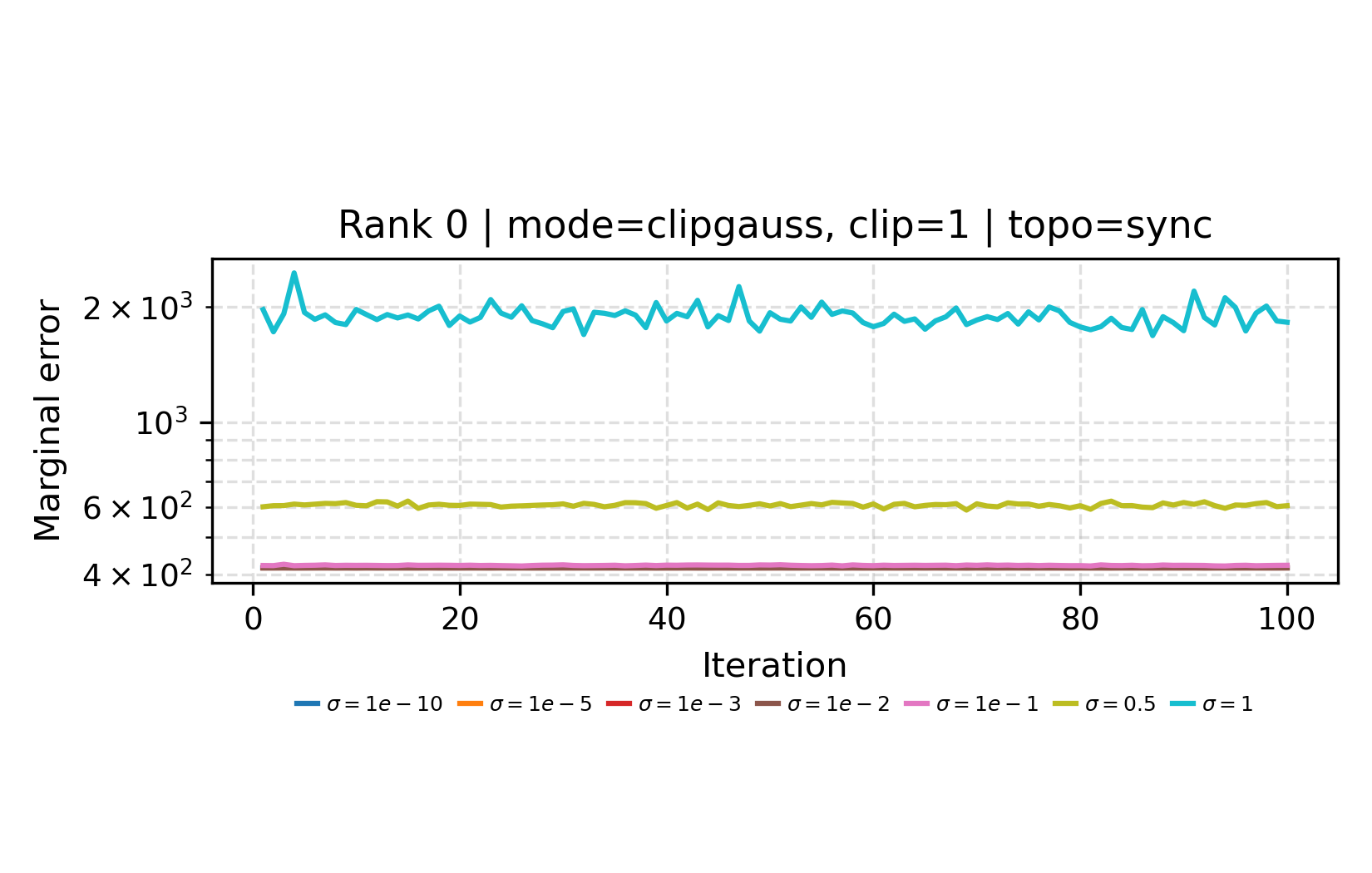}
    \caption{$S=1$}
  \end{subfigure}
  \hfill
  \begin{subfigure}[b]{0.32\linewidth}
    \centering
    \includegraphics[width=\linewidth]{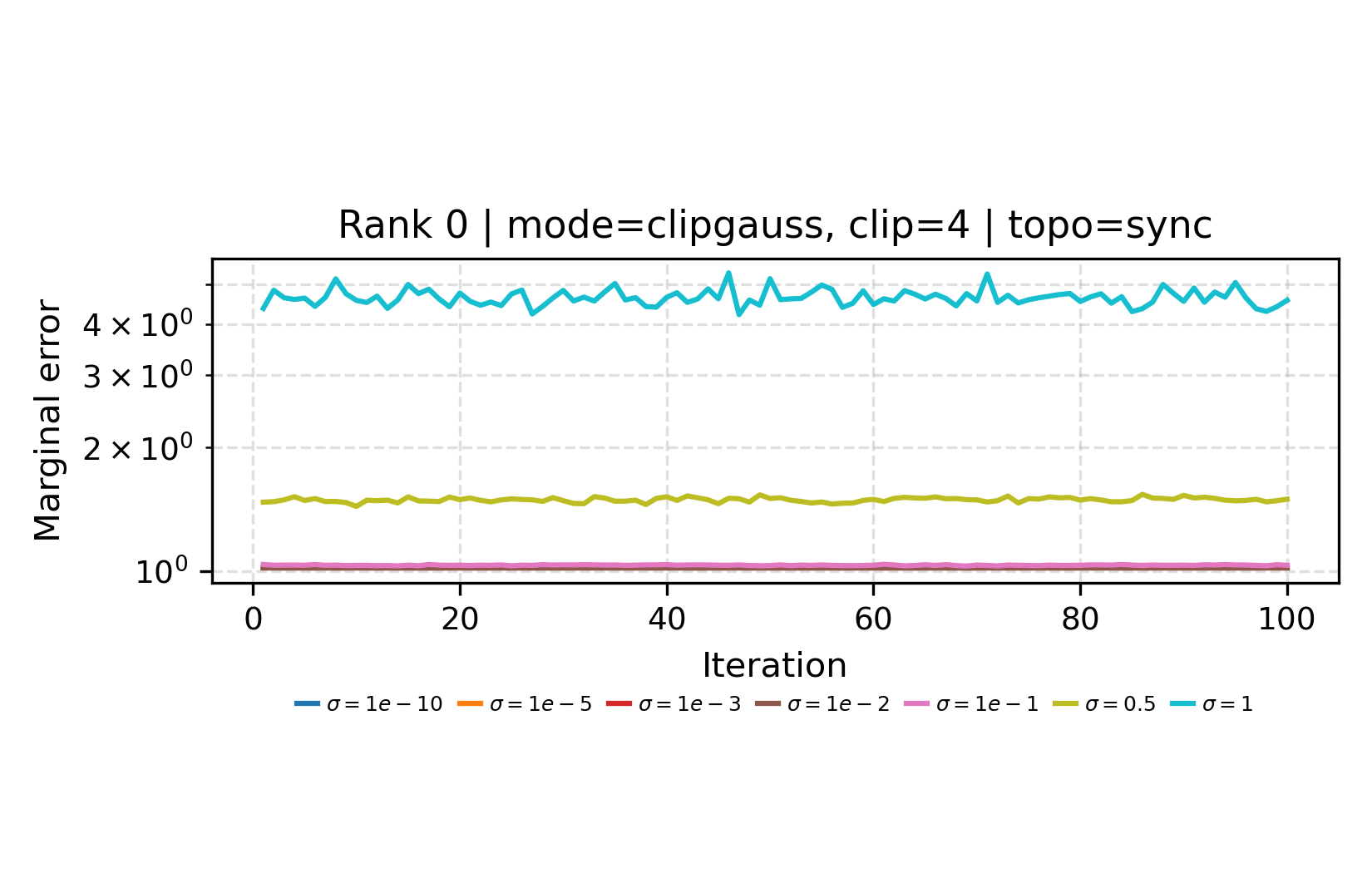}
    \caption{$S=4$}
  \end{subfigure}
  \hfill
  \begin{subfigure}[b]{0.32\linewidth}
    \centering
    \includegraphics[width=\linewidth]{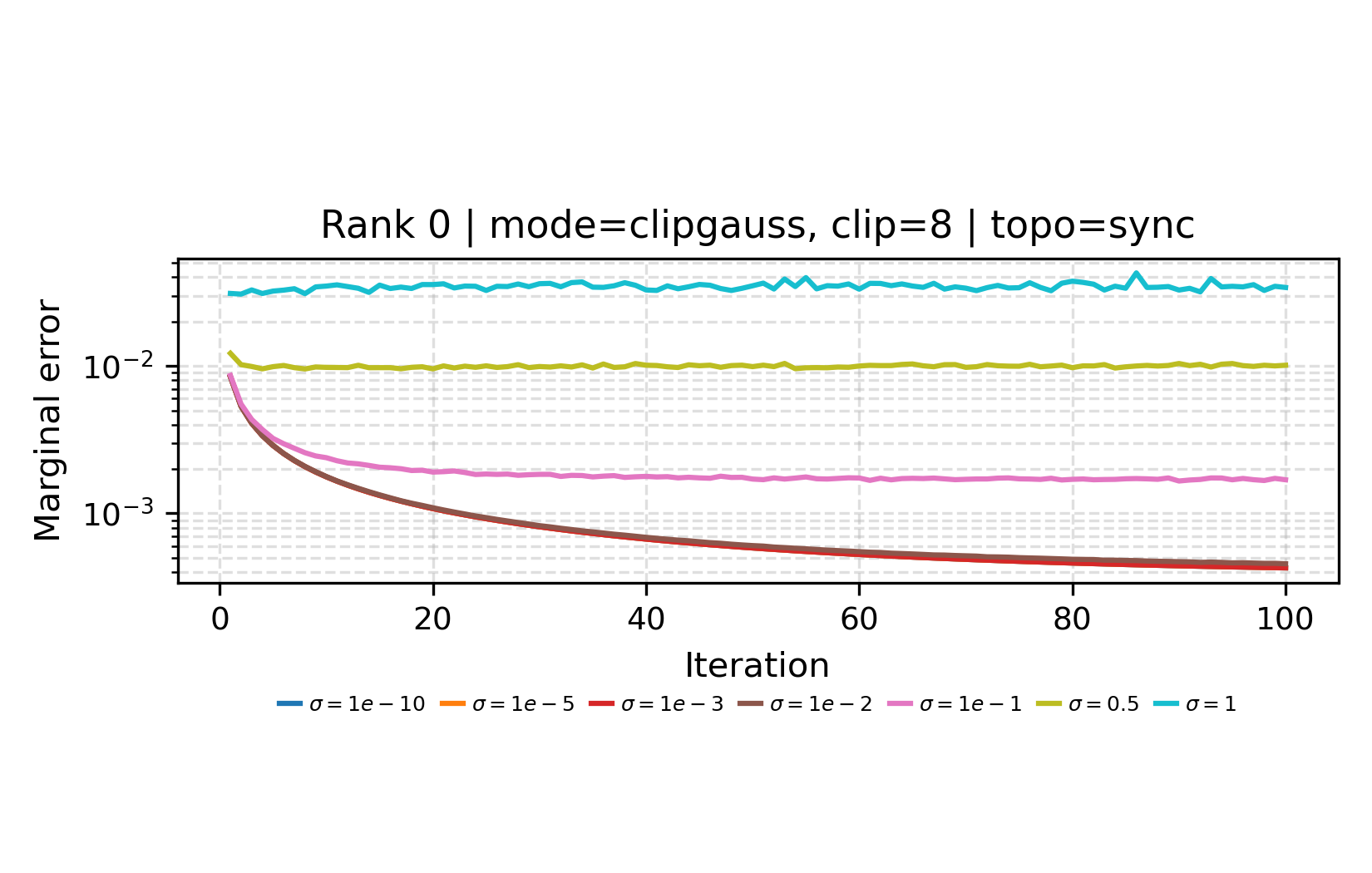}
    \caption{$S=8$}
  \end{subfigure}
  \\
  \vspace{0.5em}
  \begin{subfigure}[b]{0.32\linewidth}
    \centering
    \includegraphics[width=\linewidth]{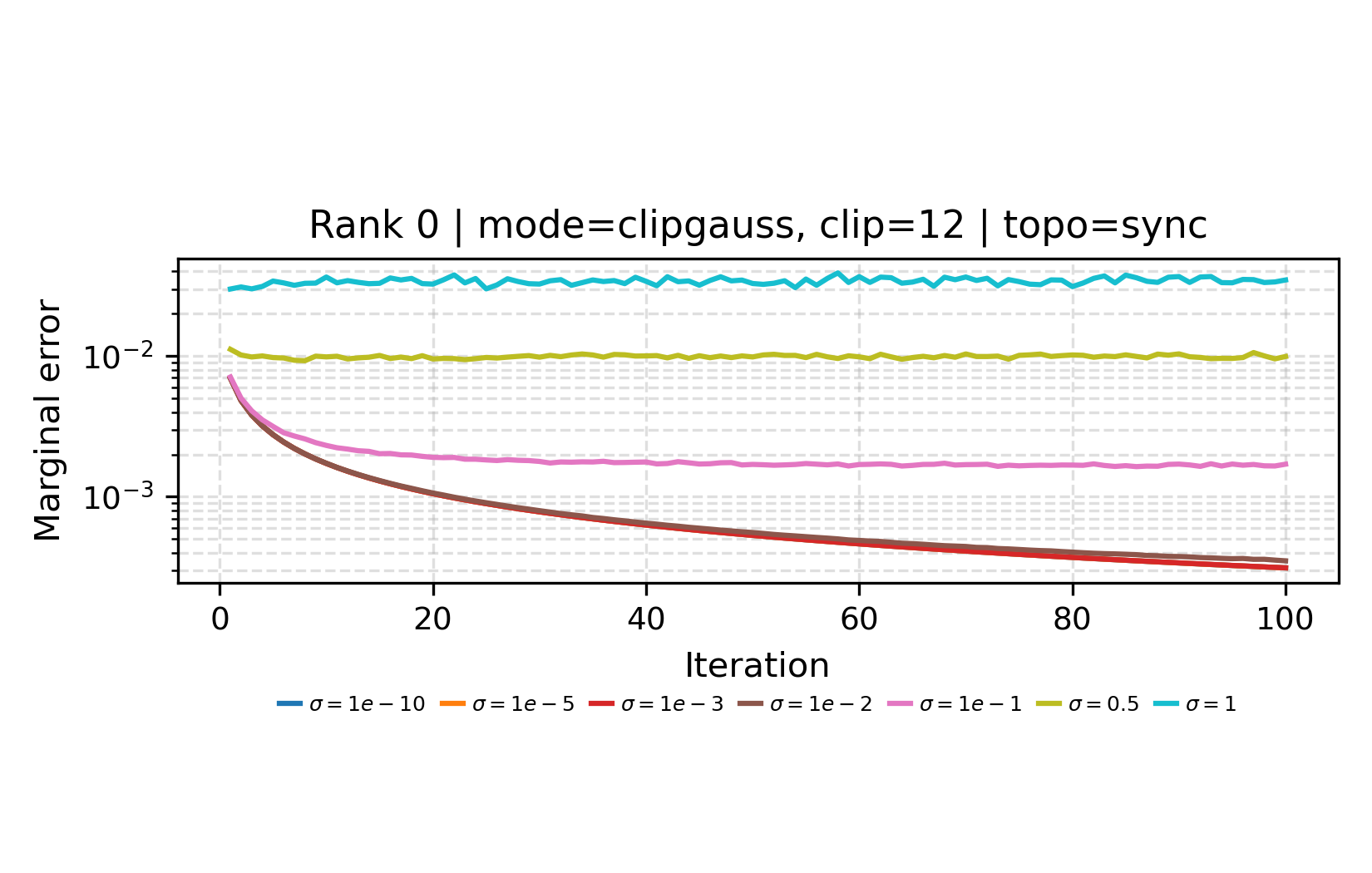}
    \caption{$S=12$}
  \end{subfigure}
  \hfill
  \begin{subfigure}[b]{0.32\linewidth}
    \centering
    \includegraphics[width=\linewidth]{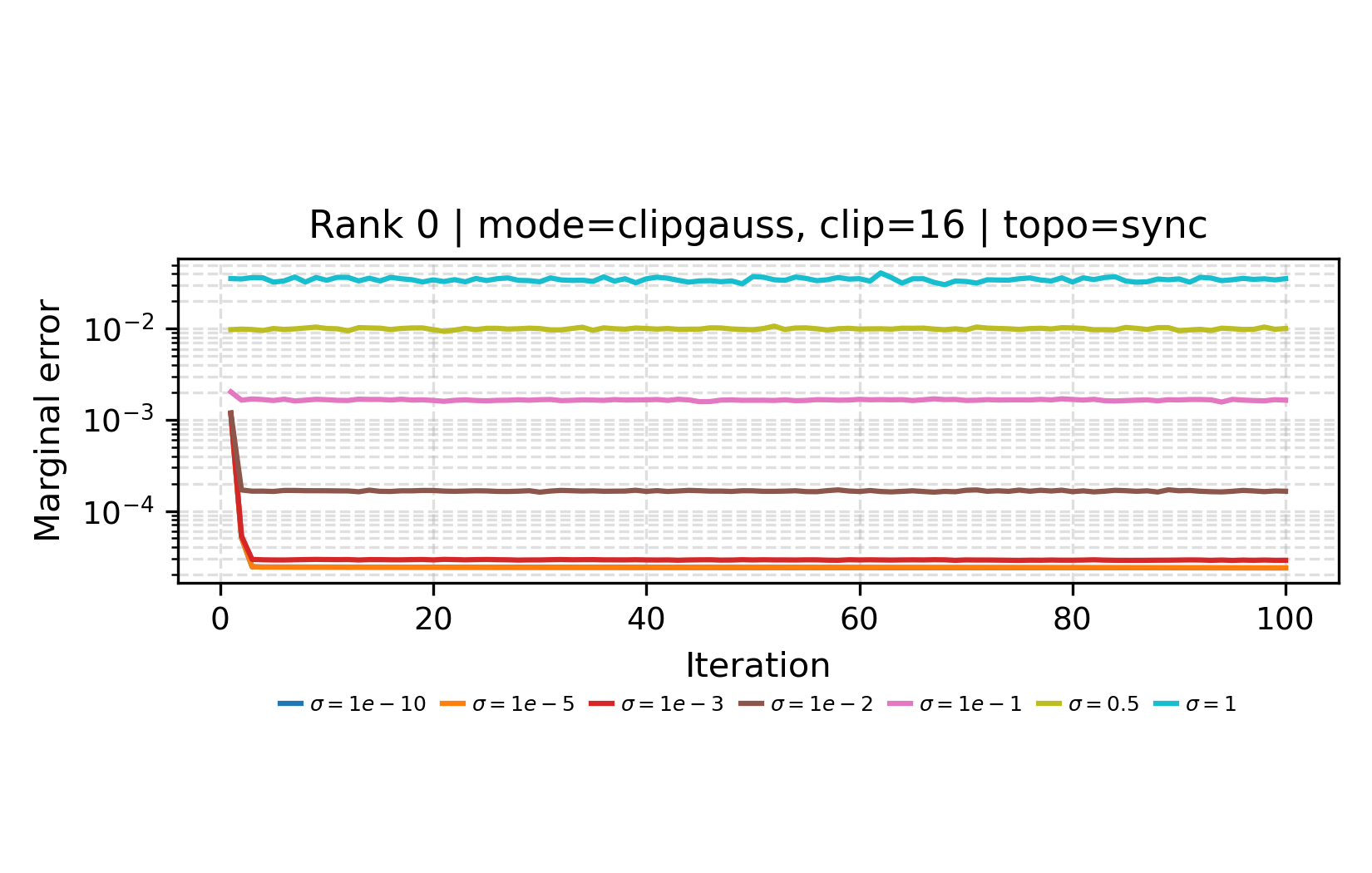}
    \caption{$S=16$}
  \end{subfigure}
  \hfill
  \begin{subfigure}[b]{0.32\linewidth}
    \centering
    \includegraphics[width=\linewidth]{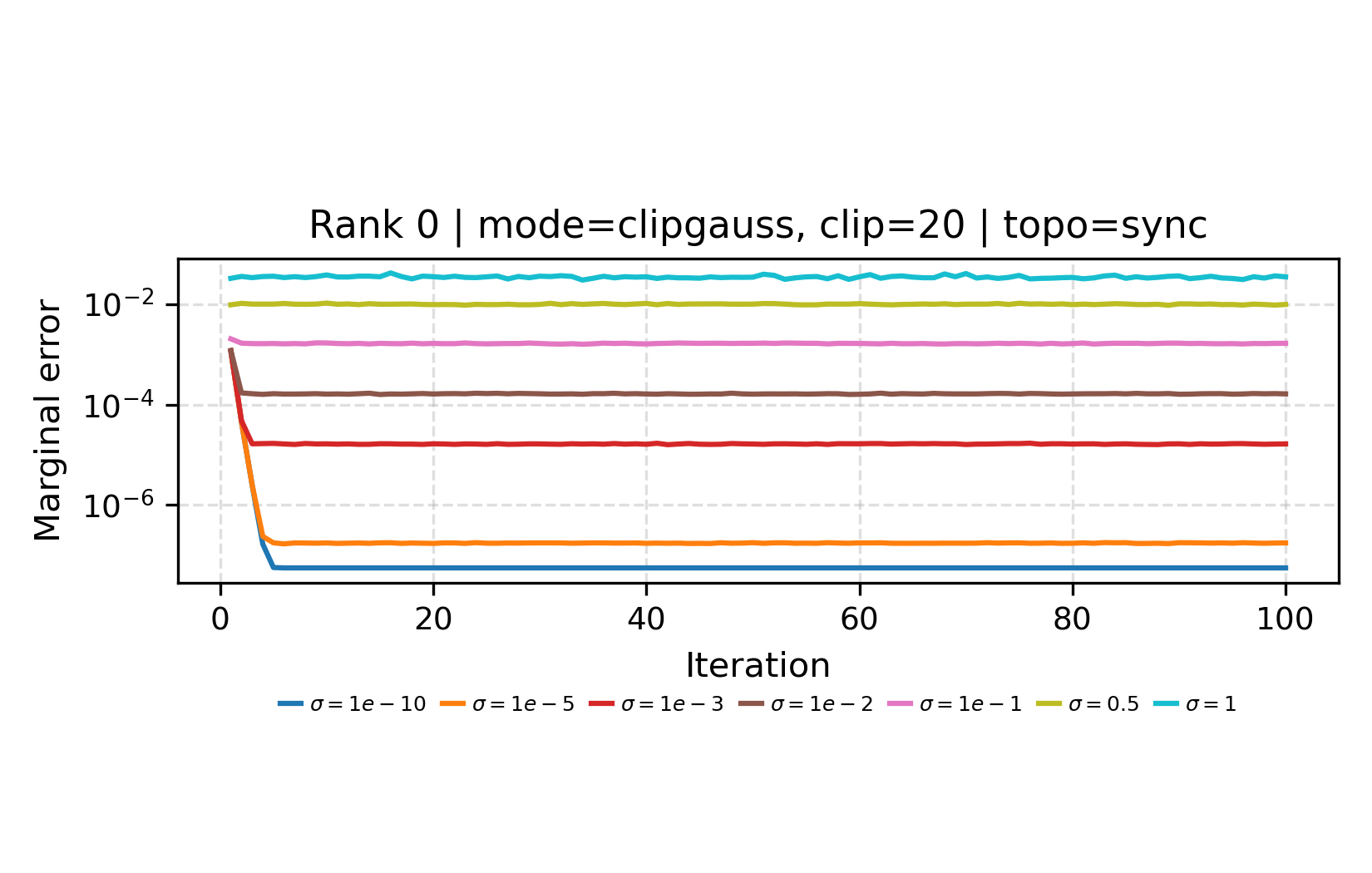}
    \caption{$S=20$}
  \end{subfigure}
  \caption{Marginal error as a function of iteration for the clip+Gaussian mechanism with different clipping radii $S_u=S_v$ and several noise scales $\sigma$ (synchronous All-to-All topology, rank~0). For small radii the curves are almost flat and dominated by clipping bias; for $S\ge 16$ they closely track the Gaussian-only baseline in Figure~\ref{fig:gauss-sweep-sync} for small~$\sigma$.}
  \label{fig:clip-sweep-sync}
\end{figure}

\begin{figure}[!htbp]
  \centering
  \includegraphics[width=0.7\linewidth]{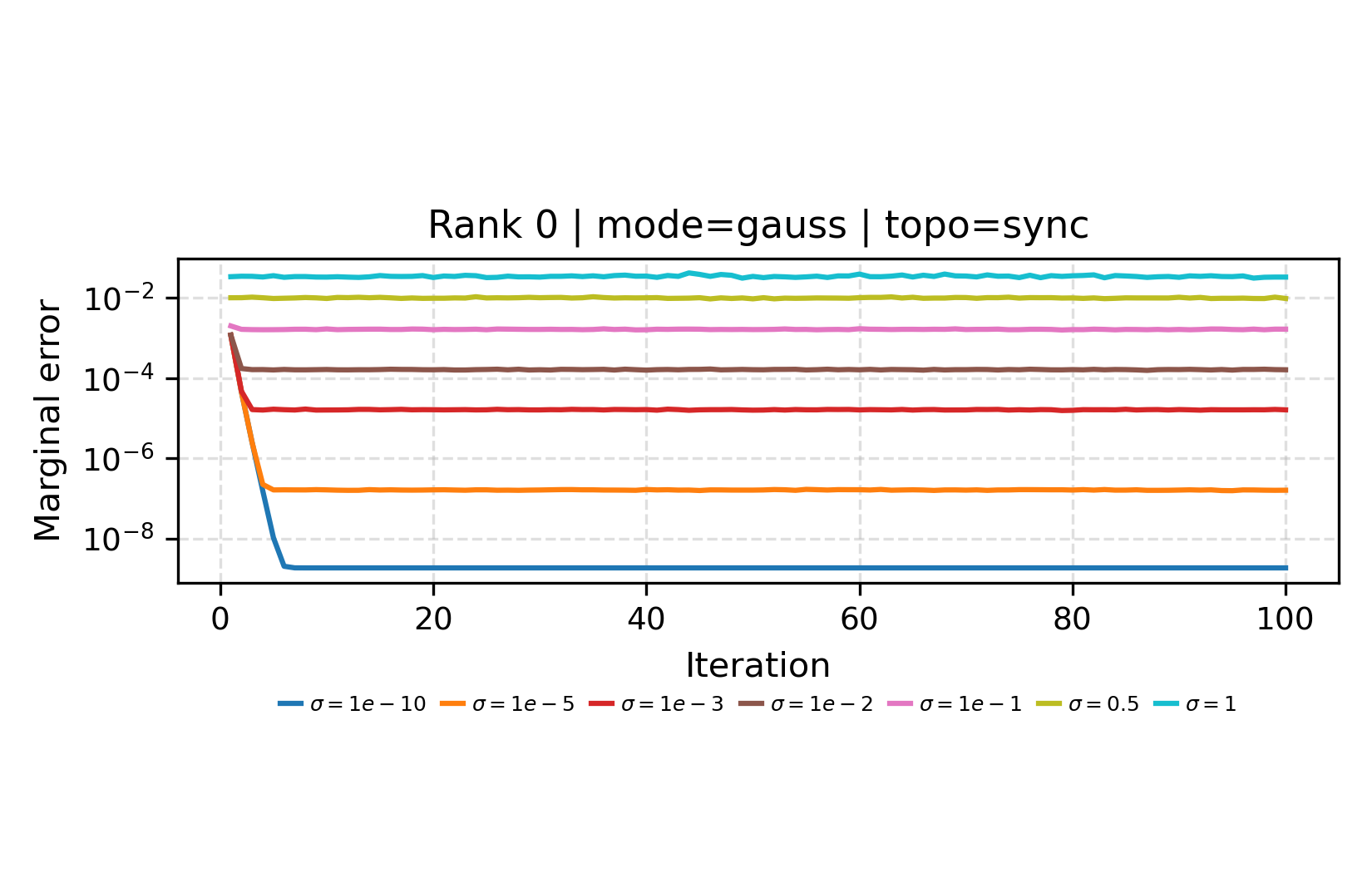}
  \caption{Gaussian-only ablation: marginal error as a function of iteration for several noise scales $\sigma$ without clipping (synchronous All-to-All topology, rank~0). This serves as a reference for Figure~\ref{fig:clip-sweep-sync}.}
  \label{fig:gauss-sweep-sync}
\end{figure}

\subsection{Marginal error vs iteration for $\sigma=10^{-10}$}

\begin{figure}[!htbp]
  \centering
  \begin{subfigure}[b]{0.49\linewidth}
    \centering
    \includegraphics[width=\linewidth]{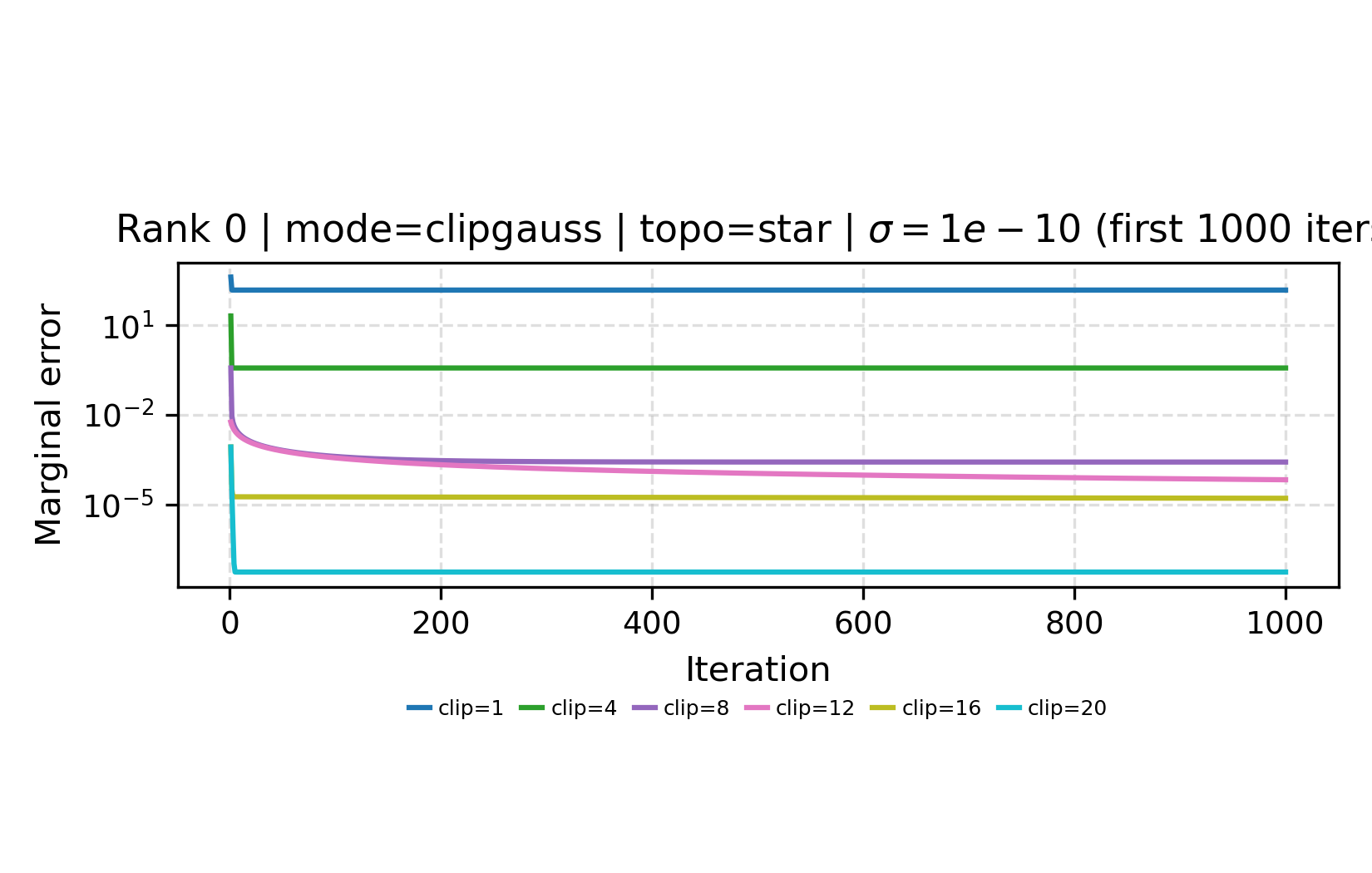}
    \caption{Star topology, rank~0.}
    \label{fig:iters_sigma_star_rank0}
  \end{subfigure}
  \hfill
  \begin{subfigure}[b]{0.49\linewidth}
    \centering
    \includegraphics[width=\linewidth]{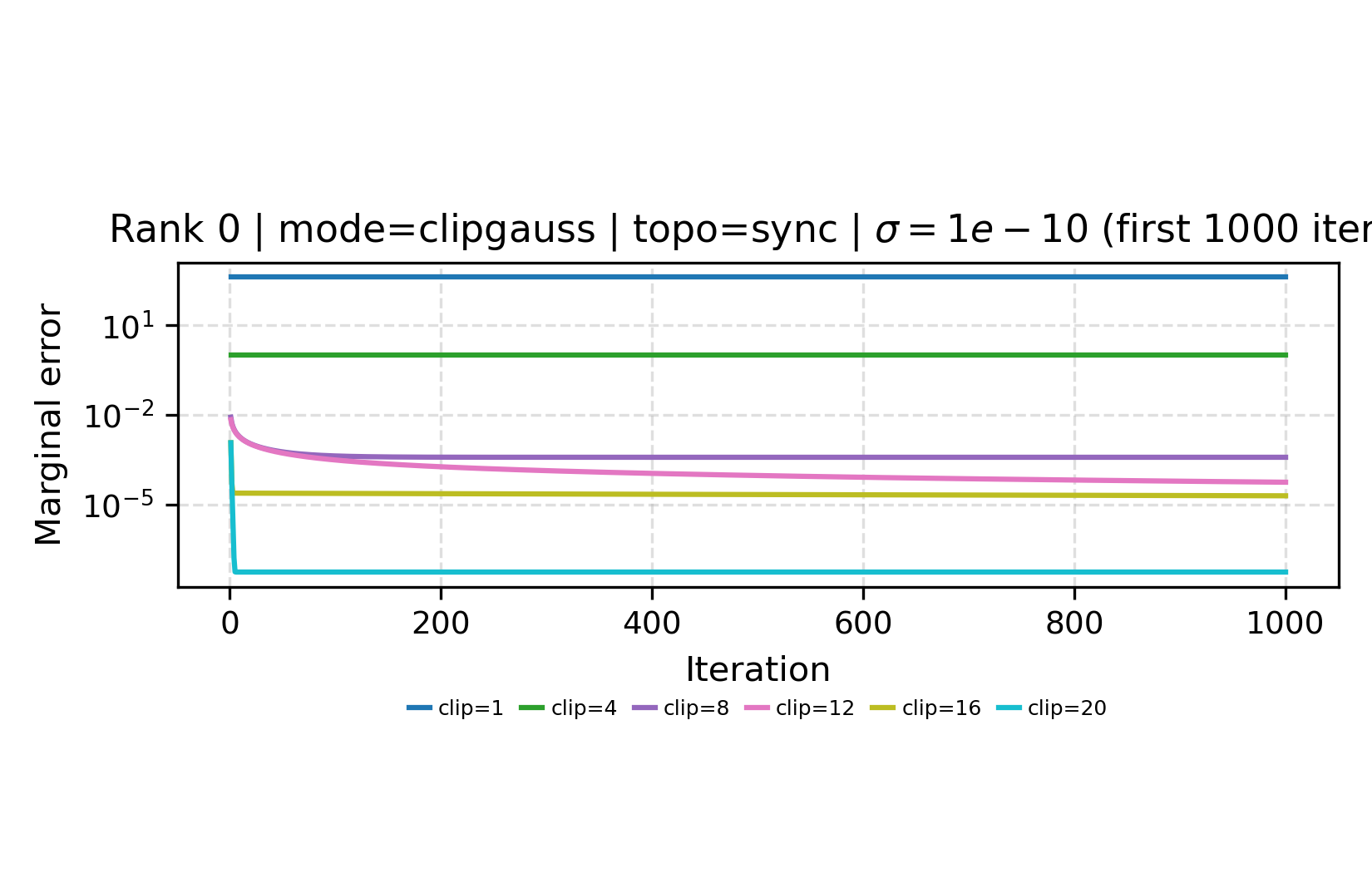}
    \caption{All-to-All topology, rank~0.}
    \label{fig:iters_sigma_sync_rank0}
  \end{subfigure}
  \caption{Marginal error on $a$ during the first 1000 iterations for log-domain noise scale $\sigma = 10^{-10}$ and several clipping radii $S$ in the clip+Gaussian mechanism. The plateau level is controlled almost entirely by $S$, and the star and All-to-All topologies (and different ranks) exhibit nearly identical behavior.}
  \label{fig:iters_sigma1e-10}
\end{figure}

\subsection{Error vs $\varepsilon$}

\begin{figure}[!htbp]
  \centering
  \includegraphics[width=0.7\linewidth]{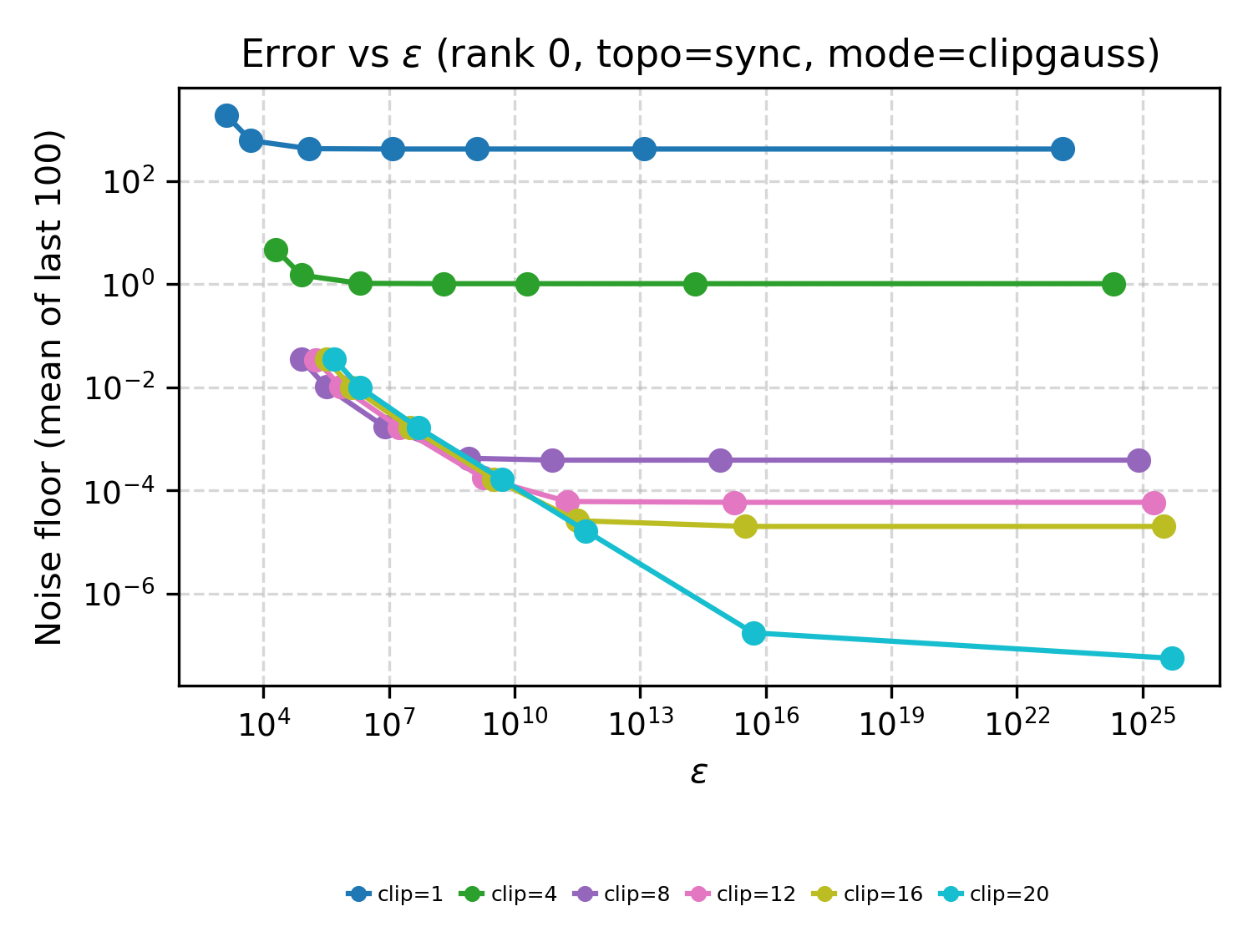}
  \caption{Noise floor (mean marginal error over the last $100$ iterations) against the corresponding R\'enyi-DP $\varepsilon_{\scriptscriptstyle\mathrm{DP}}$ for the clip+Gaussian mechanism (synchronous All-to-All topology, rank~0). All configurations lie in the regime $\varepsilon_{\scriptscriptstyle\mathrm{DP}}\gg 1$, indicating weak formal privacy guarantees at the hyperparameters used.}
  \label{fig:error_vs_eps_clip}
\end{figure}

\subsection{Synchronous star plots}
\label{subsec:privacy_star_plots}

\subsubsection{Effect of log-domain Gaussian noise}

\begin{figure}[!htbp]
  \centering
  \begin{subfigure}[b]{0.49\linewidth}
    \centering
    \includegraphics[width=\linewidth]{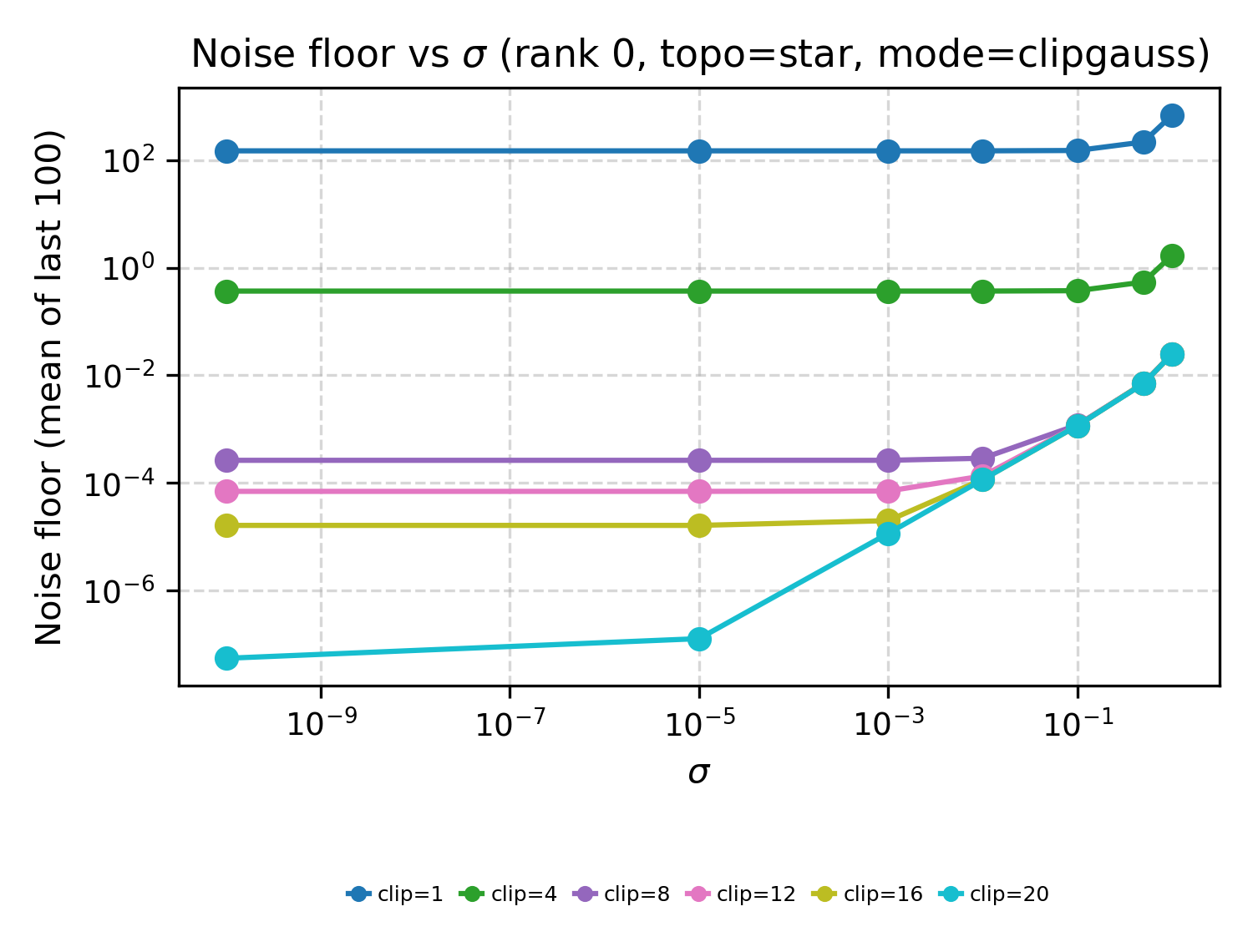}
    \caption{Clip+Gaussian mechanism.}
    \label{fig:noisefloor_sigma_clip_star}
  \end{subfigure}
  \hfill
  \begin{subfigure}[b]{0.49\linewidth}
    \centering
    \includegraphics[width=\linewidth]{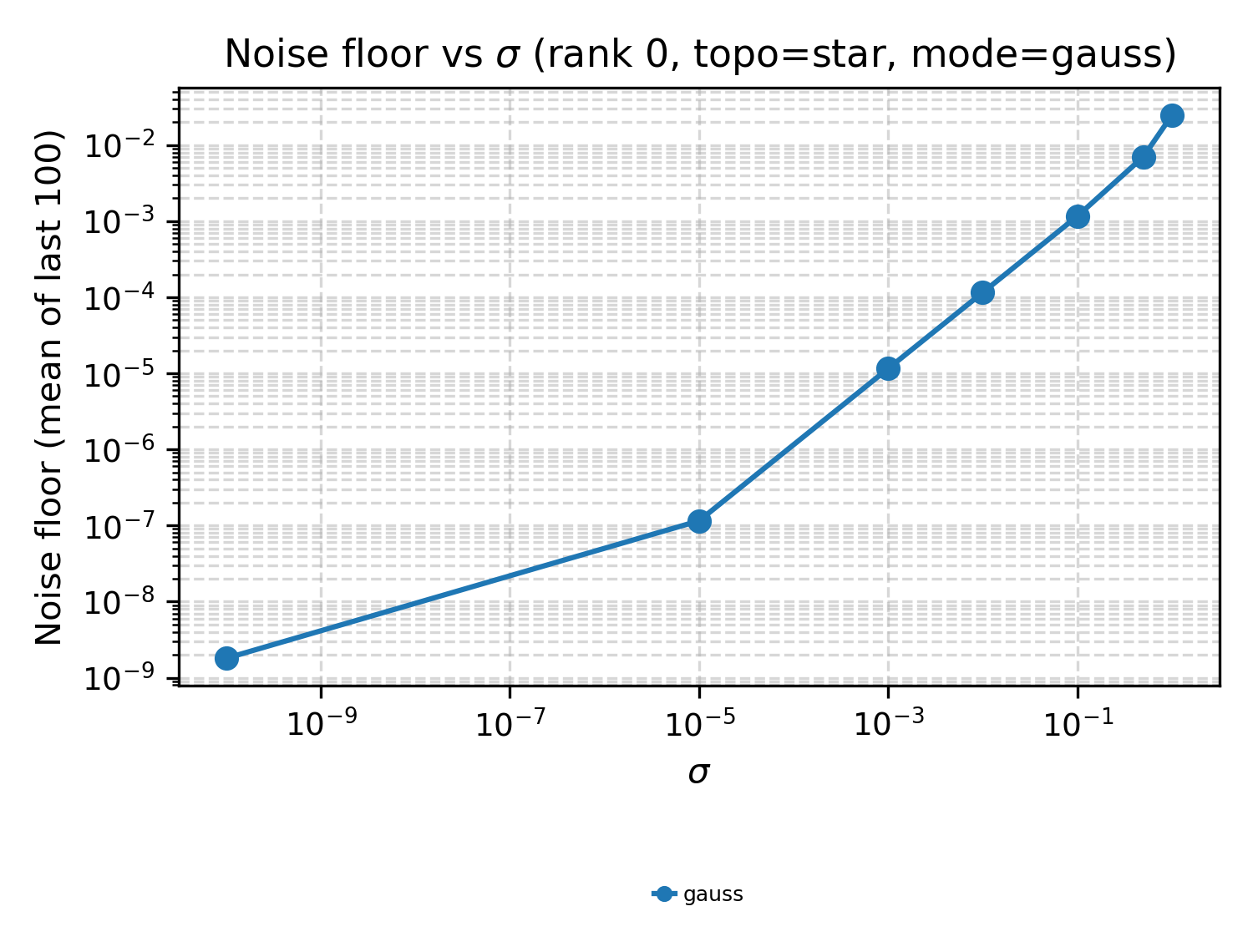}
    \caption{Gaussian-only ablation (no clipping).}
    \label{fig:noisefloor_sigma_gauss_star}
  \end{subfigure}
  \caption{Noise floor (mean marginal error over the last 100 iterations) as a
  function of log-domain noise scale $\sigma$ on the synchronous star
  topology. Each curve in panel~(a) corresponds to a different
  clipping radius $S$; panel~(b) uses no clipping and is shown for robustness
  comparison only (no formal DP guarantee).}
  \label{fig:noisefloor_sigma_star}
\end{figure}

\subsubsection{Utility--privacy tradeoff}

\begin{figure}[!htbp]
  \centering
  \includegraphics[width=0.6\linewidth]{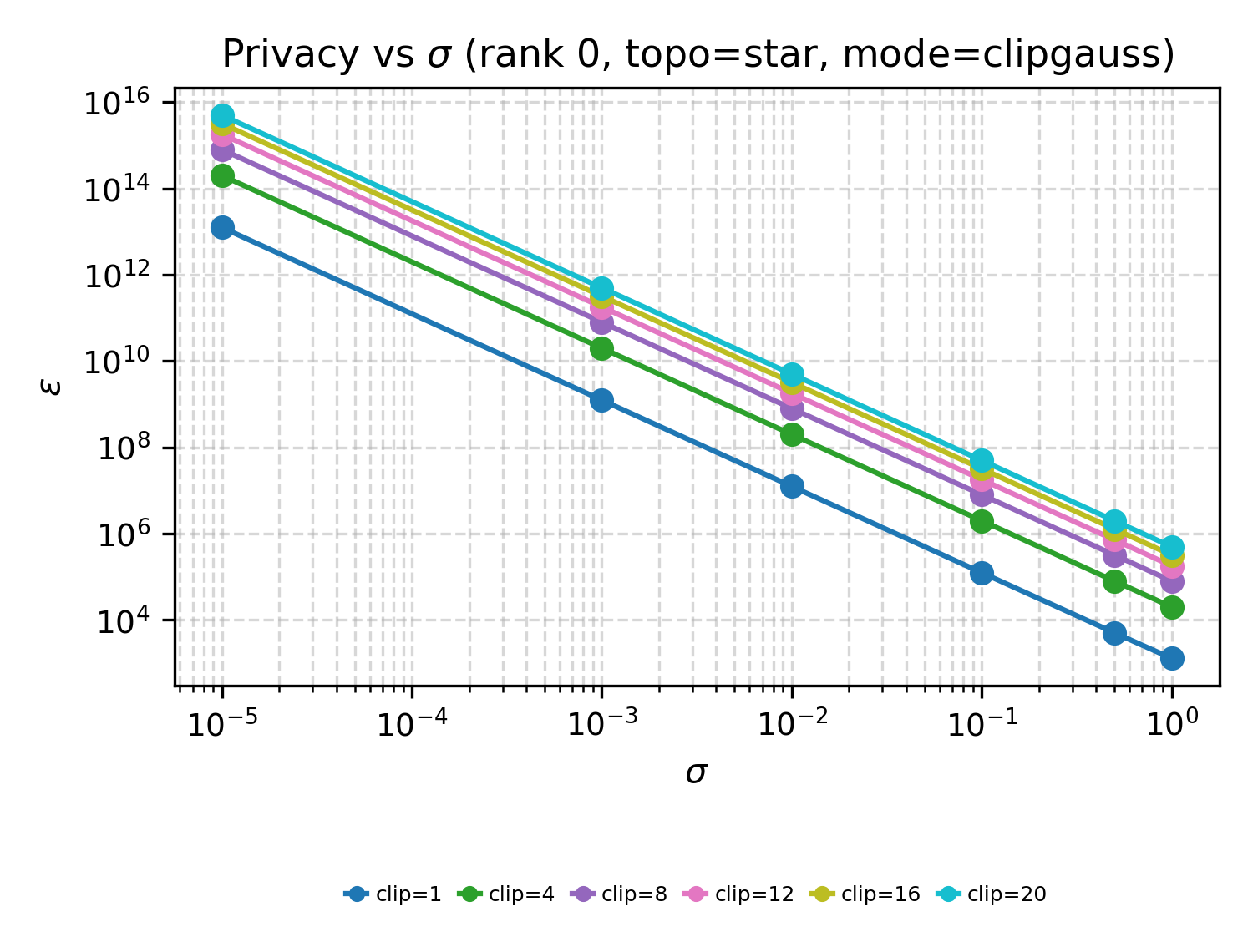}
  \caption{RDP-based privacy parameter
  $\varepsilon_{\scriptscriptstyle\mathrm{DP}}$ (for $\delta=10^{-5}$) as a
  function of log-domain noise scale $\sigma$ for the clip+Gaussian mechanism
  on the synchronous star topology. Each curve corresponds to a
  different clipping radius $S$. Larger $\sigma$ improves the formal privacy
  bound but, as shown in Figure~\ref{fig:noisefloor_sigma_star}, also increases the
  optimization error.}
  \label{fig:eps_sigma_clip_star}
\end{figure}

\subsubsection{Full clip sweep: error vs iteration for each clip value (star)}

\begin{figure}[!htbp]
  \centering
  \begin{subfigure}[b]{0.32\linewidth}
    \centering
    \includegraphics[width=\linewidth]{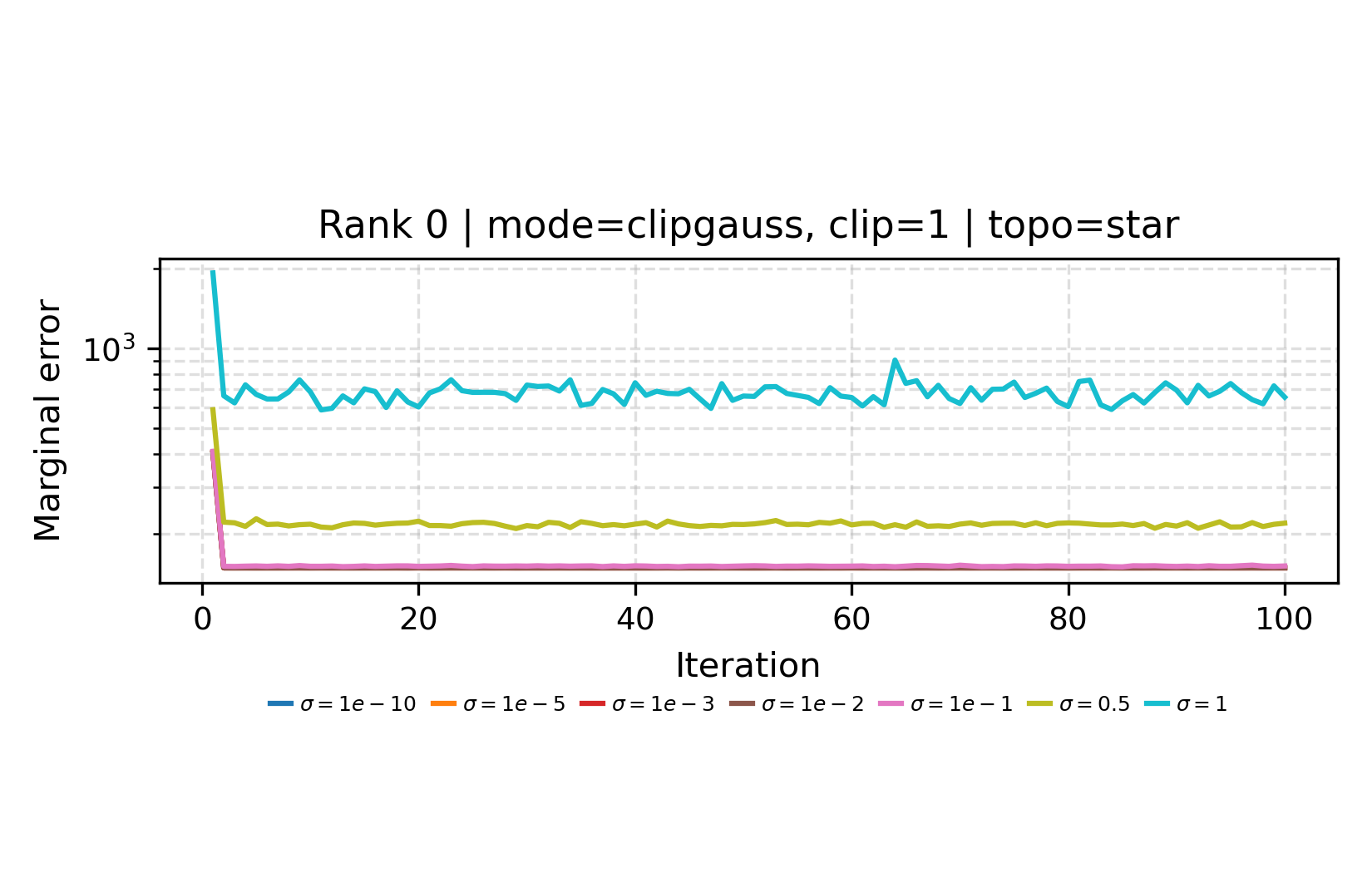}
    \caption{$S=1$}
  \end{subfigure}
  \hfill
  \begin{subfigure}[b]{0.32\linewidth}
    \centering
    \includegraphics[width=\linewidth]{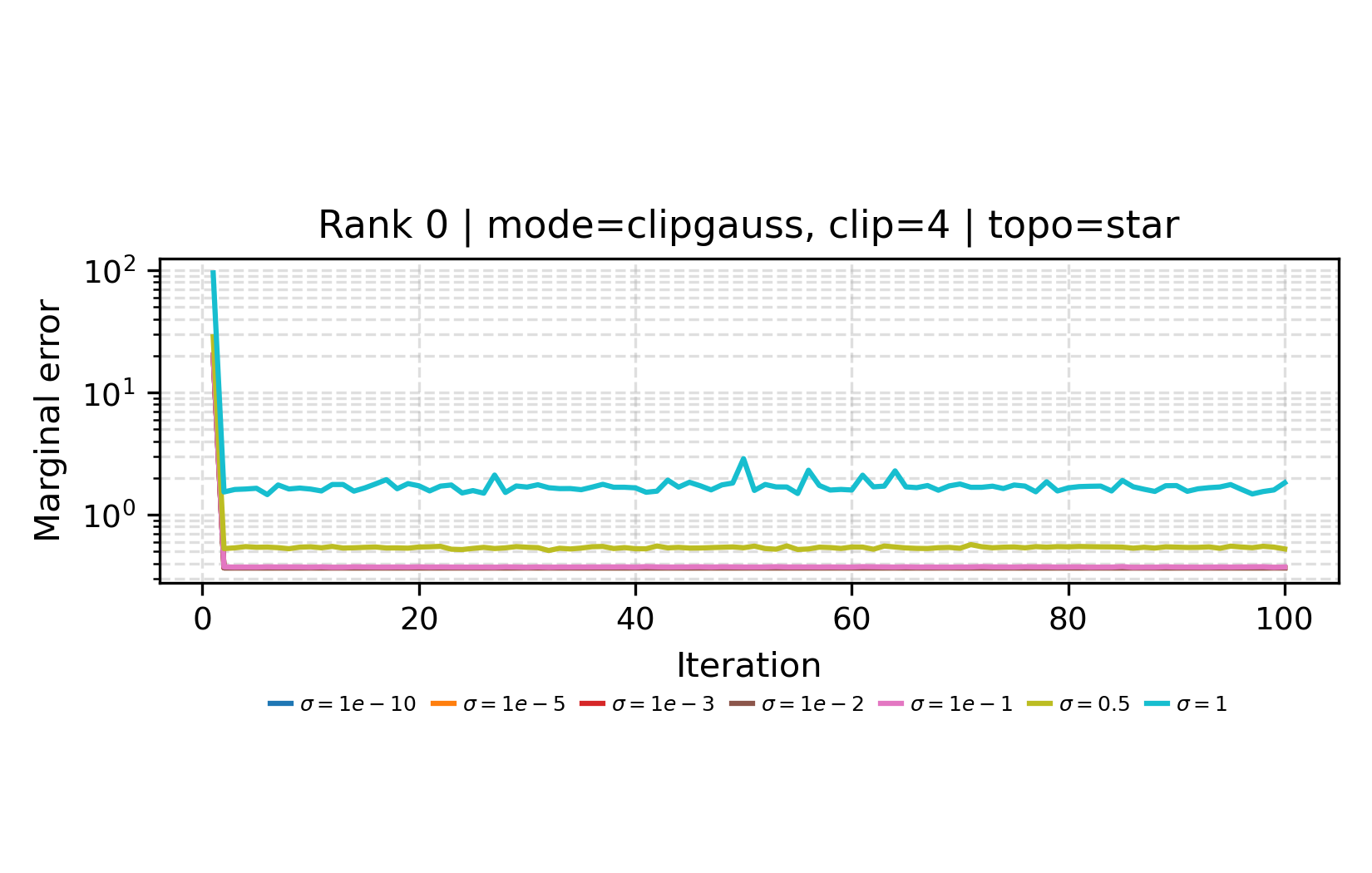}
    \caption{$S=4$}
  \end{subfigure}
  \hfill
  \begin{subfigure}[b]{0.32\linewidth}
    \centering
    \includegraphics[width=\linewidth]{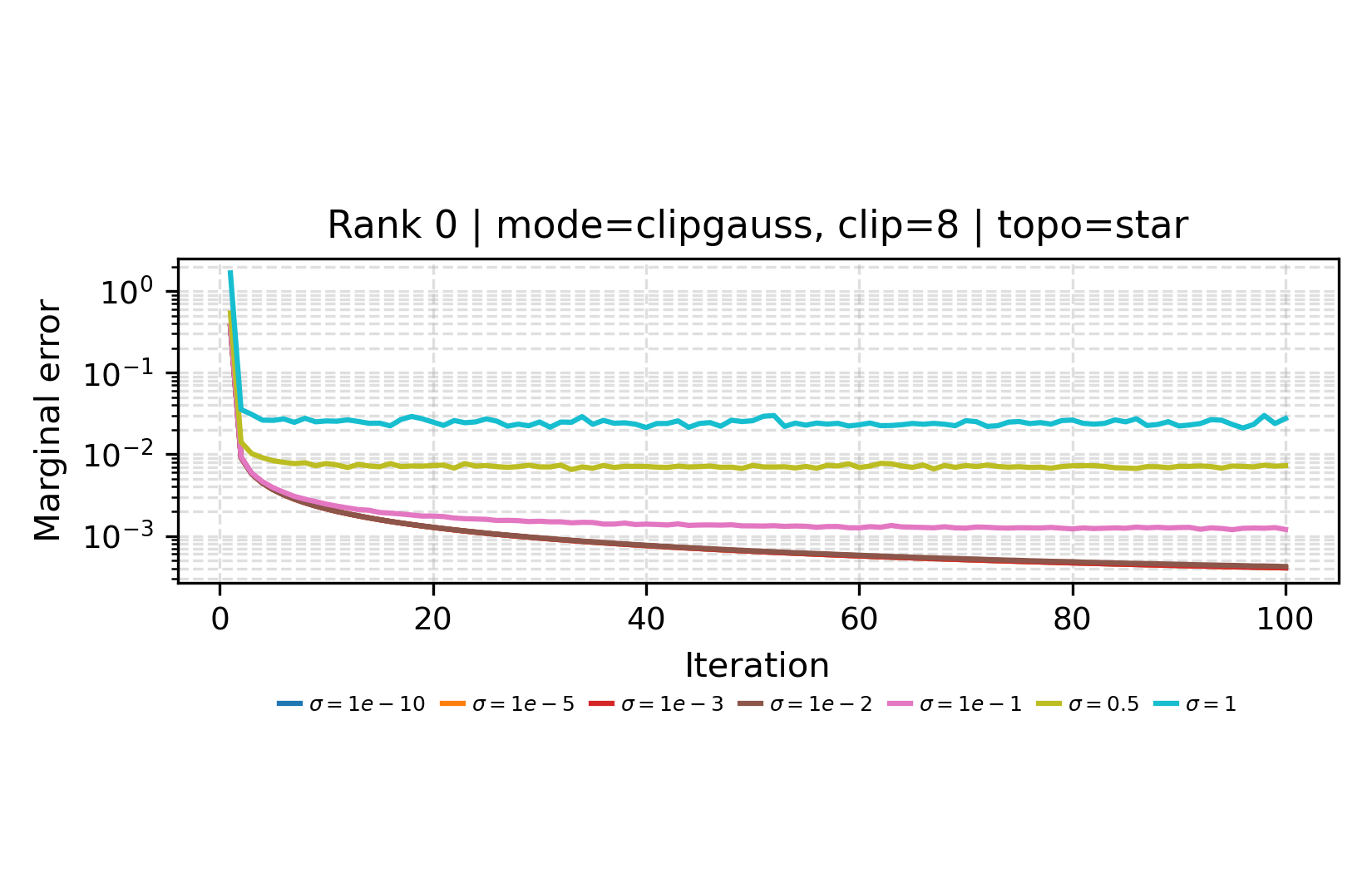}
    \caption{$S=8$}
  \end{subfigure}
  \\
  \vspace{0.5em}
  \begin{subfigure}[b]{0.32\linewidth}
    \centering
    \includegraphics[width=\linewidth]{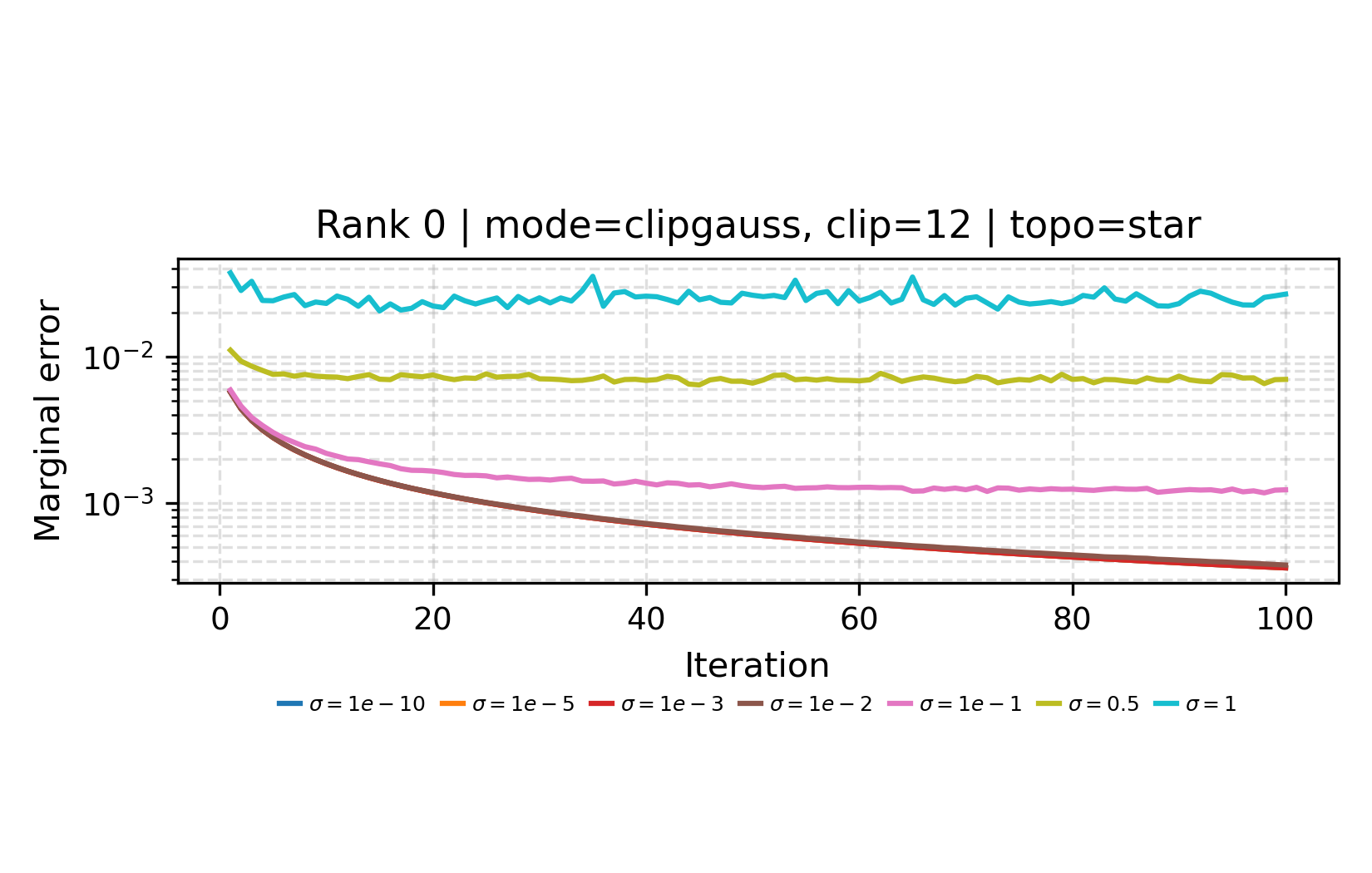}
    \caption{$S=12$}
  \end{subfigure}
  \hfill
  \begin{subfigure}[b]{0.32\linewidth}
    \centering
    \includegraphics[width=\linewidth]{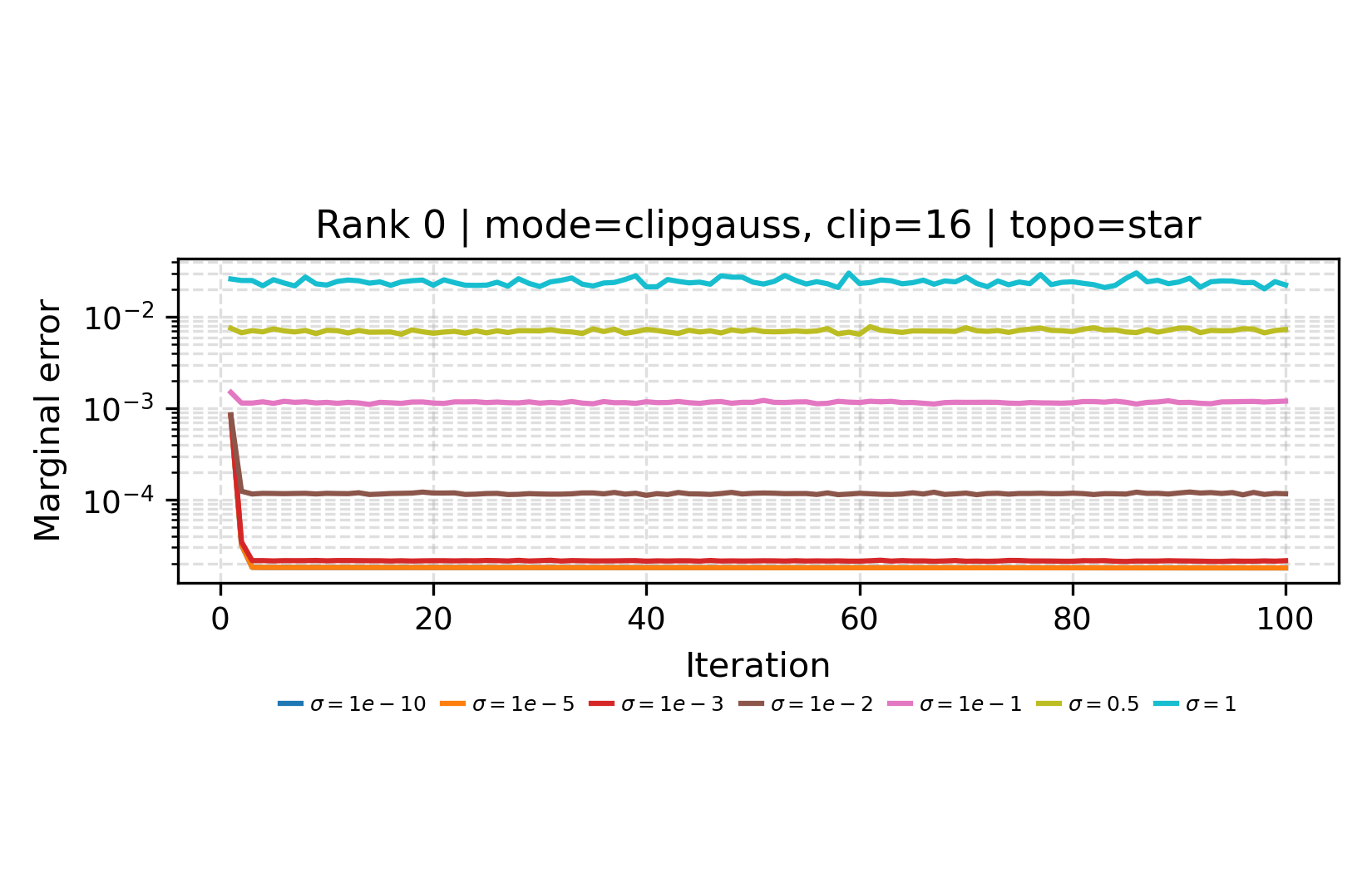}
    \caption{$S=16$}
  \end{subfigure}
  \hfill
  \begin{subfigure}[b]{0.32\linewidth}
    \centering
    \includegraphics[width=\linewidth]{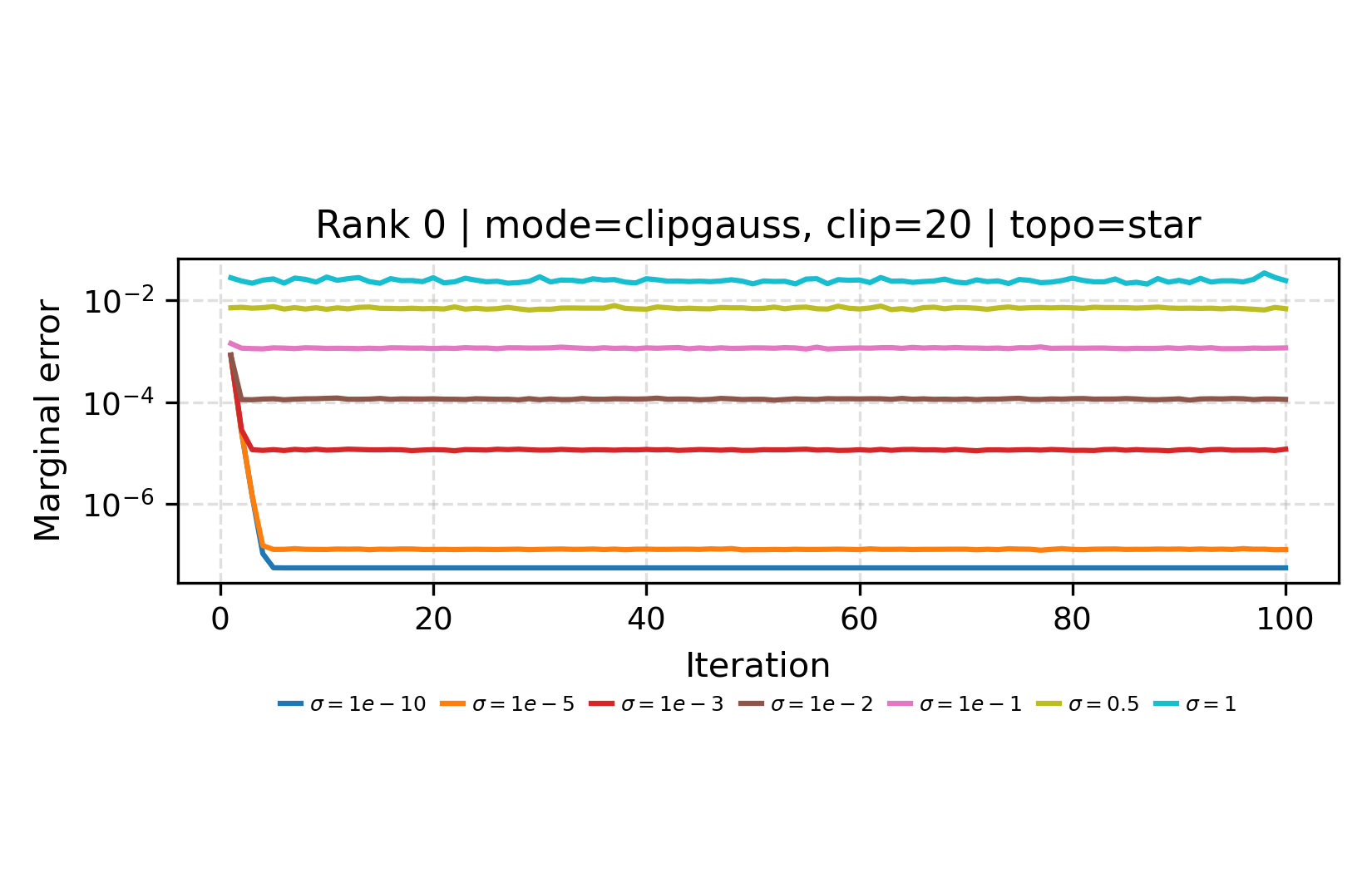}
    \caption{$S=20$}
  \end{subfigure}
  \caption{Marginal error as a function of iteration for the clip+Gaussian mechanism with different clipping radii $S_u=S_v$ and several noise scales $\sigma$ (synchronous star topology). For small radii the curves are almost flat and dominated by clipping bias; for $S\ge 16$ they closely track the Gaussian-only baseline in Figure~\ref{fig:gauss-sweep-star} for small~$\sigma$.}
  \label{fig:clip-sweep-star}
\end{figure}

\begin{figure}[!htbp]
  \centering
  \includegraphics[width=0.7\linewidth]{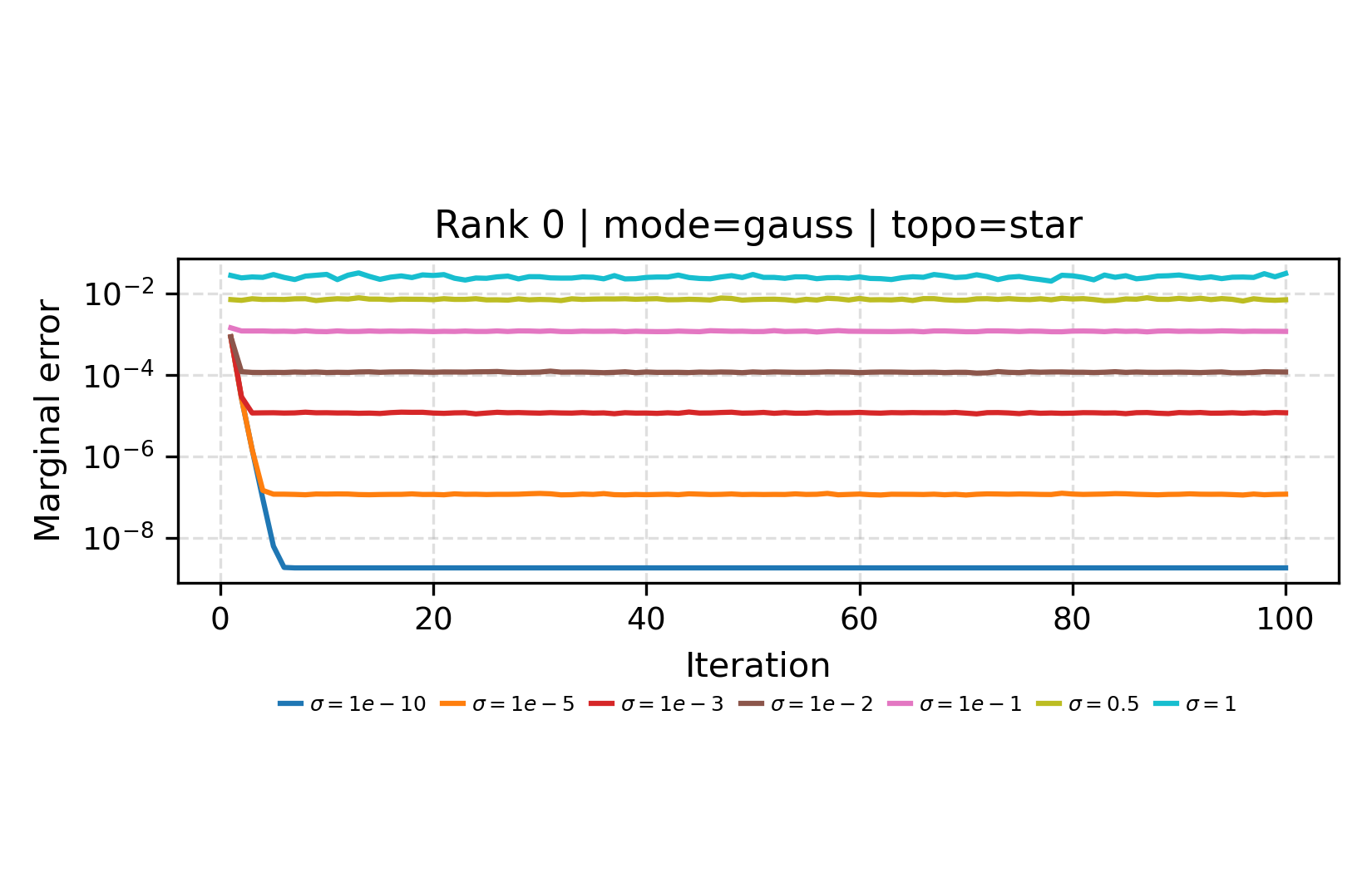}
  \caption{Gaussian-only ablation: marginal error as a function of iteration for several noise scales $\sigma$ without clipping (synchronous star topology). This serves as a reference for Figure~\ref{fig:clip-sweep-star}.}
  \label{fig:gauss-sweep-star}
\end{figure}

\subsubsection{Error vs $\varepsilon$}

\begin{figure}[!htbp]
  \centering
  \includegraphics[width=0.7\linewidth]{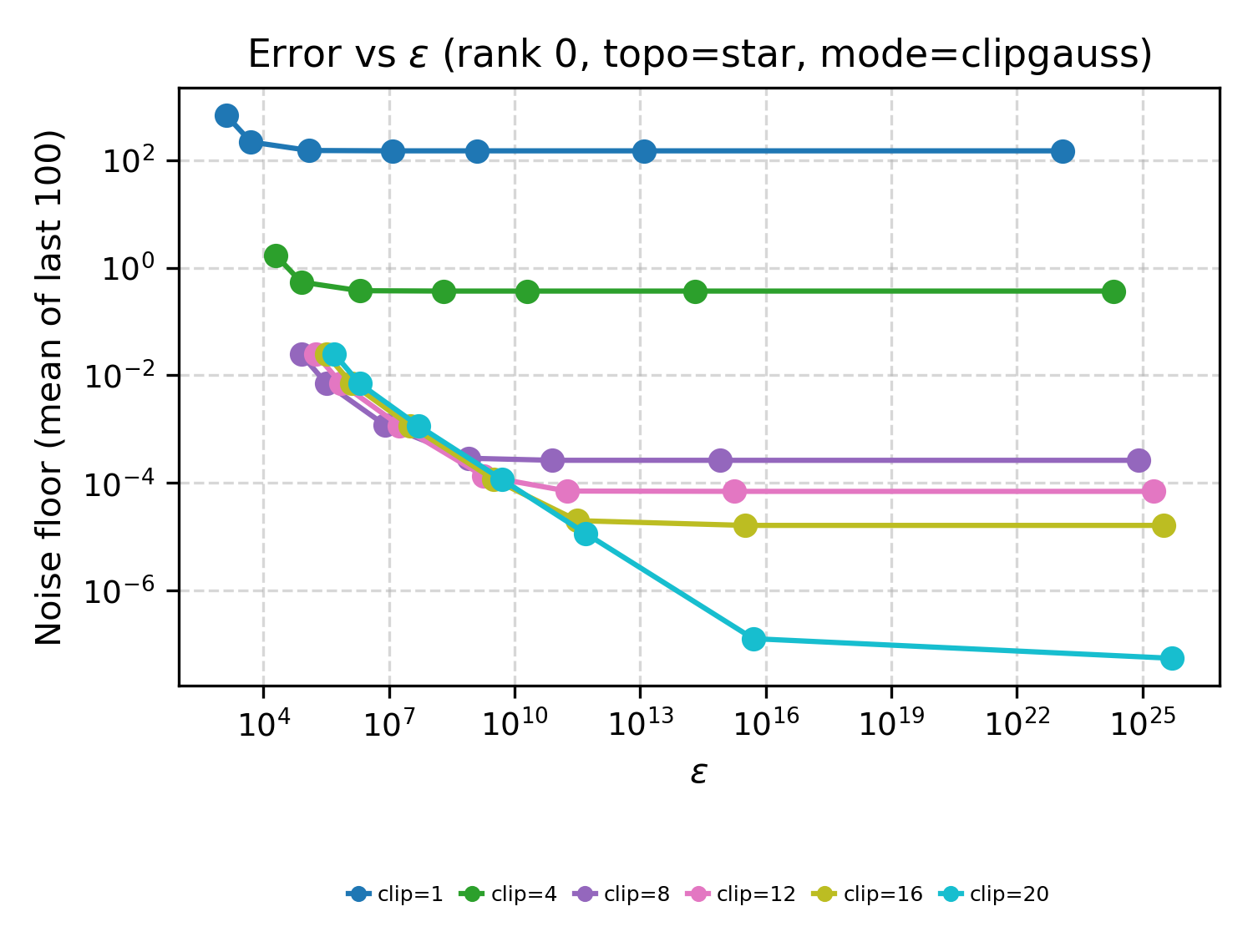}
  \caption{Noise floor (mean marginal error over the last $100$ iterations) against the corresponding R\'enyi-DP $\varepsilon_{\scriptscriptstyle\mathrm{DP}}$ for the clip+Gaussian mechanism (synchronous star topology). All configurations lie in the regime $\varepsilon_{\scriptscriptstyle\mathrm{DP}}\gg 1$, indicating weak formal privacy guarantees at the hyperparameters used.}
  \label{fig:error_vs_eps_clip_star}
\end{figure}

\section{Going further with the financial risk formulation}
\label{app:finance}

\subsection{Blanchet--Murthy ambiguity set and dual form}

Following \cite{blanchet2019quantifying}, we consider a Wasserstein ambiguity
set around the empirical distribution $\hat P$ of historical returns:
\[
  \mathcal{P}_\delta
  = \bigl\{ P\in\mathcal{P}(\mathbb{R}^d) :
      W_c(P,\hat P)\le\delta \bigr\},
\]
where $W_c$ is the Wasserstein distance with ground cost $c$, and $\delta\ge 0$
controls the size of the ambiguity set (a Wasserstein ball of radius $\delta$).

Given a loss function $l:\mathbb{R}^d\to\mathbb{R}$, the worst-case expected
loss is
\[
  \rho_{\text{worst}}
  = \sup_{P\in\mathcal{P}_\delta} \mathbb{E}_P[l(X)].
\]
\cite{blanchet2019quantifying} show that this can be written in dual form as
\begin{equation}
  \rho_{\text{worst}}
  = \inf_{\lambda\ge 0}\left\{ \lambda\delta
      + \mathbb{E}_{\hat P}\bigl[h_\lambda(X)\bigr] \right\},
  \qquad
  h_\lambda(x)
    = \sup_{x'}\bigl(l(x')-\lambda c(x,x')\bigr),
\end{equation}
which corresponds to \cref{eq:bm_dual} in the main text.

\subsection{Discretization and connection to optimal transport}

Assume $\hat P$ is supported on $n$ historical return vectors
$\{x_i\}_{i=1}^n$, with empirical weights $\hat p_i=1/n$, and that we restrict
the supremum in $h_\lambda$ to a finite set of candidate outcomes
$\{x'_j\}_{j=1}^m$ (in our experiments we take $m=n$ and $x'_j$ as stress
scenarios produced by analysts). For each $x_i$ we have
\[
  h_\lambda(x_i)
  = \max_{1\le j\le m}\bigl\{ l(x'_j)-\lambda c(x_i,x'_j)\bigr\}.
\]

Introduce a nonnegative transport plan $P\in\mathbb{R}_+^{n\times m}$ and
consider, for fixed $i$,
\[
  h^{\text{OT}}_\lambda(x_i)
  = \max_{P_{ij}\ge 0,\ \sum_j P_{ij}=\hat p_i}
    \sum_{j=1}^m P_{ij}\bigl(l(x'_j)-\lambda c(x_i,x'_j)\bigr).
\]
Because the objective is linear in $P_{ij}$ and the row sum is fixed, the
maximiser puts all mass on the $j$ achieving the maximum of
$l(x'_j)-\lambda c(x_i,x'_j)$, so
\[
  h^{\text{OT}}_\lambda(x_i)
  = \hat p_i \max_j\bigl(l(x'_j)-\lambda c(x_i,x'_j)\bigr)
  = \hat p_i h_\lambda(x_i).
\]
Summing over $i$ shows that $\sum_i \hat p_i h_\lambda(x_i)$ can be written as
a linear objective in $P$:
\[
  \sum_{i=1}^n \hat p_i h_\lambda(x_i)
  = \sum_{i=1}^n\sum_{j=1}^m P_{ij}\bigl(l(x'_j)-\lambda c(x_i,x'_j)\bigr),
\]
subject to the marginal constraint $\sum_j P_{ij}=\hat p_i$. If we introduce a
target marginal $b\in\mathbb{R}^m_+$ and enforce
$\sum_i P_{ij}=b_j$, this becomes a standard OT coupling with feasible set
\[
  \mathcal{U} = \Bigl\{ P\ge 0 : P\mathbf{1}=a,\ P^\top\mathbf{1}=b\Bigr\},
  \quad a_i = \hat p_i.
\]

Up to a sign flip, we obtain the discrete OT problem
\[
  \min_{P\in\mathcal{U}}
    \sum_{i,j} P_{ij}\bigl(\lambda c(x_i,x'_j)-l(x'_j)\bigr)
  = \min_{P\in\mathcal{U}} \langle P, C_\lambda\rangle,
\]
where $C_{\lambda,ij} = \lambda c(x_i,x'_j)-l(x'_j)$ is the combined cost
matrix. Adding entropic regularization yields
\[
  \min_{P\in\mathcal{U}} \left\{
    \langle P, C_\lambda\rangle
    + \varepsilon_{\scriptscriptstyle\mathrm{OT}}\sum_{i,j}P_{ij}(\log P_{ij}-1)\right\},
\]
which is precisely \cref{eq:finance_entropic_ot}.

\subsection{Enforcing the Wasserstein radius}

So far $\lambda$ is a free dual parameter. The original ambiguity set
constraint $W_c(P,\hat P)\le\delta$ corresponds, in the discrete setting, to a
constraint on the transport cost
\[
  \langle P, c\rangle
  = \sum_{i,j} P_{ij} c(x_i,x'_j) \le \delta.
\]
At optimality the worst-case distribution typically lies on the boundary,
$\langle P^\star, c\rangle = \delta$. Since the cost matrix $C_\lambda$ depends
on $\lambda$, the optimal plan $P^\star(\lambda)$ changes with $\lambda$ and we
can search for $\lambda^\star$ such that
$\langle P^\star(\lambda^\star),c\rangle = \delta$. In practice we perform:

\begin{enumerate}
  \item Fix $\lambda$ and solve \cref{eq:finance_entropic_ot} with
    (federated) Sinkhorn to obtain $P^\star(\lambda)$.
  \item Compute $g(\lambda) = \langle P^\star(\lambda),c\rangle-\delta$.
  \item Update $\lambda$ using a 1D root-finding method (e.g., bisection or
    Newton) until $|g(\lambda)|$ is below a tolerance.
\end{enumerate}

The output is $(\lambda^\star, P^\star)$, and the corresponding worst-case
loss is
\[
  \rho_{\text{worst}}
  = \mathbb{E}_{P^\star}[l(X)]
  = \sum_{i,j} P^\star_{ij} l(x'_j),
\]
which matches the dual value
$\lambda^\star\delta + \sum_i \hat p_i h_{\lambda^\star}(x_i)$.

\subsection{Toy portfolio example: numerical details}

We now give the details underlying the toy experiment summarized in~\ref{sec:numt}. Consider three technology stocks with one-day
historical returns (in percent)
\[
  x = [-0.51,\ -0.66,\ 4.34],
\]
weights
\[
  w = \Bigl[\frac{2}{5},\ \frac{1}{10},\ \frac{1}{2}\Bigr],
\]
and analyst-predicted next-day returns
\[
  x' = [0.43,\ -0.8,\ 3.86].
\]
We take $l(x)=w^\top x$ and a squared Euclidean cost
$c(x_i,x'_j)=\|x_i-x'_j\|_2^2$, leading to a combined cost
$C_{\lambda,ij}=\lambda c(x_i,x'_j)-l(x'_j)$.

To construct OT marginals, we first shift $x$ and $x'$ by a constant $k$ so
that all entries are positive, with
\[
  k = \max(|\min(x)|,|\min(x')|)+0.01 = 0.81.
\]
This yields
\[
  x_{\text{shifted}} = x + k = [0.30, 0.15, 5.15], \qquad
  x'_{\text{shifted}} = x' + k = [1.24, 0.01, 4.67].
\]
We then normalize:
\[
  a_i = \frac{x_{\text{shifted},i}}{\sum_j x_{\text{shifted},j}},\qquad
  b_j = \frac{x'_{\text{shifted},j}}{\sum_j x'_{\text{shifted},j}}.
\]
For the hyperparameters we choose
$\lambda=0.1$, $\delta=0.01$, and entropic regularization
$\varepsilon_{\scriptscriptstyle\mathrm{OT}}=0.01$.

The cost matrix $C$ used in the experiment is
\[
C = 
\begin{bmatrix}
0.164 & 0.163 & 0.214 \\
0.163 & 0.161 & 0.232 \\
0.214 & 0.232 & 0.163
\end{bmatrix}.
\]
We simulate a federated setting with three offices,
each holding one row (and the corresponding column) of $C$ and the local
components of the marginals and portfolio weights. Applying the federated
Sinkhorn solver with the synchronous all‑to‑all, synchronous star, and
asynchronous all‑to‑all communication patterns yields the optimal coupling
\[
P^{\star} =
\begin{bmatrix}
1.40 \times 10^{-1} & 6.78 \times 10^{-4} & 7.45 \times 10^{-38} \\
6.94 \times 10^{-2} & 1.02 \times 10^{-3} & 1.19 \times 10^{-46} \\
3.95 \times 10^{-8} & 6.19 \times 10^{-19} & 7.89 \times 10^{-1}
\end{bmatrix}
\]
(up to numerical precision). The corresponding worst-case expected loss is
\[
  \rho_{\text{worst}}
  = \sum_{i,j} P^\star_{ij} l(x'_j) \approx -0.48,
\]
i.e., approximately a $48\%$ one-day loss under the model.

Figure~\ref{fig:example_cvg} in the main text reports the marginal error
versus time for the three communication patterns. All runs converge in less
than $0.5$\,s on our hardware, and they yield essentially identical values of
$\rho_{\text{worst}}$, confirming that the federated implementation reproduces
the Blanchet--Murthy risk in this toy setting.

\end{document}